\documentclass[twocolumn, twocolappendix]{aastex631}

\usepackage{amsmath} 

\usepackage{rotating}              

\PassOptionsToPackage{hypertexnames=false}{hyperref}

\shorttitle{AASTeX v6.3.1 Sample article}
\shortauthors{Zeng et al.}
\graphicspath{{./}{figures/}}

\defcitealias{paperiv}{Paper IV}

\begin{document}

\title{Deep Submillimeter and Radio Observations in the SSA22 Field. III.
Multiwavelength Identifications and Properties of the 850 $\micron$-selected Submillimeter Galaxies}

\author[0009-0000-6036-7320]{Xin Zeng}
\affiliation{Purple Mountain Observatory, Chinese Academy of Sciences, Nanjing 210023, China}
\affiliation{School of Astronomy and Space Sciences, University of Science and Technology of China, Hefei 230026, China}

\author[0000-0003-3139-2724]{Yiping Ao}
\correspondingauthor{Yiping Ao} 
\email{ypao@pmo.ac.cn}
\affiliation{Purple Mountain Observatory, Chinese Academy of Sciences, Nanjing 210023, China}
\affiliation{School of Astronomy and Space Sciences, University of Science and Technology of China, Hefei 230026, China}

\author{Zongfei Lyu}
\affiliation{Purple Mountain Observatory, Chinese Academy of Sciences, Nanjing 210023, China}
\affiliation{School of Astronomy and Space Sciences, University of Science and Technology of China, Hefei 230026, China}

\author[0000-0001-9773-7479]{Daizhong Liu}
\affiliation{Purple Mountain Observatory, Chinese Academy of Sciences, Nanjing 210023, China}
\affiliation{School of Astronomy and Space Sciences, University of Science and Technology of China, Hefei 230026, China}

\author[0000-0002-3331-9590]{Emanuele Daddi}
\affiliation{Université Paris-Saclay, Université Paris Cité, CEA, CNRS, AIM, 91191 Gif-sur-Yvette, France}

\author[0000-0003-4985-0201]{Ken Mawatari}
\affiliation{Waseda Research Institute for Science and Engineering, Faculty of Science and Engineering, Waseda University, 3-4-1 Okubo, Shinjuku, Tokyo 169-8555, Japan}
\affiliation{Department of Pure and Applied Physics, School of Advanced Science and Engineering, Faculty of Science and Engineering, Waseda University, 3-4-1 Okubo, Shinjuku, Tokyo 169-8555, Japan}

\author[0000-0003-1937-0573]{Hideki Umehata}
\affiliation{Department of Physics, Graduate School of Science, Nagoya University, Furocho, Chikusa, Nagoya 464-8602, Japan}
\affiliation{Institute for Advanced Research, Nagoya University, Furocho, Chikusa, Nagoya 464-8602, Japan}



\begin{abstract}
We present a multiwavelength analysis of 850 $\micron$-selected submillimeter galaxies (SMGs; deblended S$_{\rm 850} \gtrsim$ 1 mJy) in the SSA22 field, using JCMT/SCUBA-2 data reaching $\sigma_{850} \sim$ 0.80 mJy beam$^{-1}$.
We identify 248 SMG candidates for 192 SCUBA-2 sources, with a multiplicity rate of $\sim$ 26\% that increases with flux.
After quality cuts based on SED fitting reliability, our final sample contains 221 SMGs associated with 186 SCUBA-2 sources.
The SSA22 SMGs have a median infrared luminosity of (2.25 $\pm$ 0.25) $\times$10$^{12}$ L$_{\odot}$, with $\sim$ 63\% ($\sim$ 8\%)  of the sample classified as ULIRGs (HLIRGs).
The median redshift is $z = 2.00 \pm 0.08$ (higher, $\sim 2.20$, for optically faint galaxies). 
SMGs trace large-scale structure, showing a significant overdensity at $z \sim 3.09$.
The comoving SMG number density rises by a factor of $\sim 6$ at $z \lesssim 4$, plateauing at $\sim 1.8-3.2\times10^{-5}$ cMpc$^{-3}$ over $z \sim$ 1-3 (including the overdensity).
The sample has median stellar mass $\sim (1.55\pm0.22)\times10^{11}$ M$_\odot$, SFR $\sim 166\pm25$ M$_\odot$ yr$^{-1}$, dust mass $\sim (1.95\pm0.14)\times10^{9}$ M$_{\odot}$, and $A_V \sim 3.09\pm0.07$ mag.
We detect a clear “downsizing” signature: $z \lesssim 2$, massive SMGs quench, while lower-mass systems dominate later star formation. 
The dust-to-stellar mass ratio, median $(1.4 \pm 0.2) \times 10^{-2}$, evolves significantly below $z \sim 2$.
Although $A_V$ remains stable over $z \sim 1-4$, combined literature data suggest a sharp drop at $z > 4$, indicating a transition to less obscured star formation likely due to low metal enrichment in the early universe.

\end{abstract}



\keywords{Submillimeter astronomy (1647), High-redshift galaxies (734), Galaxy properties (615), Ultraluminous infrared galaxies (1735)}




\section{Introduction}
\label{sec:intro}

Massive dusty star-forming galaxies represent crucial probes for addressing fundamental questions about our universe: when and how did most stars and galaxies form and evolve? How did cosmic metals enrich? What was the assemble history and dynamic evolution of the galaxy? How did the large-scale structure form and interact with galaxies?
Following cosmic reionization, dust-obscured star formation gradually dominated the cosmic star formation rate density, contributing approximately 50-80\% of the total \citep{bourne2017, dunlop2017, zavala2021}. However, such activity—hidden within dense dust clouds—often evades detection at ultraviolet and optical wavelengths \citep{wang2019, hyun2023}, making direct observation of dust heated by intense star formation a key method for its study \citep{hughes1998}.

Ultraviolet photons from massive stars are absorbed by dense dust and re-emitted in the far-infrared \citep[peak wavelength $\simeq$ 100 $\micron$;][]{lim2020}, resembling modified blackbody radiation \citep{dud2020}. Submillimeter observations lie on the Rayleigh–Jeans tail of this emission. As redshift increases, the observed band shifts toward the peak, counteracting cosmological dimming via the negative $k$-correction \citep{blain2002, casey2014}. 
Longer wavelengths enhance this effect, leading to higher typical redshifts for SMGs selected at longer wavelengths \citep{chapin2009, chen2013a, roseboom2013, casey2014, zavala2014, wang2017}. This work focuses primarily on 850 $\micron$-selected SMGs.
Observational constraints and atmospheric windows have historically confined submillimeter surveys to around 850 $\micron$. At this wavelength, the negative $k$-correction nearly balances cosmic dimming, keeping the observed flux density relatively constant over $z \sim 1–6$ for galaxies of similar luminosity \citep{blain2002}. Thus, submillimeter surveys allow the construction of samples of dust-obscured star-forming galaxies across a wide cosmic timeframe \citep{dud2020}.

The study of SMGs has advanced considerably over the past three decades. Initial detections with JCMT/SCUBA in the late 1990s \citep{smail1997, barger1998, hughes1998, ivison1998, lilly1999} revealed a population of galaxies hidden by dust. Large-area surveys like the SCUBA-2 Cosmology Legacy Survey (S2CLS; \citealt{geach2017}) and the SCUBA-2 Large eXtragalactic Survey (S2LXS; \citealt{garratt2023}), covering 5-10 deg$^2$, along with deep fields such as STUDIES in COSMOS field \citep{wang2017, gao2024}, have expanded our understanding. 
Single-dish submillimeter/millimeter surveys are limited by confusion noise and sensitivity, typically resolving only 10–40\% of the cosmic infrared background (CIB) \citep{wang2017, gao2024}. 
Gravitational lensing has enabled the study of faint SMGs \citep{hsu2016}, with targeted lensing-field surveys resolving 40–60\% of the CIB—and in some cases, up to $\sim$80\% \citep{chen2013a, chen2013b}.
We anticipate that next-generation submillimeter facilities, such as the Xue-shan-mu-chang 15-meter SubMillimeter Telescope (XSMT-15m, \citealp{xsmt2025}), will improve constraints on both the cosmic infrared background and the cosmic star formation rate density (CSFRD).

Facilities like \textit{Spitzer}, \textit{Herschel}, IRAM 30-m, ASTE, APEX, and SPT have greatly contributed, though their coarse angular resolution \citep{hodge2020} has been a limitation. Interferometers like SMA, NOEMA, and especially ALMA have revolutionized the field with high resolution and sensitivity, enabling detailed studies of individual high-redshift SMGs, including their gas dynamics \citep{hodge2020}. \citet{umehata2017a,umehata2018, franco2018, hatsukade2018, gomez2022} mapped GOODS-S and SSA22 in a wide region utilizing ALMA, while the Ex-MORA survey has assembled the largest homogeneous ALMA continuum dataset to date in the COSMOS field \citep{casey2021, long2024}.
Typical ALMA surveys achieve CIB resolution fractions of $\gtrsim$ 40–50\% \citep{chen2023}, and—again leveraging gravitational lensing—the ALCS survey successfully resolved $\sim$ 80\% of the CIB \citep{fujimoto2024}.
The UDS and COSMOS fields host the largest samples of ALMA follow-up observations \citep{stach2019, dud2020, simpson2020, liu2019, adscheid2024}. Collectively, these efforts have expanded SMG studies in depth, breadth, and methodological sophistication.

These observations have revealed that SMGs are a heterogeneous population, exhibiting significant scatter relative to the star-forming main sequence \citep{shim2022}, yet they are consistently massive, dust-obscured star-forming galaxies. 850 $\micron$-selected SMGs typically possess stellar masses of $\sim 10^{10-12}$ M$_\odot$, with a median near $10^{11}$ M$_\odot$, and reside in massive dark matter halos ($\gtrsim 10^{13} h^{-1}$ M$_\odot$; \citealp{lim2020, stach2021}). 
Their star formation rates (SFRs) range from tens to thousands of solar masses per year. 
Sustaining such an active process requires substantial gas reservoirs, with typical SMGs hosting gas masses of $\sim 10^{11}$ M$_\odot$ and gas fractions $\gtrsim$ 0.5-0.9 \citep{miettinen2017, birkin2021, liao2024}, fueled either by major mergers or environmental accretion.
Their high redshift, dust obscuration, and intense star formation result in extremely red colors and high infrared luminosities $\gtrsim 10^{12-13}$ L$_\odot$, with cold dust temperatures of 20-40 K \citep{swinbank2014, dud2020}.
The fraction of SMGs hosting active galactic nuclei (AGNs) has been a subject of debate in earlier studies. Initial work suggested AGN fractions as high as 20–40\% \citep{alexander2005, pope2008, laird2010, georgantopoulos2011, johnson2013, wang2013}.  
However, studies in the ALMA era have revealed that these earlier estimates were likely inflated. More recent analyses indicate that the AGN fraction among SMGs is below 20\%, typically in the range of $\sim$ 15-20\% \citep{wang2013, an2019, stach2019, uematsu2025}.  
Moreover, brighter and more massive SMGs appear more likely to host AGNs \citep{wang2013, cowie2018, lim2020}. SMGs residing in high-redshift large-scale structures may also exhibit elevated AGN fractions \citep{umehata2015}. Additionally, SMGs hosting AGNs show a higher incidence of major mergers \citep{uematsu2025}, suggesting that major mergers play a key role in triggering black hole activity in these systems.
%
In some cases, SMGs are also identified as optically-selected quasars, where concurrent black hole growth and star formation drive rapid galaxy evolution \citep{alexander2005, alexander2005_nature}. These characteristics strongly suggest that SMGs are progenitors of present-day massive elliptical galaxies, which hosting supermassive black holes (SMBHs) at their centers \citep{genzel2003, simpson2017, lim2020b}.

Observations suggest that SMGs are often associated with overdense structures at high redshift ($z \gtrsim$ 1.5 ) \citep{men2013, dannerbauer2014, battaia2018, smail2024, zhang2024}, and broadly consistent with the general tendency for massive galaxies to reside in denser environments \citep{dressler1980}.
\citet{umehata2015} first demonstrated—free from source confusion—that SMGs anchor galaxy overdensities within cosmic web filaments, and \citet{umehata2019} later revealed the co-evolution of galaxies, gas, and black holes within these cosmic structures.
Thus, SMGs have been considered potential tracers of overdense regions \citep{tamura2009, umehata2019}, though their utility may decline below $z \sim 2.5$ due to ``downsizing" and decreasing baryon conversion efficiency \citep{chapman2009, miller2015, miller2018}, and they may incompletely trace the most massive overdensities \citep{miller2015}.

The formation mechanisms of SMGs are debated. Early views, influenced by their resemblance to scaled-up local ULIRGs, favored major gas-rich mergers \citep{swinbank2014, hyun2023}, supported by morphological, kinematic, and simulation studies \citep{toft2014}. 
Alternatively, some argue that mergers may be secondary \citep{daddi2007}. 
Many SMGs lie on the high-mass end of the star-forming main sequence \citep{cunha2015, dunlop2017, zavala2018, lim2020}. Recent ALMA and JWST studies find that $\gtrsim 20-40\%$ of SMGs possess undisturbed extended disks \citep{gillman2023, gillman2024, bail2024, chan2025, hodge2025, umehata2025, zhang2025}, suggesting that internal instabilities or minor perturbations can trigger intense star formation and dust production in isolated disks \citep{gillman2024}. Numerical models also support sustained starbursts in gas-rich disks \citep{swinbank2014}.

The S2CLS is a major JCMT/SCUBA-2 program that mapped seven extragalactic deep fields, covering a total area $\sim$ 5 deg$^2$ and a depth of $\sigma_{850} \sim 1.2$ mJy \citep{geach2017}. 
SCUBA-2 sources in fields such as COSMOS \citep{an2019, simpson2019, simpson2020}, UDS \citep{stach2018,stach2019,dud2020}, EGS \citep{zavala2018}, and NEP \citep{shim2022} have been extensively analyzed.
As one of the S2CLS target fields, SSA22 field was well observed; however, to better exploit its potential for studying SMGs, we obtained deeper SCUBA-2 integrations toward SSA22 and complemented them with coordinated VLA 3 GHz observations \citep{ao2017, zeng2024}.
The SSA22 field has been extensively studied over the past two decades across various wavelengths and cosmic structures \citep{steidel1998, steidel2000, matsuda2005, lehmer2009a, lehmer2009b, webb2009, kubo2013, kubo2016, umehata2014, umehata2019, umehata2025, kato2016, ao2017,  cooper2022, zeng2024, huang2025}, amassing rich multiwavelength data from X-ray to radio. It offers a unique laboratory for probing both early galaxy evolution and the assembly of cosmic web structures.

In this paper, we combine 850 $\micron$ and 3 GHz data with extensive multi-wavelength ancillary datasets to identify SMG counterparts, perform spectral energy distribution (SED) fitting, and investigate their physical properties.
A key challenge is the coarse resolution of single-dish submillimeter telescopes like SCUBA-2, requiring counterpart identification at higher-resolution. Although bands with positive $k$-corrections and varying depths can miss higher-redshift or warmer SMGs \citep{swinbank2014}, they successfully identify the majority within the expected redshift range. Radio and 24 $\micron$ identifications are reliable and widely used \citep{chen2016}, although these methods still miss a fraction of real SMGs \citep{hodge2013, an2018}. The ADF22 survey \citep{umehata2017a, umehata2018} provides additional counterparts for our sample.
To enhance the completeness and maximize the potential of multiwavelength data, IRAC 8 $\micron$ sources have also been widely utilized \citep{ashby2006, zavala2018}. Numerous studies employ various optical/near-infrared (NIR) wavelengths \citep{smail1999, smail2002, frayer2004} to identify high-redshift extremely red galaxies (e.g., HIEROs: \citet{wang2012}; KIEROs: \citet{wang2016}; OIRTC: \citet{chen2016}).

Far-infrared (FIR) and millimeter data also suffer from source confusion due to large beam sizes.  A critical step in deblending is constructing a prior source catalog that is both complete and optimally dense \citep{liu2018}. We use the robust ``super-deblending" technique \citep{liu2018, jin2018}, which employs SED-based flux predictions to create appropriate prior lists for each band and estimates deblended flux uncertainties via Monte Carlo simulations.

We perform SED modeling using \texttt{CIGALE}, a modular code that combines physical components. By exploring a parameter grid and enforcing energy balance, whereby dust reprocesses ultra-violet (UV) and optical energy from stars and AGN into the far-infrared, it constructs possible SEDs.
This full-SED fitting method more effectively breaks degeneracies in galaxy properties compared to using UV/optical or FIR data solely. Crucially, for SMGs lacking spectroscopic redshifts or detectable optical emission—sources often excluded in traditional analyses—full SED fitting provides robust constraints on physical properties and enables reliable photometric redshift estimation \citep{dud2020}.

This paper is structured as follows. In Section~\ref{sec:data}, we describe the submillimeter, VLA, optical/near-infrared, and ancillary data in the SSA22 field, along with data reduction and the super-deblending procedure. Section~\ref{sec:ids} details our counterpart identification methods, results, and SMG multiplicity. In Section~\ref{sec:sed-fitting}, we derive spectroscopic redshifts from a representative catalog \citep{mawatari2023}, photometric redshifts using \texttt{EAZY}, and simultaneously obtain SED-based redshifts and physical properties via \texttt{CIGALE} fitting. Finally, Section~\ref{sec:results} presents the statistical properties of SMGs in the SSA22 deep field.
Specifically, in Section~\ref{sec:redshift} we present the redshift distribution of our 850 $\micron$-selected SMG sample.
Section~\ref{sec:sfr_mstar} details the star formation rates, stellar masses, and mass-to-light ratios of these SMGs.
Section~\ref{sec:lir} shows the distribution of infrared luminosities, while Section~\ref{sec:dust} reports on dust masses and dust fractions.
In Section~\ref{sec:av}, we present the distribution of dust attenuation ($A_V$) and examine its relationship with optical colors.

We adopt the cosmological parameters: H$_0$ = 70 km$^{-1}$ s$^{-1}$ Mpc$^{-1}$, $\Omega_{\Lambda}$ = 0.70, $\Omega_m$ = 0.30. The article employs the \citet{chabrier2003} initial mass function (IMF).

\begin{figure}[ht!]
    \centering
    \begin{minipage}{\columnwidth}
    \includegraphics[width=\textwidth]{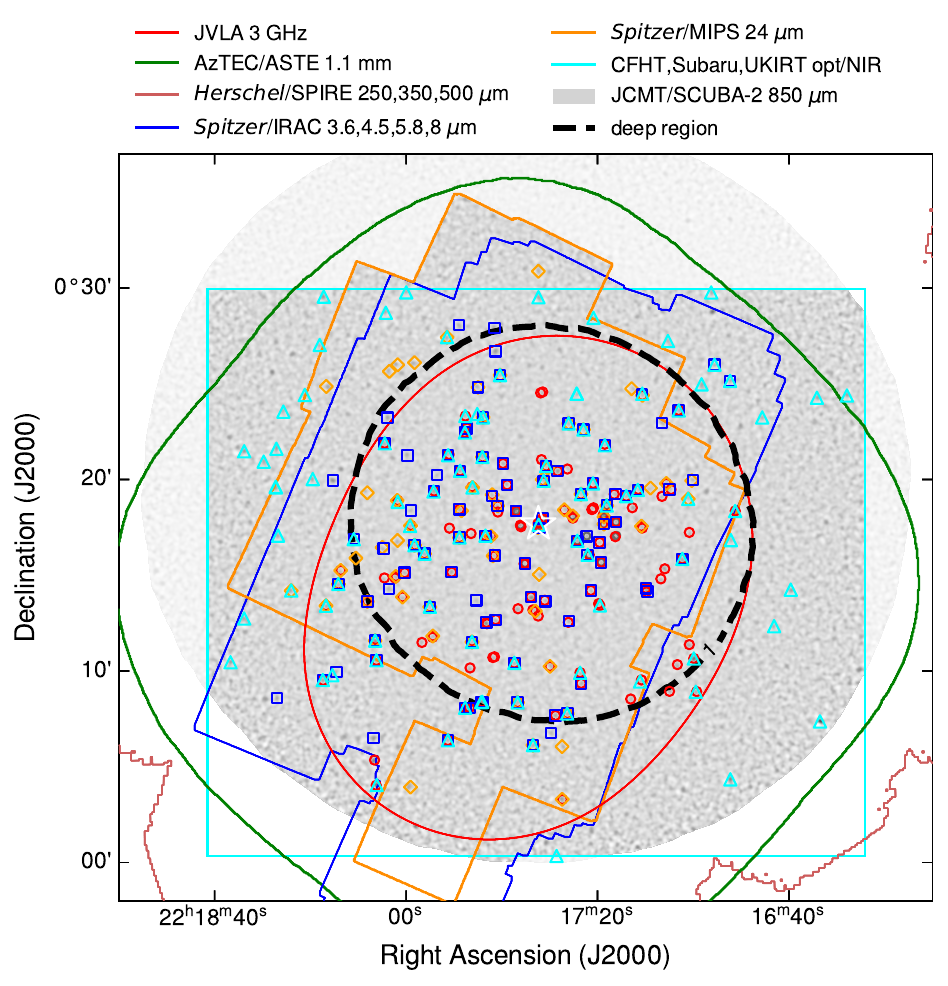}
    
    \caption{
    The gray circular base map represents the SCUBA-2 850 $\micron$ map ($\sim$ 0.34 deg$^2$), with the identified region shown in dark grey.
    The black dashed line demarcates the deep survey region ($\sigma_\text{850}$ $\leqslant$ 1 mJy).
    The multiwavelength data in the SSA22 field is depicted with colorful lines as illustrated in the legend.
    The colorful symbols represent SMGs identified using different methods, with each color matching that of the corresponding identification band coverage region.
    The white star marks the brightest SMGs, SSA22.0000, which resides at the center of SSA22 field.}
    \label{fig:fig2-0_multicontour}
  
    \end{minipage}
\end{figure}

\section{Data and Data Reduction} 
\label{sec:data}

In this section, we will introduce the multiwavelength data we have collected, spanning from optical to radio wavelengths. The majority of the data covers the main region of SSA22, specifically SSA22-Sb1 \citep{yamada2012}. The coverage areas for each wavelength are illustrated in Figure~\ref{fig:fig2-0_multicontour}, and details regarding the bands, resolutions, and observational depths for each dataset are provided in Table~\ref{tab:tab1_multiwave_depth}.


\subsection{SCUBA-2 850 $\micron$}
\label{sec:scuba2_map}

\begin{figure*}[ht!]
    \centering
    \includegraphics[width=0.95\textwidth]{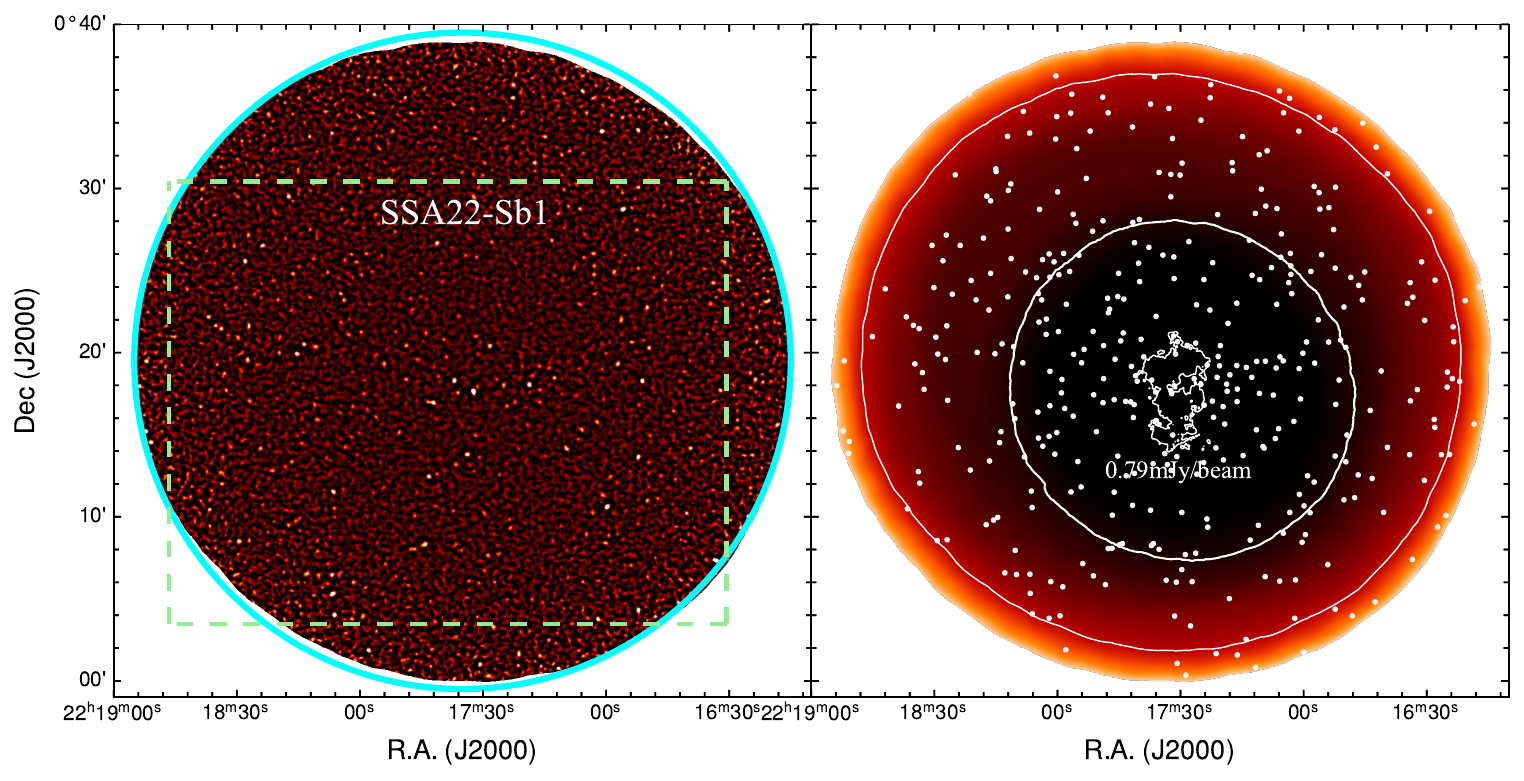}
    \caption{Left panel: The 850 $\micron$ flux density map of the SSA22 deep field observed with JCMT/SCUBA-2, covering an circular region of approximately 20$\arcmin$ in radius (cyan circle). The majority of the ``SSA22-Sb1'' region \citep{yamada2012} lies within the SCUBA-2 footprint (green rectangle). 
    Right panel: The corresponding rms noise map. Our analysis focuses on areas where the noise level is $\leqslant$ 2 mJy, encompassing $\sim 0.34$ deg$^2$. The deepest region reaches an rms of 0.79 mJy beam$^{-1}$, indicated by the innermost white isocontour, with additional contours at 1.0 and 1.5 mJy beam$^{-1}$. White dots mark submillimeter sources detected at $\geqslant$ 3.5$\sigma$ significance.}
    

    \label{fig:fig2-1_scuba2map}
\end{figure*}

Here, we provide a brief overview of the JCMT/SCUBA-2 850 $\mu$m observations. Detailed information on observations, data processing, and source number counts can be found in \citet{zeng2024}. 

The SSA22 field was part of the S2CLS \citep{geach2017}. We conducted additional exposure in the SSA22a region \citep{ao2017}, which is the core zone of the SSA22 protocluster \citep{steidel1998}. The total observations expended times of 91 hr and covered an area of 0.34 deg$^2$, reaching a noise level of 0.79 mJy beam$^{-1}$ in the overlapped area, making it the deepest map at 850 $\micron$ in the SSA22 field (see Figure~\ref{fig:fig2-1_scuba2map}).

We utilize a stacked point-spread function (PSF) and a “top-down” iterative source extraction algorithm. 
Jackknife maps are constructed, and Monte Carlo simulations are employed to estimate the completeness and false detection rates.
Compared to the S2CLS SSA22 map, our deeper SCUBA-2 map reveals 212 new sources, of which 145 lie in the central region of the field (where $\sigma_\text{850}$ $<$ 1.2 mJy) and 122 have flux densities greater than 4 mJy. 
While most of these newly detected sources are at low signal-to-noise ratios (SNR $< 4\sigma$), we still recover 93 sources at $\geqslant 4\sigma$ significance. 
Accounting for the false detection rate \citep{zeng2024}, we estimate that approximately 170 of these new detections are authentic submillimeter sources.

Finally, we construct a catalog comprising 390 SCUBA-2 sources with significance $\geqslant$ 3.5$\sigma$\footnote{Although sources below 4$\sigma$ carry an elevated risk of being spurious, statistical considerations suggest that $\sim$ 60-80\% of them are likely real.} and flux densities $\gtrsim$ 2 mJy.




\subsection{JVLA 3 GHz} 

\begin{figure}
    \centering
    \begin{minipage}{\columnwidth}
    \includegraphics[width=\textwidth]{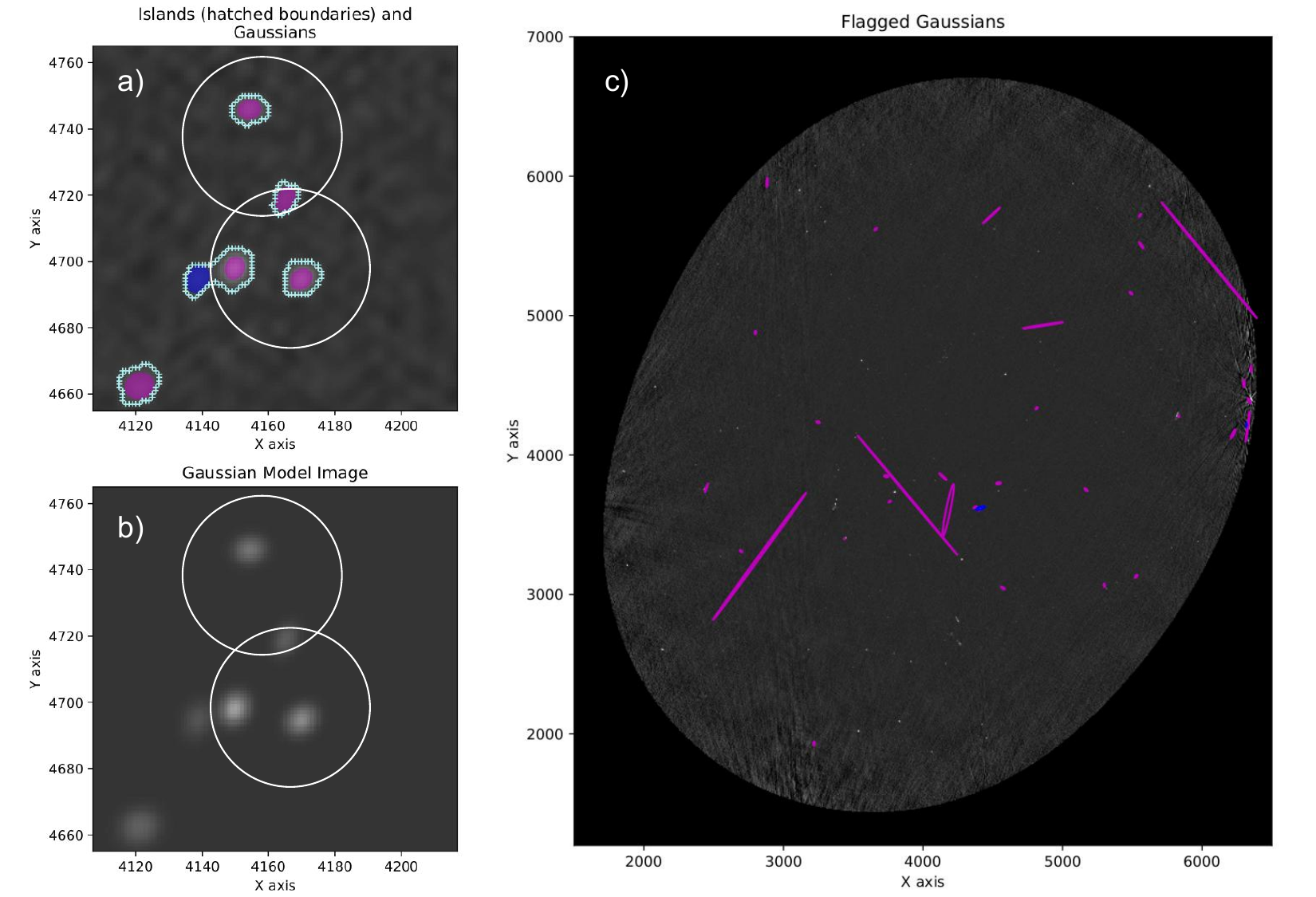}
    
    \caption{An example of the source extraction procedure implemented with \texttt{PYBDSF} on the JVLA 3 GHz map.
    Panel (a) shows a portion of the island map, centered on SSA22.0000 and SSA22.0070, with Gaussian fits overlaid.
    Each island, bounded by 3$\sigma$ pixels, is marked with "+" symbol. Pixels within islands were fitted with Gaussian functions, where colored ellipses represent Gaussian fits to peak pixels above 5$\sigma$ significance, with different colors indicating distinct sources. The two white circles indicate SCUBA-2 beam sizes, marking the extent of SCUBA-2 sources SSA22.0000 and SSA22.0070.
    Panel (b) displays the Gaussian modeling of radio sources performed by the pipeline.
    Panel (c) presents the output flagging map, where sources with Gaussian FWHM values exceeding their island size when multiplied by the \textit{flag\_maxsize\_fwhm} parameter (default=0.5) are flagged and subsequently excluded from the final catalog.
    }
    \label{fig:fig2-2_pybdsf}
    
    \end{minipage}
\end{figure}

Around our SCUBA-2 observations, we conducted JVLA 3 GHz radio band observations with B-configuration at S-band (2–4 GHz) in 2015 and 2016 \citep{ao2017}. The observations combined three pointings, and the primary beam of a single observation is about 15$\arcmin$. The final mosaic of images achieved an sensitivity of 1.5 $\mu$Jy beam$^{-1}$ before primary beam correction, with an angular resolution of 2$\farcs$3 $\times$ 2$\farcs$0.

We used the Python Blob Detection and Source Finder software \citep[\texttt{PyBDSF v1.10.3};][]{mohan2015} to detect and extract the radio sources from the JVLA 3 GHz map. Using the flux and root-mean-square (rms) maps after primary beam correction, we applied a sigma clipping threshold to extract the highly significant sources. Pixels above 3 times the background rms were identified as an island, and multiple Gaussians were fitted to the continuous pixels above detection threshold of 5$\sigma$ within each island 
(Figure~\ref{fig:fig2-2_pybdsf} panel (a) and (b)).
If all pixel values between two peak pixels exceed 3$\sigma$, and the connecting line length is less than half the sum of their FWHM values, these peak pixels are grouped as a single source and modeled with a Gaussian profile. 		The total flux of each source was estimated by integrating all pixel values within the Gaussian, and the positions of the sources were computed using moment analysis \citep{andrade2019, vlugt2021}. 

Considering that SMGs are generally at high redshifts and have relatively compact starburst regions \citep{simpson2015}, we did not consider extended emissions. The parameter \textit{flag\_maxsize\_fwhm} was set to default value of 0.5, meaning any fitted Gaussian profile with an FWHM exceeding half the island size would be flagged\footnote{https://pybdsf.readthedocs.io/en/latest/process\_image.html} (Figure~\ref{fig:fig2-2_pybdsf} panel (c)).

Additionally, we observed artifacts around some bright sources caused by projection effects, contamination from nearby faint sources, or noise. These appeared as unnatural extended shadows around high-significance detections. 

After visual inspection of the modeled images and flagged sources (Figure~\ref{fig:fig2-2_pybdsf} panel (b),(c)), we excluded both the flagged sources and two spurious detections, resulting in a total of 1989 radio sources with a significance $\textgreater$ $\sigma$, which were later used to identify deblended SMGs.


\subsection{Optical/Near-infrared}
\label{sec:opt_nir}

\begin{figure}
    \centering
    \begin{minipage}{\columnwidth}
    \includegraphics[width=\textwidth]{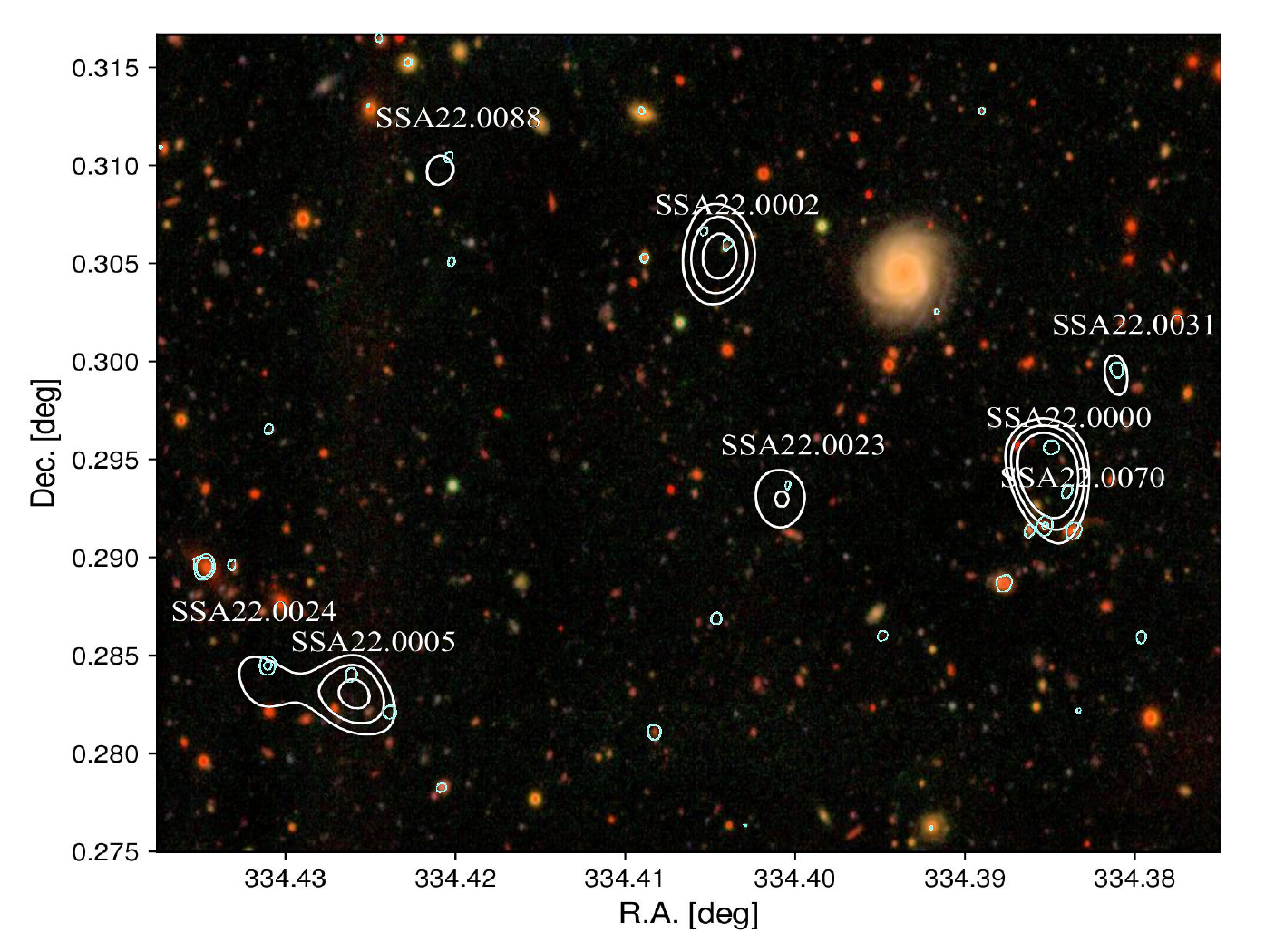}
    \caption{A window within the central region of the SSA22 deep field.
    The base image displays the false-color map composed of g-r-i bands, with white contours indicating submillimeter emission at 4.5, 7, 10$\sigma$ significance, and cyan contours showing 3 GHz emission at 5 and 30$\sigma$ levels.
    Due to the tight correlation between infrared and radio emission, submillimeter sources are commonly associated with radio counterparts. 
    Robust SMG identifications typically appear as red point sources or are undetected in the optical/near-infrared bands. }
    \label{fig:fig2-3-1}
    \end{minipage}
\end{figure}

\begin{figure}
    \centering
    \begin{minipage}{\columnwidth}
    \includegraphics[width=\textwidth]{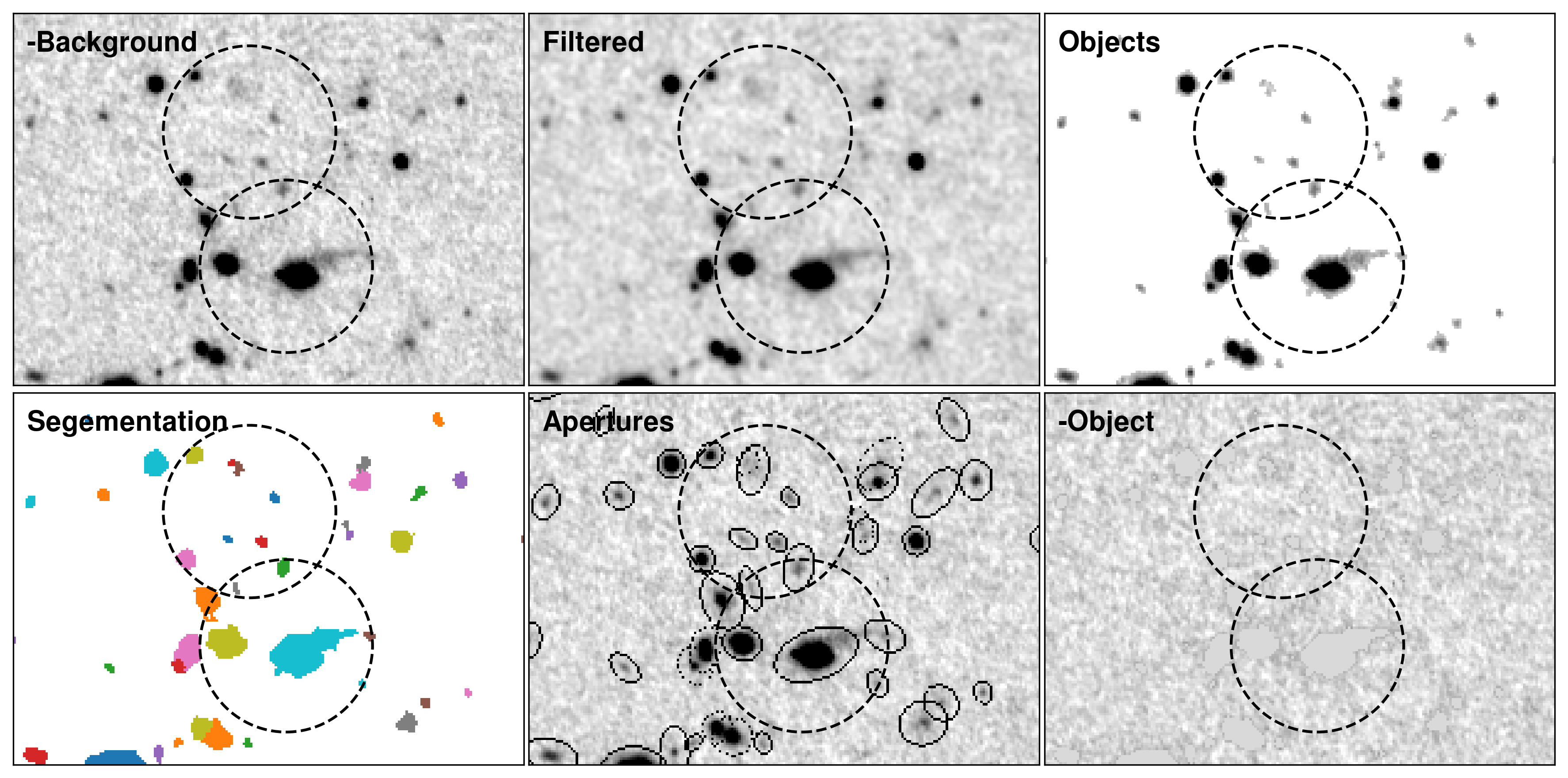}
    \caption{An example of the source extraction procedure implemented with \texttt{SExtractor},  centered on SSA22.0000 and SSA22.0070. 
    The two dashed circles indicate the SCUBA-2 beams for these sources, each with a radius of 7$\arcsec$.
    The panels show, form left to right and top to bottom: the background map, the filtered image, the detection image, the segmentation map, the photometric aperture, and the residual image after source subtraction.}
    \label{fig:fig2-3-2}
    \end{minipage}
\end{figure}

The optical and near-infrared (opt/NIR) multi-band images are centered on the Lyman-$\alpha$ emitters (LAEs) overdensity peak region \citep{steidel1998, steidel2000, hayashino2004, matsuda2005, yamada2012} (Figure~\ref{fig:fig2-3-1}). For a detailed description and data processing of the dataset, please refer to \citet{mawatari2023}.

The dataset includes the following multi-band data: the CFHT/MegaPrime \citep{boulade2003} \textit{u*}-band data; the Subaru/Suprime-Cam \citep{miyazaki2002} \textit{B, V, R, i$^{\prime}$, z$^{\prime}$} band images \citep{hayashino2004, matsuda2004}; the Subaru/HSC \citep{miyazaki2012, miyazaki2018} \textit{g, r, i, z, Y} band images collected from the Subaru Strategic Program \citep[HSC-SSP;][]{aihara2018}; and the Subaru/MOIRCS \citep{ichikawa2006} \textit{J, H, Ks} band data \citep{uchimoto2008, uchimoto2012}. Additionally, the dataset includes the UKIRT/WFCAM \citep{casali2007} \textit{J, K} band data from the deep extragalactic survey, which was part of the UKIRT Infrared Deep Sky Survey \citep[UKIDSS/DXS;][]{lawrence2007}. The two groups of near-infrared bands are similar, but the footprints of WFCAM are more comprehensive, and the MOIRCS bands are deeper.


We used \texttt{SExtractor} \citep{bertin1996} for source extraction and photometry on the optical and near-infrared images. Only a brief description is provided here; more detailed procedures are given in the Appendix~\ref{sec:app_opt_reduction}.  
The image is divided into a grid, and the median and standard deviation of the pixel value distribution are computed for each cell to estimate the local background and RMS. After smoothing and applying spline interpolation, a continuous model of the sky background is constructed and subtracted from the original image, while the resulting RMS map is used for source detection via sigma clipping.
Signals are identified as potential sources only if they exceed twice the background noise level and contain at least five connected pixels \citep{hayashino2004, umehata2014}. 
\texttt{SExtractor} can also deblends overlapping sources by estimating the contribution from neighboring objects—a step particularly useful for faint sources contaminated by bright foreground objects.  
We adopted the Kron aperture (MAG\_AUTO) for automatic photometry to measure source fluxes \citep{uchimoto2012}, which is regarded as one of the most accurate, adaptive, and robust flux estimation methods and is widely used in extragalactic surveys \citep{bertin1996}.
\texttt{SExtractor} further refines photometry by estimating the mean local background residual around each signal based on neighboring pixels. 
The final source catalog is restricted to detections at $\geqslant 3\sigma$ significance.
The diagnostic images for each procedural step are presented in Figure~\ref{fig:fig2-3-2}.

The photometry was corrected for Galactic extinction, with \textit{$R_{V}$} = 3.1 applied.
The color excesses at each position were obtained using the python package \texttt{dustmap}\footnote{https://dustmaps.readthedocs.io/en/latest/index.html} \citep{green2018}, which provides 2D reddening maps of interstellar dust based on the ``SFD model" \citep{schlegel1998}, with a factor of 0.86 also applied to the reddening values \citep{schlafly2011}.
The extinction curves derived by \citet{fitzpatrick1999} and improved by \citet{indebetouw2005} in the infrared were selected to derive the extinction at each band \citep{rodrigo2024}\footnote{http://svo2.cab.inta-csic.es/svo/theory/fps3/index.php?mode=browse}.

In Section~\ref{sec:irac8_ikcolor_ids}, We identified galaxies with extreme red colors: $i-K$ $\textgreater$ 2 as potential counterparts to SMGs. 
The color measurement was conducted on PSF-matched images that were smoothed to the largest PSF within the dataset, all having a Full Width at Half Maximum (FWHM) of approximately 1.1$\arcsec$.		
Given the reduced impact of dust extinction in the near-infrared bands, we employed a 2.2$\arcsec$ aperture for photometry on the $K$-band images (WFCAM/$K$ and MOIRCS/$Ks$ bands), while processing the Scam $i^{\prime}$ band using the dual mode of \texttt{SExtractor} with identical configuration. To ensure the reliability of this approach, we focused on sources detected at 4$\sigma$ significance in the $K$ band, simultaneously reducing the spatial density of potential candidates.

Finally, we estimate the depth of the opt/NIR imaging data.
However, on small scales, pixel-to-pixel correlations may arise from either source confusion or systematic biases introduced during data reduction.
Similarly, on large scale, pixel correlations can result from imperfect background subtraction, contamination by extended sources, undetected faint sources, or systematic errors such as flat-fielding inaccuracies \citep{straatman2016}.
Hence, we perform aperture photometry 10000 times using a 2$\arcsec$ radius on background-subtracted and detected sources masked images in each band, generated by \texttt{SExtractor}.
The photometry values represent noise distribution, from which we calculate their standard deviation and convert it to magnitudes as the 1$\sigma$ depth of the images, referred to as the 1$\sigma$ limiting magnitudes \citep{matsuda2004, an2018, weaver2022}.
Based on this, we estimate the corresponding 3$\sigma$ depth of the images for each band, which are summarized in Table~\ref{tab:tab1_multiwave_depth}, along with additional multi-band data depths, resolutions.




\begin{deluxetable*}{ccccc}
\tablecaption{The summary of multiwavelength data, includes the image band, angular resolution, and 3$\sigma$ limiting magnitude.}
\label{tab:tab1_multiwave_depth}
\tablewidth{0pt}

\addtocounter{table}{0}    
\tablehead{
    \colhead{Instrument} & \colhead{Band} & \colhead{Angular resolution} & \colhead{5$\sigma$ limiting mag.}    & \colhead{Reference} \\
    \colhead{} & \colhead{} & \colhead{(arcsec)} & \colhead{(AB mag)} & \colhead{} }


\startdata
CFHT/MagePrime     & \textit{u$^{\ast}$}   & 1.04 & 26.8 & \citet{kousai2011}  \\
Subaru/Suprime-Cam & \textit{B}            & 0.78 & 27.0 & \citet{nakamura2011} \\ 
                   & \textit{V}            & 0.82 & 27.2 & \citet{nakamura2011} \\
                   & \textit{R}            & 1.08 & 27.1 & \citet{hayashino2004} \\
                   & \textit{i$^{\prime}$} & 0.76 & 26.6 & \citet{nakamura2011} \\
                   & \textit{z$^{\prime}$} & 0.76 & 26.1 & \citet{nakamura2011} \\
Subaru/HSC & \textit{g} & 0.63 & 26.2 & \citet{aihara2018} \\
           & \textit{r} & 0.74 & 25.6 & \citet{aihara2018} \\
           & \textit{i} & 0.63 & 25.6 & \citet{aihara2018} \\
           & \textit{z} & 0.84 & 24.7 & \citet{aihara2018} \\
           & \textit{Y} & 0.56 & 24.2 & \citet{aihara2018} \\
Subaru/MOIRCS &       \textit{J} & 0.68 & 24.6 & \citet{uchimoto2012} \\
              &       \textit{H} & 0.68 & 23.9 & \citet{uchimoto2012} \\
              & \textit{K$_{s}$} & 0.55 & 23.9 & \citet{uchimoto2012} \\
UKIRT/WFCAM   &       \textit{J} & 0.91 & 23.8 & \citet{lawrence2007} \\
              &       \textit{K} & 0.86 & 23.2 & \citet{lawrence2007} \\
\textit{Spitzer}/IRAC & 3.6 $\micron$ & 1.7 & 26.00/22.49 \tablenotemark{c} & \citet{seip, timlin2016} \\
                      & 4.5 $\micron$ & 1.7 & 25.24/22.56 \tablenotemark{c} & \citet{seip, timlin2016} \\
                      & 5.8 $\micron$ & 1.9 & 23.68 \tablenotemark{d} & \citet{seip} \\
                      & 8.0 $\micron$ & 2.0 & 22.08 \tablenotemark{d} & \citet{seip} \\
\textit{Spitzer}/MIPS &  24 $\micron$ & 5.9 & 45.6 $\mu$Jy beam$^{-1}$ \tablenotemark{d} & \citet{seip} \\
\textit{Herschel}/SPIRE & 250 $\micron$ & 17.9 & 2.6 mJy beam$^{-1}$ \tablenotemark{e} & \citet{hhli} \\
                        & 350 $\micron$ & 24.2 & 2.2 mJy beam$^{-1}$ \tablenotemark{e} & \citet{hhli} \\
                        & 500 $\micron$ & 35.4 & 2.7 mJy beam$^{-1}$ \tablenotemark{e} & \citet{hhli} \\
JCMT/SCUBA-2 & 850 $\micron$ & 13.9 & 0.8–2.0 mJy beam$^{-1}$ \tablenotemark{f} & \citet{zeng2024} \\
AzTEC/ASTE   & 1.1 mm        & 28   & 0.7–1.3 mJy beam$^{-1}$ \tablenotemark{f} & \citet{umehata2014} \\
ALMA & 1.14 mm & 0$\farcs$53 $\times$ 0$\farcs$52 \tablenotemark{g} &  75 $\mu$Jy beam$^{-1}$ & \citet{umehata2018} \\
JVLA & 3 GHz   & 2$\farcs$3  $\times$ 2$\farcs$0  \tablenotemark{h} & 1.5 $\mu$Jy beam$^{-1}$ \tablenotemark{i} & \citet{ao2017} \\         
\enddata

\tablenotetext{a}{ The sky-projected ACIS pixel size is $\approx$ 0.492 arcsec.}
\tablenotetext{b}{ The 3 count sensitivity limit.}
\tablenotetext{c}{ The mean 3$\sigma$ noise levels of SEPI and SpIES.}
\tablenotetext{d}{ The mean 3$\sigma$ noise level of SEPI.}
\tablenotetext{e}{ The median 1$\sigma$ depth of HHLI data.}
\tablenotetext{f}{ The 1$\sigma$ depth range of image.}
\tablenotetext{g}{ The resulting synthesized beam size of ALMA band 6.}
\tablenotetext{h}{ The final synthesized beam of VLA S-band.}
\tablenotetext{i}{ The typical sensitivity before primary beam correction.}

\end{deluxetable*}

\subsection{Ancillary Data}


\subsubsection{Mid-infrared}
The $\textit{Spizter}$ data \citep[see also][]{webb2009} from the Spitzer Enhanced Imaging Products \citep[SEIP;][]{seip} and the Spitzer IRAC Equatorial Survey \citep[SpIES;][]{timlin2016, spies} were utilized. 

The SEIP ``Super Mosaics'' were created by combining multiple observation requests, containing the four channels of IRAC (3.6, 4.5, 5.8, 8 $\micron$) and the 24 $\micron$ channel of MIPS. 	
This program produced a photometry source catalog for compact sources and implemented stringent criteria for the selection of extracted sources to ensure high reliability. 		
The mean 3$\sigma$ depths of the four IRAC band enhanced images were approximately 26.00, 25.24, 23.68, 22.08 mag, and the 5$\sigma$ depth for MIPS was approximately 19.20 mag (see Table~\ref{tab:tab1_multiwave_depth}),  whereas the SEIP images covered about half of our SCUBA-2 map footprint ($\sim$ 0.17 and 0.15 deg$^2$ for IRAC and MIPS). 

The SpIES is a large-area survey in the equator, fully covering the SSA22 field, thus serve as a supplement to the SEIP. 
However, the SpIES products only contain the IRAC 3.6 and 4.5 $\micron$ data, and the mapped sensitivities are much shallower than the former. The 3$\sigma$ noise levels for the 3.6 and 4.5 $\micron$ bands are approximately 22.49 and 22.56 AB mag. 

We utilized the primary fluxes of the SEIP and SpIES source list, which are the aperture fluxes in a 3.8$\arcsec$ diameter of IRAC sources applied aperture correction for point sources, and the PSF fitting fluxes of MIPS 24 $\micron$ sources. To ensure robust flux densities, we excluded sources that could be affected by nearby saturated sources or a nearby extended source and by soft saturation\footnote{https://irsa.ipac.caltech.edu/docs/knowledgebase/spitzer\_seip.html}. Additionally, SNR cuts were applied, with a SNR cut of $\geqslant$ 3 for IRAC and $\geqslant$ 5 for MIPS.  \citep[FLUX/FLUXERR;][]{timlin2016}.

\subsubsection{Far-infrared and Millimeter}
\label{sec:data_fir}
The $\textit{Herschel}$ data \citep{kato2016} from the Herschel High Level Images \citep[HHLI;][]{hhli}, a data subset of the Herschel Science Archive (HSA), included only SPIRE data in the SSA22 field with the median 1$\sigma$ sensitivities were 2.6, 2.2, and 2.7 mJy beam$^{-1}$ for 250, 350, and 500 $\micron$, respectively. Due to the large beam of the SPIRE, the three FIR images suffered from severe confusion noise and source blending. The flux densities of SPIRE sources were deblended by the ``super deblending'' method described in section~\ref{sec:super_deblend}, and the final 3$\sigma$ deblended sources were used for SED fitting.

\citet{tamura2009, umehata2014} successively conducted the AzTEC/ASTE 1.1 mm survey in the SSA22 field, with a total observation time of 74 hr. The resulting map covered an area of 0.27 deg$^2$, with the rms ranging from 0.7-1.3 mJy beam$^{-1}$. Considering the coarse resolution of 28$\arcsec$ of AzTEC, we also employed the same ``super-deblended'' program to deblend AzTEC sources to derive 1.1 mm fluxes for identifications for SED fitting. 

The ALMA deep field in SSA22 \citep[ADF22 survey;][]{umehata2017a, umehata2018} mosaicked a 20 arcmin$^2$ core region at 1.1 mm (band 6), composed of two contiguous observations. The total on-source observation time was approximately 13.1 hr, resulting in a map with a typical sensitivity level of approximately 0.07 mJy beam$^{-1}$ and resolutions of approximately 0.5$\arcsec$ to 1$\arcsec$. A total of 35 ALMA SMGs were detected at $\textgreater$ 5$\sigma$, and the high resolution and precise location of the ALMA observation will be used to identify the positions of the SCUBA-2 sources.

\subsection{``Super-deblending''} \label{sec:super_deblend}

\begin{figure*}
    \centering
    \includegraphics[width=0.95\textwidth]{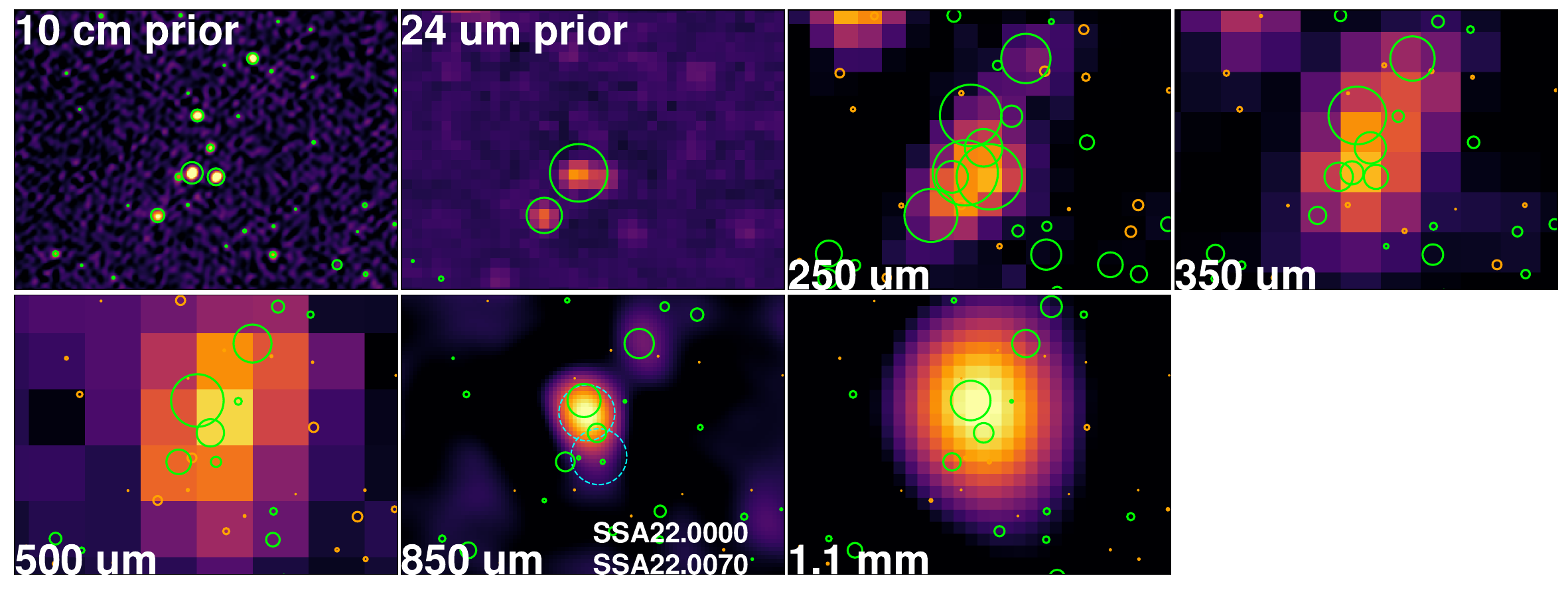}
    \caption{The multiwavelength deblending workflow, illustrated using sources SSA22.0000 and SSA22.0070 (cyan dashed circles) as an example, the most significant submillimeter sources in detection region.
    The first two panel sequentially display 3 GHz and 24 $\micron$  images, with prior sources marked by green circles.
    The algorithm processes images from shorter to longer wavelengths.
    At each steps, prior sources whose SED-predicted flux density ($f_{\text{SED}}+2\sigma_{\text{SED}}$) falls below the threshold for the current band are temporarily excluded from fitting catalog and labeled with orange circles.
    The circle sizes for both fitted and excluded sources scale with their flux density at each band, with brighter flux es represented by larger circles.}
    \label{fig:fig2-5-3_spdb_allband}
\end{figure*}

Accurate FIR to millimeter photometry is essential for measuring star formation rates and dust properties in galaxies, but coarse angular resolution ($\sim$ 14$\arcsec$ - 35$\arcsec$) in FIR/mm imaging leads to severe source confusion, especially in deep fields. 

We applied the ``super-deblending” technique \citep{liu2018, jin2018} to derive reliable deblended fluxes from 250 $\micron$ to 1.1 mm for SMGs in the SSA22 field. 
The deblending workflow is presented in Figure~\ref{fig:fig2-5-3_spdb_allband}.
We provide a introduction of the ``super-deblending" in the Appendix~\ref{sec:app_spdb}; for full technical details, please refer to \citet{liu2018, jin2018}.

The prior source catalog is constructed from relatively high SNR MIPS 24 $\micron$ and radio sources, and also includes lower-significance ($\geqslant$ 2.5$\sigma$) sources at higher redshifts ($z >$ 2), resulting in a final catalog of 2683 prior sources.

The deblending is performed individually for each band. A key aspect of the method is the use of SED-fitting predictions to optimize the prior source catalog, thereby enabling reliable flux extraction.
To obtain meaningful deblended fluxes, the source density within each FIR/mm beam must be appropriately controlled, ideally kept near or below one source per beam area. 
For each band, a flux threshold S$_\text{cut}$ is defined. Prior sources whose SED-predicted flux (accounting for fitting uncertainties) falls below the threshold considered negligible contributors at that band and are temporarily excluded. A blind extraction is also performed on the residual image to avoid missing any significant FIR emitters. 
The deblending proceeds sequentially from the shortest to the longest wavelength band. In each step, the deblended fluxes from previously processed bands are used to refine the priors for the current band, which is a crucial feature of the super-deblending method. Finally, Monte Carlo simulations are used to assess potential biases and apply flux corrections, representing a key advantage of the super-deblending approach.

\section{COUNTERPART IDENTIFICATION}
\label{sec:ids}

\begin{figure}
    \centering
    \includegraphics[width=0.45\textwidth]{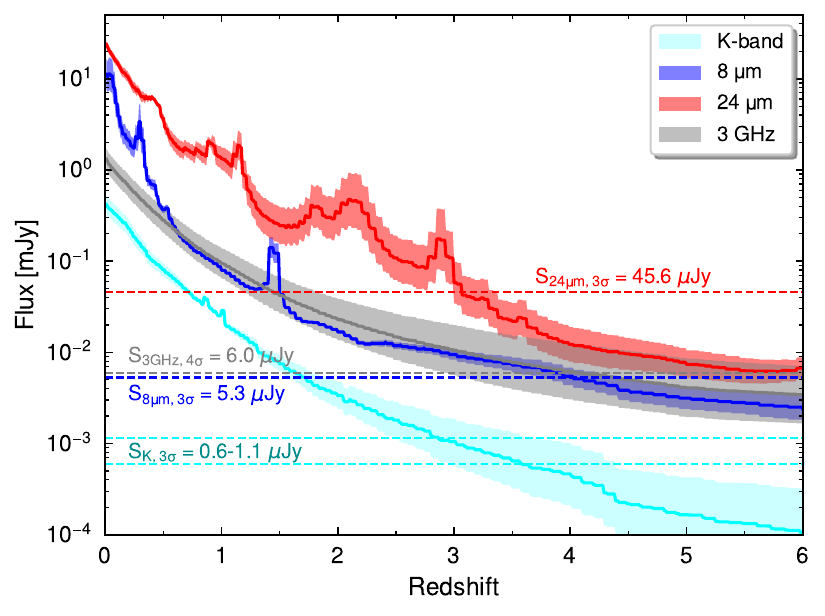}
    
    \caption{Predicted flux densities in key identification bands for a source marginally detected by SCUBA-2, as a function of redshift. Fluxes are derived from the average SED, with the shaded region indicating the 1$\sigma$ dispersion. Dashed horizontal lines mark the average depths of the respective surveys.
    The $MIPS$ 24 $\micron$ band reaches its identification limit at $z \sim 3$, which nonetheless covers the peak of the SMG redshift distribution. In contrast, the deeper IRAC 8 $\micron$ and VLA 3 GHz data can reliably identify SMGs up to $z \sim 4$, and—accounting for SED variations—may detect sources out to $z \sim 5$–6. Color-based identification, primarily limited by $K$-band depth, is effective up to $z \sim 3$–4.
    Because we adopt the flux level of a marginal SCUBA-2 detection as our baseline, most real SMGs will be detectable to even higher redshifts. Nevertheless, due to limitations in identification techniques and the heterogeneous SED properties of SMGs, our sample is likely incomplete at $z \gtrsim 4$.}
    
    \label{fig:fig3-0_fluxlimit}
\end{figure}

We primarily utilize four methods, namely VLA 3 GHz, $Spitzer$/MIPS 24 $\micron$, $Spitzer$/IRAC 8 $\micron$, and optical/NIR red-selected galaxies (hereafter ``$VLA$/radio'',``$MIPS24$'', ``$IRAC8$'', ``$iK\text{-}color$'' method), to search for potential counterpart galaxies of SCUBA-2 sources.

The radio and $MIPS24$ identification methods (Section~\ref{sec:radio+mips24}) are the most reliable and widely used approaches for pinpointing SMG counterparts \citep{barger2000, ivison2002, umehata2014, zavala2018, lim2020}, achieving positional accuracies of $\gtrsim$ 50–80\% \citep{hodge2013, chen2016, an2018}. However, their completeness is somewhat lower, typically $\sim$ 40–55\% \footnote{Notably, \citet{chen2016} point out that radio identification alone achieves a higher completeness of $\sim$ 80\%.}. 
ADF22 \citep{umehata2017a, umehata2018} contributed 5 additional SMG counterparts (two of which were also identified by the $IRAC8$ method).
In subsequent analyses, we designate sources identified through these methods as `MAIN' samples.
To fully exploit the potential of multiwavelength data, we also incorporated $IRAC8$ and $iK\text{-}color$ methods \citep{ashby2006, caputi2006,  schael2009, koprowski2015} (Section~\ref{sec:irac8_ikcolor_ids}). These approaches offer wider coverage and enable counterparts identification for marginal SCUBA-2 sources, which we classified as `SUPP'.
Section~\ref{sec:id_method} outlines the feasibility and reliability of each identification method.

We show in Figure~\ref{fig:fig3-0_fluxlimit} the predicted flux densities as a function of redshift in various identification bands for a source that is just detected by SCUBA-2 (at $0.8 \times 3.5$ mJy). 
Our identification methods are robust for sources out to at least $z \lesssim 3$–4.
Note that the majority of SMGs have 850 $\micron$ fluxes around $\sim$4 mJy—i.e., brighter than the fiducial detection threshold used here—their multi-wavelength counterparts are expected to be even more easily detectable, enabling reliable identification out to higher redshifts. Moreover, the intrinsic diversity of SMG SEDs further enhances the potential of our approach to detect dusty galaxies during the epoch ending reionization ($z \sim 5$–6).

Several studies have selected sole counterparts based on the minimum $p$-value criterion \citep{zavala2018}, demonstrating  that this approach typically recovers the majority of submillimeter flux \citep{simpson2015}. However, we adopt a  more conservative approach by retaining all potential counterparts.

\begin{figure*}
    \centering
    \includegraphics[width=0.9\textwidth]{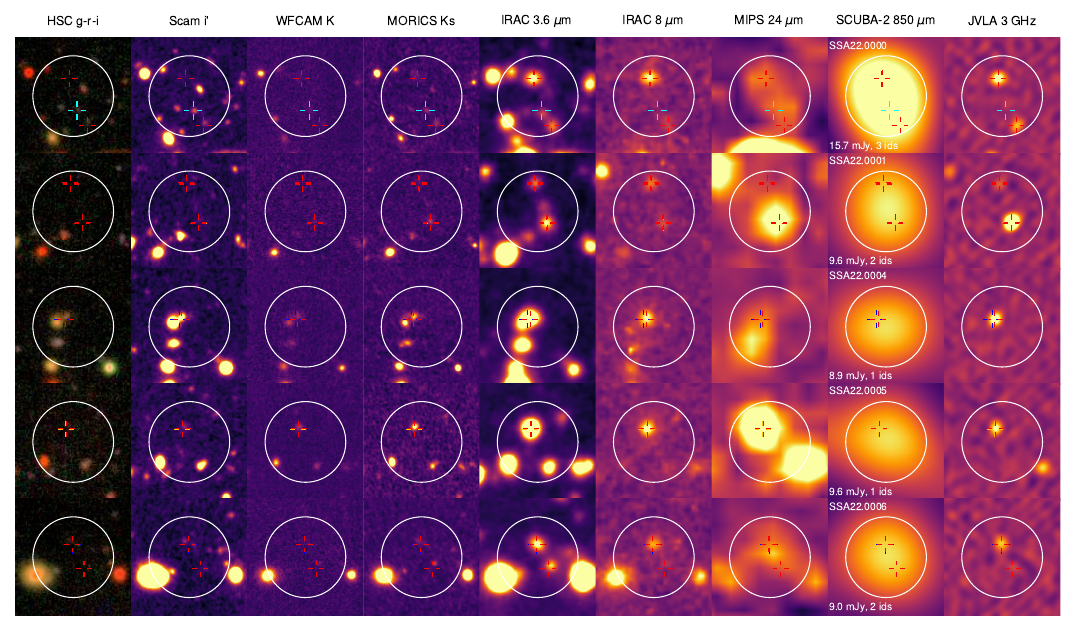}
    
    \caption{Multiwavelength images show the identified counterparts of SCUBA-2 sources, with column headers indicating the observational instrument and wavelength.
    Red crosses mark radio or ALMA counterparts, yellow crosses indicate SMGs identified via the ``$MIPS24$" method, while blue and cyan crosses denote sources found through the ``$IRAC8$" and ``$iK\text{-}color$" methods, respectively.
    Counterparts identified by different methods are considered a single source if their separation is less than 1.5$\arcsec$, accounting for the astrometric errors and source extraction uncertainties.
    White circles represent the SCUBA-2 beam are with a radius of 7$\arcsec$.
    The label in the upper left corner of each 850 $\micron$ panel gives the SCUBA-2 source name, while the lower left text lists the source's flux density and the number of identified counterparts.
    Complete images are provided in the supplementary material.    }
    
    \label{fig:fig3-0_ids}
\end{figure*}

By combining all identification methods, we matched 248 counterparts to 192 SCUBA-2 sources.
The vast majority ($\gtrsim 84\%$) of these identified galaxies have deblended 850 $\micron$ flux densities greater than 1 mJy.
Notably, for the 212 newly detected SCUBA-2 sources (Section~\ref{sec:scuba2_map}), we successfully identified 136 SMG counterparts for 101 SCUBA-2 sources, representing more than 50\% of the total confirmed counterparts.
Figure~\ref{fig:fig3-0_ids} presents representative examples of multiwavelength counterpart identification.
A more detailed breakdown of the identifications is provided in Section~\ref{sec:id_result}.

We consider these methods to be generally robust and represent our best-effort approach. However, we emphasize that they are not perfect: some genuine SMGs are inevitably missed, while a fraction of spurious sources may also be included.
We offer an in-depth discussion of the reliability of these identification methods and their potential limitations in Section~\ref{sec:caveat}.

Table~\ref{tab:tab2_photo_catalog} presents the photometric data for the first ten SMG sources in our sample. The complete catalog, available in the online supplementary material, includes continuum photometry spanning from the optical/NIR to the radio bands; here we list only the measured flux densities in each band.
Table~\ref{tab:tab3_catalog} lists the first ten entries of our identified source catalog, including information on the SCUBA-2 sources, their galaxy counterparts, and derived physical properties. 
When a source is identified by multiple methods, we report only the most reliable counterpart according to the following hierarchy: $VLA > MIPS24 > IRAC8 > iK\text{-}color$.
The two complete catalogs are available in the supplementary material.

\movetabledown=50mm
\begin{rotatetable*}

\begin{deluxetable*}{lccccccccccccccc }
\tablecaption{Photometry catalog of the SSA22 SMGs}
\label{tab:tab2_photo_catalog}
\tablewidth{0pt}

\addtocounter{table}{0}    
\tablehead{
    \colhead{id ID} & \colhead{id RA} & \colhead{id DEC} & \colhead{$i^{\prime}$} & \colhead{$Ks$} & \colhead{S$_\text{3.6\micron}$} &   \colhead{S$_\text{8\micron}$} &    \colhead{S$_\text{24\micron}$} &    \colhead{S$_\text{250\micron}$ \tablenotemark{a} } & \colhead{S$_\text{850\micron}$ \tablenotemark{a} } & \colhead{S$_\text{1100mm}$ \tablenotemark{a} } &             \colhead{S$_\text{3GHz}$} \\
    \colhead{} & \colhead{[deg]} & \colhead{[deg]} & \colhead{[$\mu$Jy]} & \colhead{[$\mu$Jy]} & \colhead{[$\mu$Jy]} & \colhead{[$\mu$Jy]} & \colhead{[mJy]} &    \colhead{[mJy]} & \colhead{[mJy]} & \colhead{[mJy]} &    \colhead{[$\mu$Jy]} \\
         }

\startdata
SSA22.0000a & 334.385 & 0.295 &  & 2.25$\pm$0.42 & 8.95$\pm$0.06 &  &  & 10.47$\pm$5.63 & 10.87$\pm$1.35 & 6.79$\pm$2.50 & 37.46$\pm$2.96 \\
SSA22.0000b & 334.385 & 0.294 & 0.33$\pm$0.01 & 1.56$\pm$0.32 &  &  &  & 0.69$\pm$5.49 & 7.15$\pm$1.81 & 4.75$\pm$2.20 &  \\
SSA22.0000c & 334.384 & 0.293 & 0.25$\pm$0.01 & 2.65$\pm$0.45 & 5.46$\pm$0.06 &  &  & 13.90$\pm$5.85 & 0.00$\pm$1.85 & 3.86$\pm$1.76 &  \\
SSA22.0001a & 334.328 & 0.301 & 0.10$\pm$0.01 & 4.13$\pm$0.55 & 7.65$\pm$0.06 &  & 0.22$\pm$0.02 & 21.27$\pm$6.25 & 7.17$\pm$0.92 & 3.44$\pm$1.23 & 93.31$\pm$3.13 \\
SSA22.0001b & 334.328 & 0.303 &  &  & 4.00$\pm$0.05 &  &  & 23.30$\pm$4.35 & 4.47$\pm$0.83 & 3.08$\pm$1.23 & 10.68$\pm$3.34 \\
SSA22.0002a & 334.404 & 0.306 & 2.67$\pm$0.01 & 3.61$\pm$0.50 & 10.07$\pm$0.06 & 29.97$\pm$1.78 &  & 7.39$\pm$10.80 &  &  & 21.45$\pm$4.17 \\
SSA22.0002b & 334.404 & 0.306 &  & 3.07$\pm$0.49 & 10.07$\pm$0.06 & 29.97$\pm$1.78 &  & 2.75$\pm$11.45 & 4.71$\pm$1.63 & 3.01$\pm$1.29 &  \\
SSA22.0002c & 334.403 & 0.305 & 0.23$\pm$0.02 & 2.62$\pm$0.48 & 3.83$\pm$0.05 &  &  & 13.29$\pm$5.50 & 2.30$\pm$1.13 & 3.11$\pm$1.56 &  \\
SSA22.0003 & 334.250 & 0.178 & 0.65$\pm$0.01 &  &  &  &  & 27.70$\pm$6.81 & 11.09$\pm$1.26 & 5.59$\pm$1.45 & 122.12$\pm$10.37 \\
SSA22.0004 & 334.391 & 0.231 &  & 13.71$\pm$0.95 & 21.44$\pm$0.06 & 39.77$\pm$1.78 &  & 15.54$\pm$4.92 & 7.24$\pm$1.15 & 6.44$\pm$1.21 & 39.84$\pm$2.36 \\
\enddata

\tablecomments{The first ten entries of the SSA22 SMG photometric source catalog; the full catalog is available in the supplementary material.}

\tablenotetext{a}{The far-infrared and (sub)millimeter flux densities are derived using the super-deblending technique, as described in Section~\ref{sec:data_fir} and Section~\ref{sec:super_deblend}.}

\end{deluxetable*}
\end{rotatetable*}

\movetabledown=30mm
\begin{rotatetable*}
\begin{deluxetable*}{lccccccccccc}
\tablecaption{Identifications and Properties catalog of the SSA22 SMGs}
\label{tab:tab3_catalog}

\tablewidth{0pt}

\addtocounter{table}{0}    

\tablehead{
    \colhead{id ID} & \colhead{S$_\text{850\micron,spdb}$ \tablenotemark{a} } & \colhead{id method \tablenotemark{b} } & \colhead{$p$-value} & \colhead{$z$} & \colhead{z-flag \tablenotemark{c} } & \colhead{log $M_{star}$} & \colhead{log SFR} & \colhead{log $L_{IR}$}   & \colhead{log $M_{dust}$} & \colhead{$A_V$} & \colhead{$\Delta$MS \tablenotemark{d} } \\  
    \colhead{} & \colhead{[mJy]} & \colhead{} & \colhead{} & \colhead{} & \colhead{} &    \colhead{[M$_{\odot}$]} & \colhead{[M$_{\odot}$ yr$^{-1}$]} &  \colhead{[L$_{\odot}$]} & \colhead{[M$_{\odot}$]} & \colhead{[mag]} & \colhead{} }

\startdata
SSA22.0000a & 10.87$^{+1.35}_{-1.35}$ & alma($VLA$) & 0.02 & 3.09$^{+0.00}_{-0.00}$ & spec & 11.93$^{+0.11}_{-0.11}$ & 2.84$^{+0.14}_{-0.14}$ & 12.95$^{+0.08}_{-0.08}$ & 9.78 & 3.97 & -0.27 \\
SSA22.0000b & 7.15$^{+1.81}_{-1.81}$ & $iK\text{-}color$ & 0.04 & 0.59$^{+0.26}_{-0.02}$ & eazy & 8.97$^{+0.51}_{-0.51}$ & 1.76$^{+0.06}_{-0.06}$ & 11.56$^{+0.06}_{-0.06}$ & 9.95 & 5.78 & 1.70 \\
SSA22.0000c & & alma & 0.05 & 3.09$^{+0.00}_{-0.00}$ & spec & 10.85$^{+0.13}_{-0.13}$ & 3.04$^{+0.14}_{-0.14}$ & 12.83$^{+0.14}_{-0.14}$ & 8.94 & 2.79 & 0.79 \\
SSA22.0001a & 7.17$^{+0.92}_{-0.92}$ & alma($VLA$) & 0.01 & 3.07$^{+0.00}_{-0.00}$ & spec & 10.71$^{+0.23}_{-0.23}$ & 3.25$^{+0.05}_{-0.05}$ & 13.08$^{+0.04}_{-0.04}$ & 9.19 & 3.15 & 1.12 \\
SSA22.0001b & 4.47$^{+0.83}_{-0.83}$ & alma($VLA$) & 0.04 & 3.10$^{+0.55}_{-2.55}$ & cigale & 11.91$^{+0.12}_{-0.12}$ & 2.55$^{+0.28}_{-0.28}$ & 12.89$^{+0.14}_{-0.14}$ & 9.00 & 5.09 & -0.53 \\
SSA22.0002a &  & alma($VLA$) & 0.00 & 0.58$^{+0.00}_{-0.00}$ & spec & 7.99$^{+1.61}_{-1.61}$ & 0.82$^{+0.03}_{-0.03}$ & 10.59$^{+0.02}_{-0.02}$ & 9.35 & 2.32 & 1.39 \\
SSA22.0002b & 4.71$^{+1.63}_{-1.63}$ & alma($IRAC8$) & 0.01 & 3.09$^{+0.00}_{-0.00}$ & spec & 11.73$^{+0.09}_{-0.09}$ & 2.64$^{+0.11}_{-0.11}$ & 12.75$^{+0.07}_{-0.07}$ & 8.62 & 3.17 & -0.31 \\
SSA22.0002c & 2.30$^{+1.13}_{-1.13}$ & alma & 0.02 & 3.09$^{+0.00}_{-0.00}$ & spec & 10.71$^{+0.09}_{-0.09}$ & 1.94$^{+0.55}_{-0.55}$ & 12.89$^{+0.04}_{-0.04}$ & 7.77 & 1.84 & -0.21 \\
SSA22.0003 & 11.09$^{+1.26}_{-1.26}$ & $VLA$ & 0.00 & 0.56$^{+0.01}_{-0.03}$ & eazy & 9.59$^{+0.12}_{-0.12}$ & 1.80$^{+0.10}_{-0.10}$ & 11.64$^{+0.08}_{-0.08}$ & 10.04 & 5.10 & 1.37 \\
SSA22.0004 & 7.24$^{+1.15}_{-1.15}$ & $VLA$ & 0.01 & 2.56$^{+0.00}_{-0.00}$ & spec & 10.98$^{+0.07}_{-0.07}$ & 2.79$^{+0.02}_{-0.02}$ & 12.81$^{+0.02}_{-0.02}$ & 9.63 & 2.00 & 0.52 \\
\enddata

\tablecomments{The first ten entries of our identified source catalog, the full catalog is available in the supplementary material.}

\tablenotetext{a}{For some sources, the deblended fluxes are poorly constrained due to source blending or low signal-to-noise.}
\tablenotetext{b}{Identification method.
Most sources with ALMA counterparts are also recovered by our identification method. If additional identifications exist, the next most reliable method—ranked as $VLA > MIPS24 > IRAC8 > iK\text{-}color$—is indicated in parentheses.
Owing to the superior astrometric precision of ALMA, we adopt ALMA-derived coordinates for source positions whenever available.}
\tablenotetext{c}{Redshift types include spectroscopic redshifts, EAZY photometric redshifts, and CIGALE SED-fitting redshifts.
When multiple redshift values are available, we adopt the most reliable one according to the hierarchy: $z_{\text{spec}} > z_{\text{EAZY}} > z_{\text{CIGALE}}$ (Section~\ref{sec:cigale_fitting}).}
\tablenotetext{d}{The offset of SMGs from the main sequence, defined as $\Delta$MS = log$_{10}$(sSFR/sSFR$_\text{MS}$), where the main sequence relation is adopted from \citet{speagle2014}. See also Paper IV.}

\end{deluxetable*}
\end{rotatetable*}

\subsection{Identification Methods}
\label{sec:id_method}

\subsubsection{Radio+$MIPS24$ Methods}
\label{sec:radio+mips24}

Radio and mid-infrared (MIR) emission serve as reliable tracers of star formation activity in galaxies, exhibiting strong correlations with submillimeter emission. 

Due to the radio-FIR correlation spanning several orders of magnitude in luminosity \citep{condon1992}, it remains well-preserved even at high redshifts. Additionally, with its relatively low spatial density \citep{koprowski2015}, high-precision radio interferometric observations are widely used for counterpart studies of submillimeter sources \citep{dud2020}.
Although the origin of the radio-FIR correlation is debatable, it is widely believed to trace the activity of massive stars ($M \gtrsim 5 M_\odot$) in galaxies \citep{helou1985, condon1992, dud2020}.
Submillimeter radiation arises from the re-emission of ultraviolet photons absorbed by dust from star-forming activities in galaxies \citep{blain2002}.
Radio emissions in galaxies arise from synchrotron and free-free radiation. Due to their flat spectrum, free-free emissions are weaker than synchrotron emissions at wavelengths longer than $\sim$ 1 cm (frequencies $\textless$ 30 GHz). Therefore, radio emissions are generally attributed to the synchrotron acceleration of relativistic electrons produced by HII regions of massive stars and the supernova remnants left after their death \citep{condon1992}.

Similar to the former, the MIR emissions also exhibit a strong correlation with FIR emissions \citep{appleton2004, takeuchi2005}, serving as a reliable indicator of the infrared luminosity of dust-obscured galaxies  \citep{toba2017, shim2022}.
A significant portion of the MIR emissions from SMGs originates from thermal emissions, generated by dust grains heated by star-forming regions, AGN, or nuclear starburst regions. Moreover, MIR emissions also stem from atomic and molecular lines, primarily induced by the bending and stretching of polycyclic aromatic hydrocarbons (PAHs), excited by UV photons associated with nearby star formation \citep{menendez2009, casey2014, boquien2019}. The dominant PAH feature lines are emitted at rest-frame wavelengths of 3.3, 6.2, 7.7, 8.6, 11.3, 12.7, and 17 $\micron$ \citep{dl07, dud2020}, falling within the observed 24 $\micron$ band at typical redshifts of SMGs.
We should note that it is precisely due to these prominent PAH emission and absorption features \citep{huang2009} that the 24 $\micron$ flux in SMGs can exhibit significant variations.

We employ the corrected-Poissonian probability, $p$-value, to calculate the probability of chance matches between the identified counterparts and submillimeter sources, as outlined by \citet{downes1986}:
\begin{equation}
    p = 1 - exp( -\pi n \theta^2),
\end{equation}
where $n$ is the surface density of VLA 3 GHz (or 24 $\micron$) sources, and $\theta$ is the angular distance between identifications and the SCUBA-2 source positions.

\citet{hodge2013} argued that, even in single-dish submillimeter maps where source confusion is significant, the positional offset between the submillimeter source and true SMG counterpart does not appear to vary systematically with SNR. 
\citet{an2018} adopted a statistical threshold of $p =$ 0.065 for reliable radio identifications, though they noted that a small fraction of confirmed radio counterparts still lie beyond this threshold—corresponding to $\gtrsim$ 5.5$\arcsec$.
Similarly, \citet{chen2016} found that using a fixed search radius for counterpart identification could be a better strategy, and suggested that adopting $p \leqslant$ 0.1 balances accuracy and completeness.

We therefore adopt a fixed search radius of 7$\arcsec$ (half of the SCUBA-2 beam size) and consider all VLA and MIPS 24 $\micron$ counterparts with $p <$ 0.1 as potential associations. 
Specifically, all VLA sources within a 5.4$\arcsec$ radius of a SCUBA-2 centroid have $p <$ 0.1. 
For example, in Figure~\ref{fig:fig3-2-1_misid_adf}, the sources SSA22.0000, SSA22.0070, and SSA22.0002 are affected by source blending. Our identification method did not associate the radio counterparts located near the edge of the SCUBA-2 beam with these sources; however, ALMA detections confirm that these radio sources are indeed physically associated SMGs.
For MIPS 24 $\micron$, due to its lower source surface density, all counterparts within the full SCUBA-2 beam (7$\arcsec$) satisfy $p <$ 0.065.
Taking into account SCUBA-2 beam effect, source blending, uncertainties in source extraction, astrometric errors, and telescope pointing accuracy ($\sim$ 1–2$\arcsec$ for JCMT; \citealp{geach2017}), we regard the $p <$ 0.1 criterion as well justified for radio counterparts. Given that 24 $\micron$ band has a resolution of $\sim$ 6$\arcsec$, we similarly consider this threshold appropriate for 24 $\micron$ identifications.
For the IRAC 8 $\micron$ and $iK$-color identifications used in the next section, we adopt a stricter threshold of $p <$ 0.05, corresponding to angular radii of approximately 2.4$\arcsec$ and 2.8$\arcsec$, respectively. We provide a more detailed justification for these choices in Section~\ref{sec:caveat}.

For convenience, we will continue the tradition of labeling sources in the catalog as `robust' for $p \leqslant$ 0.05 and `tentative' for $p >$ 0.05.
After visually inspecting all counterparts and candidate sources, we defined sources with coordinate separations less than 1.5$\arcsec$ as the same counterpart. 
This threshold is appropriate as it avoids misclassifying sources that are too far apart as the same, or splitting the same source into different counterparts. 
Multiple identifications for the same SCUBA-2 source were sorted in ascending order of $p$-values and assigned labels a, b, c, and so forth.
Additionally, we found that despite the larger FWHM of 24 $\micron$ sources compared to high-precision VLA detections, their coordinates are still relatively accurate.

\subsubsection{ $IRAC8$+$iK\text{-}color$ Methods}
\label{sec:irac8_ikcolor_ids}

We also employ IRAC 8 $\micron$ and red opt-NIR color as a method to search for potential counterparts. However, for reliability considerations, we only include those counterparts with $p$ $\textless$ 0.05 in the source catalog.

Similar to the MIR-FIR correlation mentioned earlier, exploration in the IRAC 8 $\micron$ band is often conducted to enhance the completeness of identifying submillimeter sources \citep{ashby2006, koprowski2015, koprowski2017, zavala2018}.
SED analysis of SMGs suggested that their 8 $\micron$ emission primarily traces ongoing star formation \citep{ashby2006}. The evolved stellar populations have been discovered in massive SMGs, the expected spectral signatures of evolved stellar populations should peak at rest-frame $\sim$ 2-3 $\micron$ \citep{koprowski2015}, corresponding to observed 8 $\micron$.

Some studies have used a NIR color to select high-redshift, extremely red galaxies, such as HIEROs \citep{wang2012}, KIEROs \citep{wang2016}, OIRTC \citep{chen2016}. 
Similarly, due to the highly dust-obscured nature of SMGs often displaying red optical/NIR colors, many researchers have used color-selected galaxies to identify DSFGs or SMGs \citep{cowie2017, an2018, zhang2022, mckay2025}.
Following \citet{michalowski2012}, we employed a red optical/NIR color criteria ( $i - K$ $\textgreater$ 2 \& $K$ $\textless$ 23 mag) to identify potential SMGs counterparts. The detailed measurement description could be seen in the Section~\ref{sec:opt_nir}.

SMGs typically exhibit redshifts in the range $z \sim 2$–3. In this redshift interval, the observed $i$- and $K$-bands correspond approximately to the rest-frame near-UV and $R$-band, respectively, straddling the Balmer/4000$\text{\AA}$ break—a key spectral feature for galaxy characterization.
The Balmer break arises at the Balmer limit (3645 $\text{\AA}$) due to bound-free absorption by hydrogen atoms with electrons in the $n = 2$ energy level. This feature is weakened in hot OB stars because most hydrogen is ionized, but it is strongest in cooler A-type stars ($1.4$–$2.1$ M$_\odot$), whose effective temperatures ($\sim$10,000 K) maximize the population of electrons in the $n = 2$ state.
This feature is particularly prominent in star-forming galaxies \citep{dud2020, wilkins2023}. Moreover, it becomes especially strong in galaxies with stellar populations of age $\sim$0.3 Gyr, when a large number of UV-bright stars begin to evolve off the main sequence \citep{bc03, vikaeus2024}.
The 4000 $\text{\AA}$ break primarily driven by metal absorption lines from cool, evolved stars \citep{mo2010, dud2020}.
Together, the Balmer/4000$\text{\AA}$ break serves as a powerful diagnostic of recent star formation history and is sensitive to dust attenuation, metallicity, and the Lyman continuum (LyC) escape fraction \citep{wilkins2023}.

These selected galaxies are referred to as Extremely Red Galaxies (ERGs), which may stem from evolved stellar components or galaxies that have reddened significantly due to dust obscuration in star-forming galaxies or starbursts (or AGN) \citep{caputi2006, schael2009, mo2010}. 
ERGs are believed to align on an evolutionary sequence with high-redshift SMGs and local massive ellipticals \citep{smail2002, schael2009}, while the most extreme ERGs represent a population of highly dust-obscured, ultraluminous galaxies at high redshift \citep{smail1999}. In the redshift range of 1.5-3, 40-60\% of high-mass ERGs are ULIRGs \citep{smail2002, caputi2006}. SMGs, characterized by severe dust obscuration and intense star formation, are expected to exhibit distinct optical/NIR features. Therefore, such selections seem to be a good criterion.

\subsection{Identification Result}
\label{sec:id_result}

Our VLA 3 GHz and MIPS 24 $\micron$ catalog includes 1989 radio sources with a significance greater than 4$\sigma$ (the image depth after primary beam correction is $\sim$ 1.5-5 $\mu$Jy), and 813 sources of $\textgreater$ 3$\sigma$ at 24 $\micron$, respectively.

Armed with the robust methods, we identified a total of 173 SMG candidates for 142 SCUBA-2 sources, with 36 candidates detected by both 3 GHz and 24 $\micron$.
These sources are predominantly located in the main SCUBA-2 map area of SSA22 (Figure~\ref{fig:fig2-0_multicontour}), including the protocluster core \citep{zeng2024}.

Specifically, of the 203 SCUBA-2 sources whose positions lie within the area covered by VLA 3 GHz observations, 119 have VLA counterparts, yielding an identification rate of 59\%. In total, the VLA data reveal 146 SMG counterparts across the full sample.
On the other hand, 218 SCUBA-2 sources are located within the MIPS 24 $\micron$ map, with 63 SCUBA-2 sources having 24 $\micron$ counterparts, resulting in an identification rate of 29\%. 
Among the 63 SCUBA-2 sources, there are 63 SMG counterparts at 24 $\micron$ sources, 27 of these counterparts are additional findings not captured by the VLA observations.
The overall identification rate for the $VLA$+$MIPS24$ method is 56\%.
Our SCUBA-2 and VLA 3 GHz maps are similar in depth to those of \citet{hyun2023}, and our radio-identification rate is approximately the same. 

The ADF22 catalog confirmed 20 SMG counterparts associated with 15 SCUBA-2 sources.
Among these SMGs, 5 were not identified by either the $VLA$ or $MIPS24$ methods, though 2 of those 5 were also detected as IRAC 8 $\micron$ sources.
Additionally, one ALMA-detected SMG serves as the unique counterpart to the SCUBA-2 source SSA22.0098.

In total, utilizing the radio+MIR method, we obtained 178 reliable SMG counterparts for 143 SCUBA2 sources, identified by ALMA, VLA 3 GHz, and MIPS 24 $\micron$ sources. 


96 candidates were identified by 8 $\micron$ sources, while 93 by the optical/NIR-red selected galaxies. These two approaches respectively added 30 and 40 new SMG counterparts, constituting 28\% of the total identifications. Among these, 18 and 29 SCUBA-2 sources were uniquely identified with counterparts by the $IRAC8$ and $iK\text{-}color$ methods, respectively.

\subsection{Identification Accuracy and Caveats}
\label{sec:caveat}

\begin{figure*}
    \centering
    \includegraphics[width=0.96\textwidth]{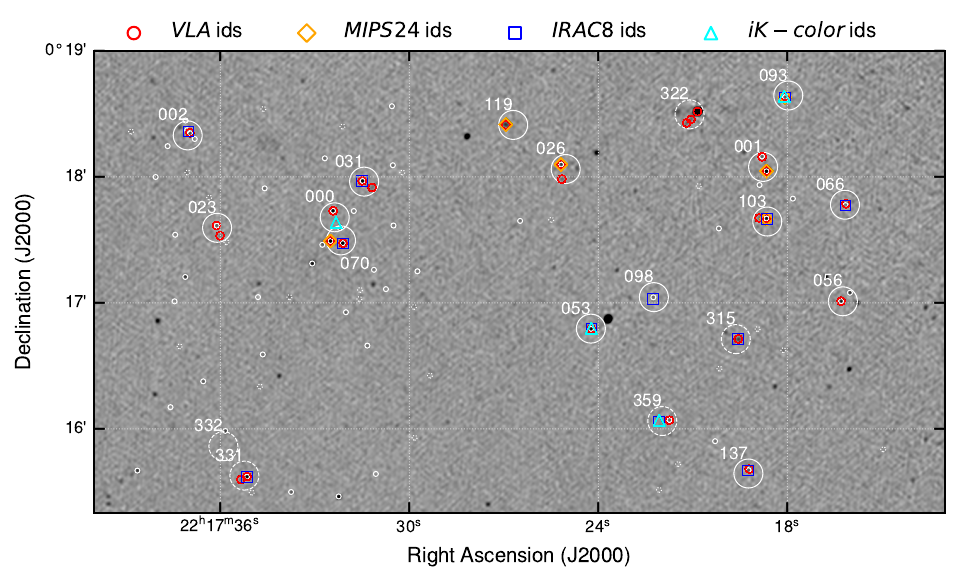}
    
    \caption{
    Comparison of multi-band identification with ADF22/ADF22+ catalog sources.
    The background image is our VLA 3 GHz mosaic covering the ADF22 survey area.  
    Large white circles indicate SCUBA-2 positions with a radius of 7$\arcsec$, labeled with corresponding source IDs. 
    Solid and dashed white contours correspond to sources detected at $\geqslant$ 4$\sigma$ and 3.5$\sigma$–4$\sigma$ significance levels, respectively. 
    Small white circles denote ALMA-detected sources, with solid and dashed contours representing sources detected at $\geqslant$ 5$\sigma$ and 4$\sigma$–5$\sigma$.  
    Symbols in different colors mark sources identified via different methods.}
    \label{fig:fig3-2-1_misid_adf}
\end{figure*}

\begin{figure*}
    \centering
    \includegraphics[width=0.96\textwidth]{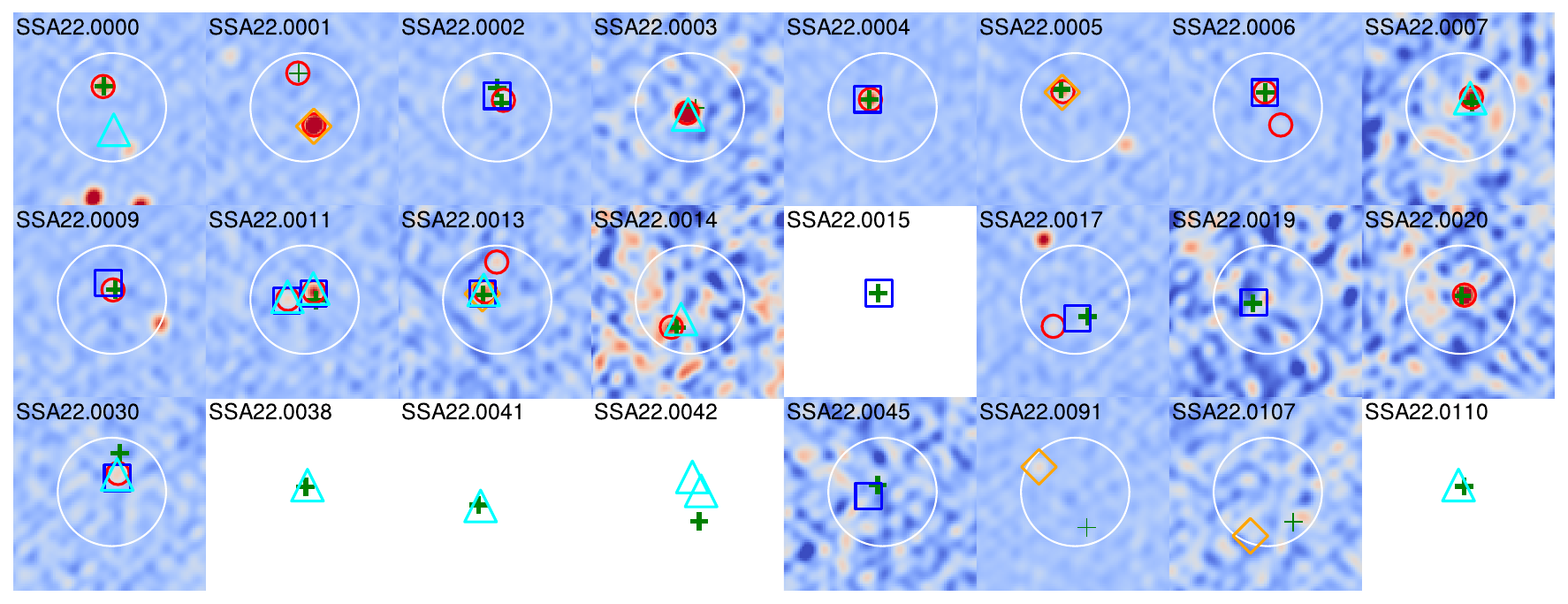}
    
    \caption{
    Comparison of multi-band identification with ALMA 2 mm catalog sources.
    The background image is our VLA 3 GHz mosaic; blank regions correspond to areas outside the VLA coverage.  
    Large white circles mark SCUBA-2 positions (radius 7$\arcsec$), with their IDs shown in the upper left corner of each stamp; these sources have relatively high flux densities and signal-to-noise ratios. 
    Green crosses indicate ALMA-detected sources: bold symbols denote $\geqslant$ 3$\sigma$ detections, while thin symbols represent sources below 3$\sigma$ \citep{cooper2022}.  
    Colored symbols correspond to sources identified via different methods, consistent with Figure~\ref{fig:fig3-2-1_misid_adf}. 
    We adopt a matching radius of 1.5$\arcsec$. The offsets between the multiwavelength counterparts of SSA22.0030, SSA22.0042, and SSA22.0045 and their nearest ALMA detections are 2.63$\arcsec$, 3.93$\arcsec$, and 1.88$\arcsec$, respectively. We therefore consider these sources unmatched to any ALMA detection. }
    \label{fig:fig3-2-2_misid_cooper}
\end{figure*}

\begin{figure*}
    \centering
    \includegraphics[width=1\textwidth]{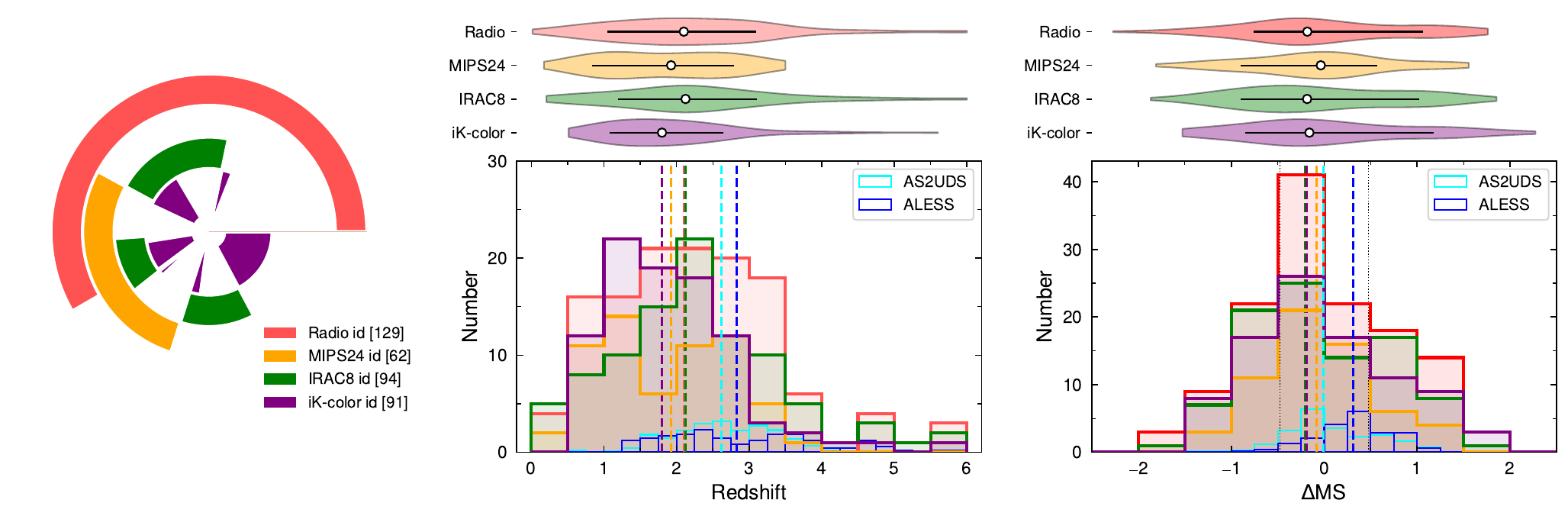}
    \caption{\textbf{Panel left}: The doughnut chart illustrates the identification statistics of SMGs using four distinct methods. Each concentric ring, color-coded by method, represents the set of galaxies identified by that technique. 
    From the center outward, the overlapping regions of different colors represent sources identified simultaneously by multiple methods.
    The numbers in the labels denote the total number of galaxies identified by each individual method.
    \textbf{Panel middle}: The redshift distributions of SMGs identified by the four methods, with vertical dashed lines indicate the median values of these distributions. 
    The accompanying violin plot shows that the redshift distribution are tightly concentrated in the range $z \sim 1$–3, with error bars indicating median redshifts (dashed lines) and the 68\% dispersion range.
    The cyan and blue bars represent the scaled distributions from AS2UDS and ALESS, respectively.
    \textbf{Panel right}: Offset of identified sources from the star-forming main sequence, defined as $\Delta$MS = $\log_{10}$(sSFR/sSFR$_\text{MS}$), where the main sequence relation is adopted from \citet{speagle2014}. The thin dotted lines demarcate different evolutionary stages: sources lying more than 0.48 dex above the main sequence are classified as starbursting, while those more than 0.48 dex below are considered quiescent (see \citetalias{paperiv}). }

    \label{fig:fig3-2-3_misid_hist}
\end{figure*}


\begin{figure}
    \centering
    \includegraphics[width=0.45\textwidth]{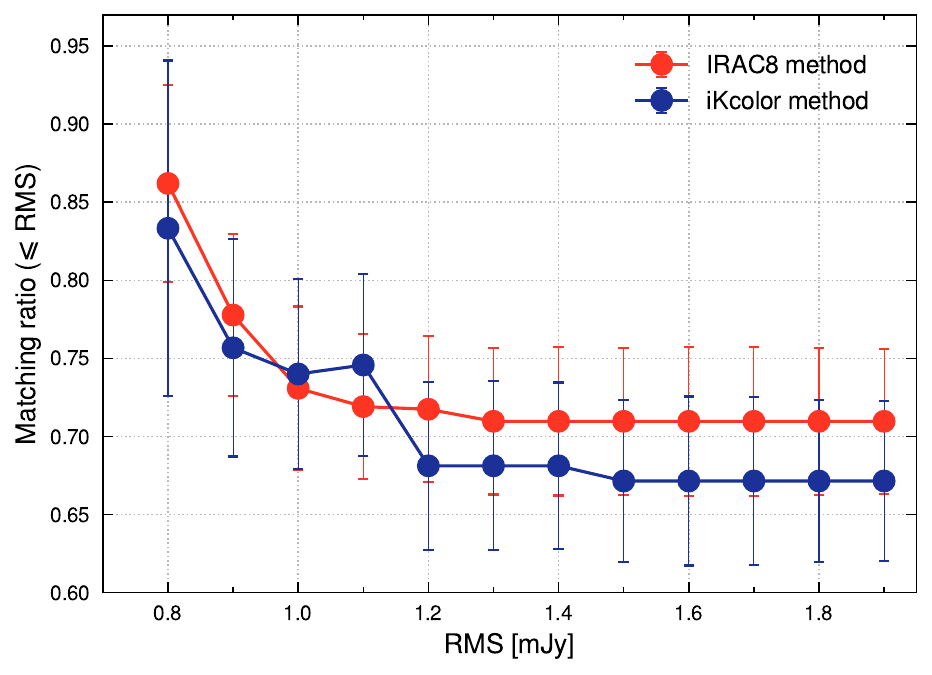}
    
    \caption{Robustness demonstration of the  ``$IRAC8$" and ``$iK\text{-}color$" selection method.
    The Y-axis shows the cumulative fraction of galaxies identified by this two method that are cross-matched with radio or 24 $\micron$ counterparts, while the X-axis represents the local noise level in the SCUBA-2 map.
    The match rates and overall trends for both methods are similar.
    The average cross-matching fraction is $\sim$ 70\%, rising to $\sim$ 85\% in the deepest region of the map. This high and consistent match ratio supports the reliability of the  ``$IRAC8$" and ``$iK\text{-}color$" method.  }
    \label{fig:fig3-2-4_reliablity}
\end{figure}

In this section, we provide a preliminary assessment of the reliability of our identification methods and offer important caveats for users of our catalog.

To evaluate the performance of our identification techniques, we cross-match our identified sources with two ALMA surveys in the SSA22 field: the ADF22 survey (including the ADF22 A/B fields; \citealp{umehata2017a, umehata2018}) and the deeper ADF22+ survey (\citealp{huang2025}), as well as the ALMA 2 mm continuum survey (\citealp{cooper2022}).

Figure~\ref{fig:fig3-2-1_misid_adf} shows the footprint of the ADF22 survey. Focusing on SCUBA-2 sources detected at $\geqslant$ 4$\sigma$, the identification accuracies of our four methods—$VLA$, $MIPS24$, $IRAC8$, $iK\text{-}color$—are 15/20 ($\sim$ 75\%), 5/6 ($\sim$ 83\%), 9/9 ($\sim$ 100\%), and 2/3 ($\sim$ 67\%), respectively. 
When including marginal detections in the range 3.5$\sigma$–4$\sigma$, the accuracies decrease modestly to 17/27, 5/6, 10/12, and 2/4.
The high fidelity for $\geqslant$ 4$\sigma$ sources reflects the fact that the ADF22 fields lie within the region of most complete multiwavelength coverage, underscoring the critical importance of deep, multi-band data for reliable source identification.  
Survey depth also plays a role: the deeper ADF22+ observations confirmed three low-significance ADF22 A detections, one of which (ADF22+ source A28) corresponds to a component of the blended SCUBA-2 source SSA22.0070. This source was robustly identified by both our $VLA$ and $MIPS24$ methods.

Comparison with the ALMA 2 mm survey yields (Figure~\ref{fig:fig3-2-2_misid_cooper}) lower—but still reasonable—identification accuracies: 13/19 ($\sim$ 68\%), 2/5 ($\sim$ 43\%), 9/12 ($\sim$ 75\%), and 8/13 ($\sim$ 62\%).
We emphasize again that these accuracy estimates depend sensitively on both multiwavelength coverage and ALMA survey depth.

We further compare sources identified via $IRAC8$ and $iK\text{-}color$ methods with those detected in radio and 24 $\micron$. 
The matching fractions are consistent between the two methods (Figure~\ref{fig:fig3-2-4_reliablity}), reaching $\sim$ 75–85\% in the central region of the field (where SCUBA-2 map RMS $\leqslant$ 1 mJy). 
Both matching rates decline with increasing SCUBA-2 map noise, yielding overall matching fractions of $\sim$ 71\% ($IRAC8$) and $\sim$ 67\% ($iK\text{-}color$)—a direct consequence of the deepest and most complete multiwavelength coverage in the field center.
Because IRAC 8 $\micron$ sources and color-selected objects have relatively high surface densities, positional association probabilities drop below $p <$ 0.05 only at angular separations of 2.4$\arcsec$–2.8$\arcsec$. 
The median offsets between our $IRAC8$ and $iK\text{-}color$ identifications and the SCUBA-2 centroids are 1.5$\arcsec$ and 1.8$\arcsec$, respectively—corresponding to physical scales of $\sim$ 10–15 kpc at $z \sim$ 2. 
This is comparable to the pointing uncertainty of JCMT \citep{geach2017}. 
Therefore, if SMGs reside near the centroid of the SCUBA-2 source, these two identification methods are unlikely to introduce significant misidentifications at such angular separations, aside from chance alignments along the line of sight.

The the $\sim$ 80\% correspondence between sources identified via the $IRAC8$/$iK\text{-}color$ methods and those detected in radio or $MIPS24$ is remarkably high—indeed, a pleasantly surprising result.  
This strong overlap suggests an underlying physical connection: even when some of the IRAC 8 $\micron$ sources or color-selected objects identifications do not correspond to genuine SMGs, they are still likely to be galaxies undergoing active star formation.  
Moreover, as shown in the earlier matching analysis, the identification accuracy of the 8 $\micron$ method reaches $\gtrsim$ 80\%, demonstrating its considerable potential—comparable to the more established $VLA$ and $MIPS24$ techniques.

For the $iK\text{-}color$ method, we perform additional tests to quantify the contamination from random alignments.  
First, within the ADF22 survey area, we mask all regions within 7$\arcsec$ of any ALMA-detected SMG or SCUBA-2 source. We then place random positions and check whether a $iK\text{-}color$ source lies within 2.78$\arcsec$ ($p \leqslant$ 0.05). 
This yields a lower limit on the contamination rate due to chance superpositions. In this small masked region, we generate 200 random positions per trial and repeat the experiment 50 times, obtaining a false-match probability of 4.02 $\pm$ 1.32\%.
Next, we expand the test to a larger rectangular area (R.A. = $22^{\mathrm{h}}16^{\mathrm{m}}50^{\mathrm{s}}$–$22^{\mathrm{h}}18^{\mathrm{m}}10^{\mathrm{s}}$, Dec. = +00$\arcdeg$07$\arcmin$00$\arcsec$–+00$\arcdeg$28$\arcmin$00$\arcsec$), which covers field center with the most complete multiwavelength coverage. Here, we draw 2000 random positions per trial over 10 repetitions, resulting in a false-match probability of 4.84 $\pm$ 0.40\%.
Overall, thanks to our strict matching criterion ($p \leqslant$ 0.05), the expected contamination from random alignments is low—below 5\%. However, it should be emphasized that this estimate represents a lower limit. 
In reality, the true contamination could be higher, because SCUBA-2 sources are preferentially located in overdense environments. Such clustering enhances the likelihood of finding a red, massive galaxy near a submillimeter position even without a physical association.

Deep, well-characterized ALMA follow-up surveys—such as ALESS \citep{cunha2015} and AS2UDS \citep{stach2019, dud2020}—contain only a small minority of sources at low redshift ($z <$ 1), with the vast majority lying on or above the star-forming main sequence.  
In contrast, our sample includes a non-negligible fraction of low-redshift or quiescent systems, regardless of the identification method used (see Figure~\ref{fig:fig3-2-3_misid_hist}). 
We acknowledge, however, that some of these apparently quiescent galaxies may be artifacts of SED fitting: When far-infrared constraints are weak or absent, it leads to increased scatter in derived physical properties and may result in spurious classifications as “quenched” systems. We briefly discuss this effect in Section~\ref{sec:sfr_mstar}.
%
We also emphasize that our multiwavelength identification approach inevitably leads to the loss of a fraction of high-redshift SMGs. To illustrate this, we apply simple selection criteria to the AS2UDS catalog \citep{dud2020}, isolating subsamples with radio detections, MIPS 24 $\micron$ counterparts, IRAC 8 $\micron$ counterparts, or those satisfying the color cut $i - K >$ 2 and $K <$ 23 mag. The median redshifts of these subsamples are 2.45, 2.31, 2.58, and 2.29, respectively—systematically lower than that of the full AS2UDS sample ($\sim$ 2.61).
This trend underscores a key limitation: multiwavelength identification methods inherently risk missing the high-redshift SMGs. This bias is particularly pronounced when relying on 24 $\micron$ or optical/NIR data. In the former case, the incompleteness is likely driven by sensitivity limits at high redshift, while in the latter, severe dust attenuation renders many high-redshift SMGs undetectable in optical/NIR bands.

We issue the following cautions:
(1) Sources in regions with incomplete multiwavelength coverage—approximately where the SCUBA-2 map RMS $\gtrsim$ 1–1.2 mJy—should be treated with care.  
(2) Identifications based solely on the $iK\text{-}color$ criterion require particular scrutiny due to higher contamination risk.  
(3) Sources with redshifts $z <$ 1 or classified as quiescent are likely contaminants and should be interpreted cautiously.
Secondly, it is important to emphasize that the current identification methods likely miss a fraction of genuine SMG counterparts, particularly at high redshifts.
Lastly, while our identification methods recover the majority of the SCUBA-2 flux, this raises a secondary concern: misidentified counterparts could lead to incorrect deblending and flux allocation. While given the relatively low multiplicity rate and the small number of $iK\text{-}color$-only identifications, we consider this risk to be modest.

In summary, cross-matching with deep ALMA data indicates that the majority of our identified sources belong to the true SMG population. 
Consequently, our sample is suitable for robust statistical analyses—a conclusion we will reiterate in subsequent sections.


\subsection{Multiplicity}
\label{sec:multiplicity}

ALMA interferometric observations have revealed that single-dish submillimeter sources are often blends of multiple SMGs \citep{karim2013}.  
In a follow-up ALMA study of 716 UDS SCUBA-2 sources, \citet{stach2018} found an overall multiplicity fraction of approximately 11\%. 
This fraction increases with submillimeter flux density: it rises to 26\% for sources with S$_{850} \gtrsim$ 5 mJy and further to 44\% for those brighter than 9 mJy. 
Among these blended systems, about 30\% consist of physically associated components \citep{stach2018}.
The measured multiplicity also depends on observational depth, as fainter SMGs are commonly found to be blended even in ALMA “blank” maps \citep{simpson2014}.
When such faint blends are included, the multiplicity fraction may increase by an additional $\sim$ 10–20\% \citep{hodge2013, simpson2015}.

In SSA22 field, the overall multiplicity rate is $26 \pm 3$\%. In the “deep region” (Figure~\ref{fig:fig2-0_multicontour}), where multiwavelength data are most complete, fraction rises to $35 \pm$ 4\%, consistent with typical multiplicity rates reported in other multiwavelength identification studies ($\gtrsim$ 20–40\%; \citealp{chen2016, an2019}).
Specifically, for SCUBA-2 sources with flux densities $<$ 8 mJy, the multiplicity rate is 24 $\pm$ 3\%. 
In contrast, for sources $\geqslant$ 8 mJy, the multiplicity rate increases significantly, reaching 47 $\pm$ 13\%. 
Within individual high-flux bins, the multiplicity fractions range from $\sim$ 40\% to 60\%, though these estimates carry large uncertainties due to small number statistics.

\begin{figure}[h]
    \centering
    \includegraphics[width=0.5\textwidth]{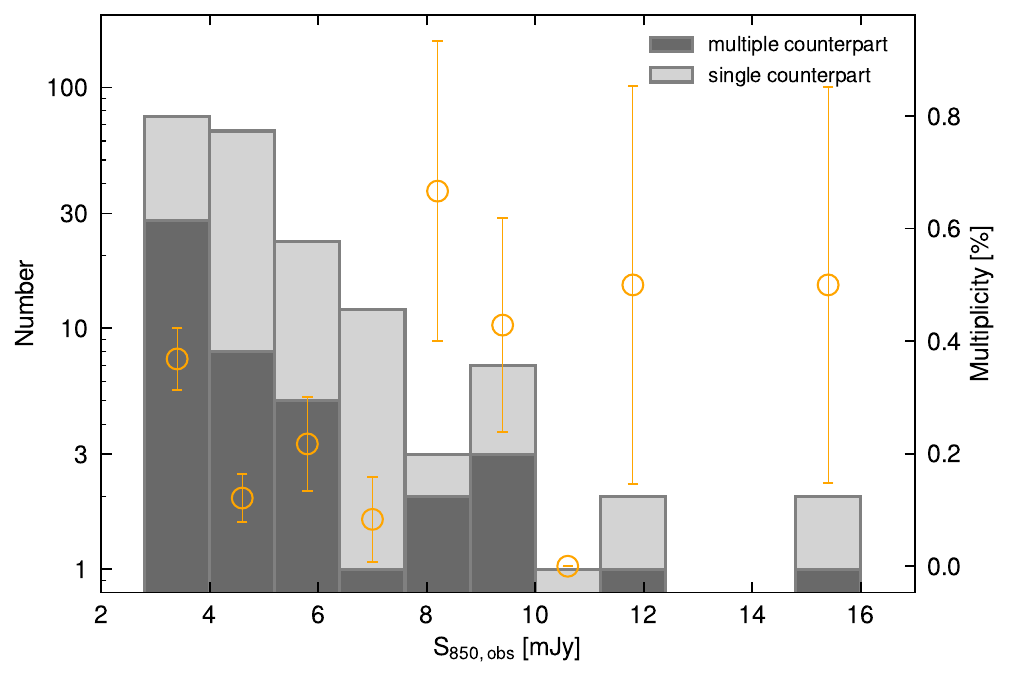}
    
    \caption{The number and multiplicity of SCUBA-2 source at each flex density bin.
    The light grey bars show the number of sources with a single counterpart.
    The dark grey bars indicate those with multiple counterpart, while the yellow points represent the corresponding multiplicity fraction plotted against the right Y-axis.
    For fluxes below 8 mJy, the multiplicity fraction ranges from approximately 10\% to 40\%. An increasing trend with higher flux is observed with a value of 46\% when considering all sources brighter than $\sim$ 8 mJy.  }
    
    \label{fig:fig3-3_multiplicity}
\end{figure}

\section{SED-fitting}
\label{sec:sed-fitting}

We present the photometric redshifts derived from \texttt{EAZY} (Section~\ref{sec:eazy}), a description of our \texttt{CIGALE} configuration (Section~\ref{sec:cigale_config}), and the resulting SED fitting outputs (Section~\ref{sec:cigale_fitting}).

\begin{figure}[!h]
    \centering
    \begin{minipage}{\columnwidth}
    \includegraphics[width=0.98\textwidth]{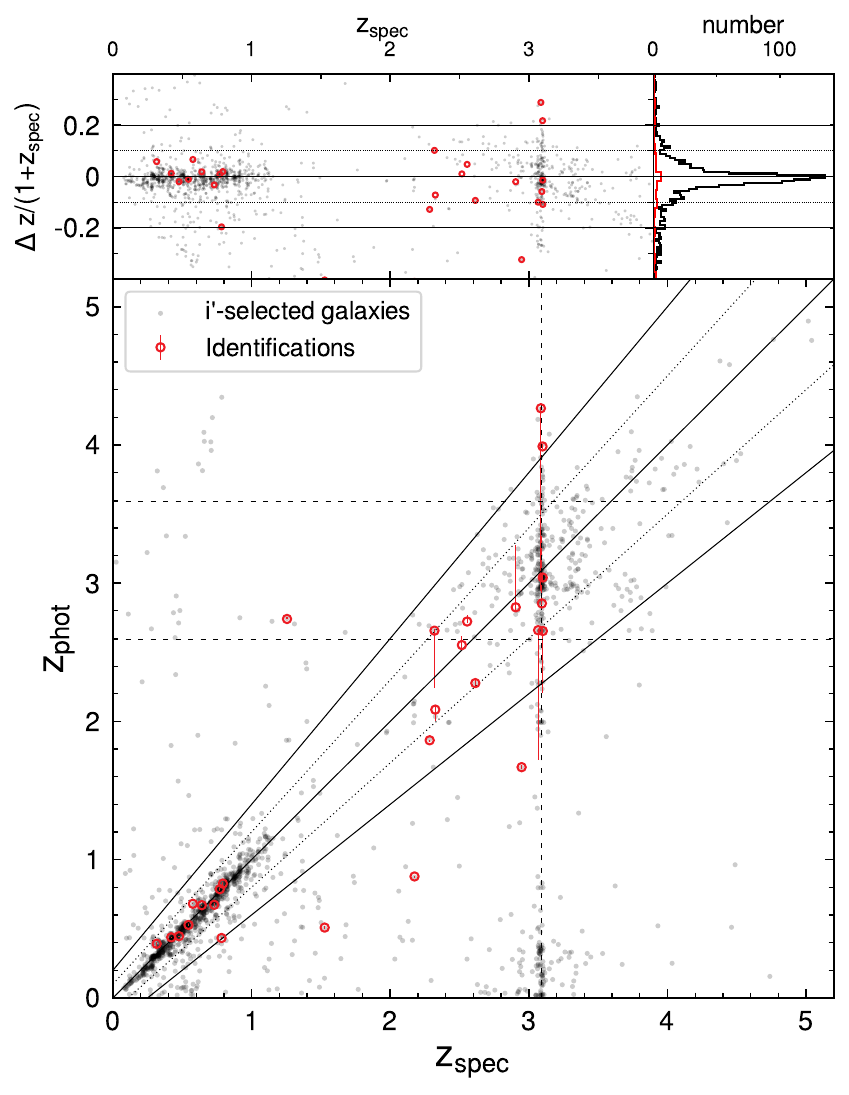}
    
    \caption{\texttt{EAZY} photometric redshift for SSA22 SMGs (red circles) vs. their spectroscopic redshifts.
    To demonstrate the robustness of \texttt{EAZY} redshift estimates, we also include  photometric redshift for i'-band selected galaxies (grey dots).
    The SMG sample shows comparable reliability to the optically select galaxies: $\sim$ 75\% sources lie within the region where the redshift residual $|\Delta z/(1+z_{spec})|$ $\leqslant$ 0.15 (bounded by dotted lines), and $\sim$ 80\% satisfy redshift residual $\leqslant$ 0.20 (within the outer solid lines).
    The vertical dashed line labels redshift of 3.09, corresponding to the protocluster, while the two horizontal dashed lines represent $z = 3.09 \pm 0.5$.
    The top panel displays the distribution of redshift residual, and the inset in the upper right corner panel shows the histogram of the residuals.}
    \label{fig:fig4-1_eazy_zphot}
    \end{minipage}
    
\end{figure}

\subsection{EAZY Photometric Redshift}
\label{sec:eazy}

We employed the Python version of \texttt{EAZY} \citep{brammer2008} to estimate optical/near-infrared ($u^*$-band to $K$-band) photometric redshifts (photo-$z$).
Additionally, we adopted an updated template set named \textit{corr\_sfhz\_13} $\footnote{https://github.com/gbrammer/eazy-photoz/tree/master/templates/sfhz}$, which incorporates redshift-dependent star-formation histories.
The transmission curves for each bandpass were obtained from the SVO filter profile service \citep{rodrigo2012, rodrigo2024}.
The program combined and scaled several spectral templates to fit the observed fluxes of identified sources, iterating three times to correct zeropoint offsets \citep{straatman2016, weaver2022} and optimize the photometric redshift. 
We adopted the output parameter `\textit{z\_ml}', representing the maximum likelihood redshift, as the final redshift determination for the counterparts. The 16th and 84th percentile probability distribution of photo-$z$ were used to calculate the 1$\sigma$ error.

To assess the reliability of our photometric redshifts, we performed a cross-match between the Subaru/Scam $i^{\prime}$-band catalog and with spectroscopically confirmed galaxies (the ``representative spectroscopic redshift catalog'' compiled by \citealt{mawatari2023}) within 1$\arcsec$, yielding a sample of 1734 optically selected galaxies for photometric redshift validation. 

Our analysis demonstrates that photometric redshifts achieve reliable accuracy when computed using at least 4 photometric bands, with 1645 out of the 1734 $i^{\prime}$-selected galaxies meeting this criterion for valid \texttt{EAZY} redshift estimates.
These validated sources are represented by gray points in Figure~\ref{fig:fig4-1_eazy_zphot}, showing excellent agreement between photometric and spectroscopic redshifts. 
The ensemble exhibits a median redshift residual ($(z_{spec}-z_{EAZY}) / (1+z_{spec})$) of -0.013 ± 0.001 (uncertainty estimated via bootstrap resampling) with a standard deviation of 0.39. The method demonstrates strong consistency, with 80.5\% and 76.5\% of sources falling within redshift residual range of 0.2 and 0.15 respectively, confirming the robustness of \texttt{EAZY} photometric redshifts for our sample.

Among the 248 identified SMGs, 27 sources have reliable `$z_{spec}$' and `$z_{EAZY}$' (red circles in Figure~\ref{fig:fig4-1_eazy_zphot}). Their median redshift residual is -0.014 ± 0.022, with a standard deviation of 0.20. Specifically, 77.8\% (74.1\%) of sources show redshift residuals smaller than 0.2 (0.15). The comparison also shows good agreement between photometric and spectroscopic redshifts.


In the end, 155 identifications (at least 4 bands available) possess `$z_{EAZY}$' values ($\sim$ 62.5\%), and the median redshift is 1.8 when considering both `$z_{EAZY}$' and `$z_{spec}$' collectively.
Our median optical/NIR photo-$z$ was noted to be lower than the redshift distribution of SMGs, with over a third of the counterparts lacking sufficient detections in the optical/NIR bands. 
Among these sources, 41 have not been detected at all.
This observation suggests that the characteristics of this population are situated in the high-redshift universe and/or are highly dust-obscured.


\begin{figure}
    \centering
    \begin{minipage}{\columnwidth}
    \includegraphics[width=0.98\textwidth]{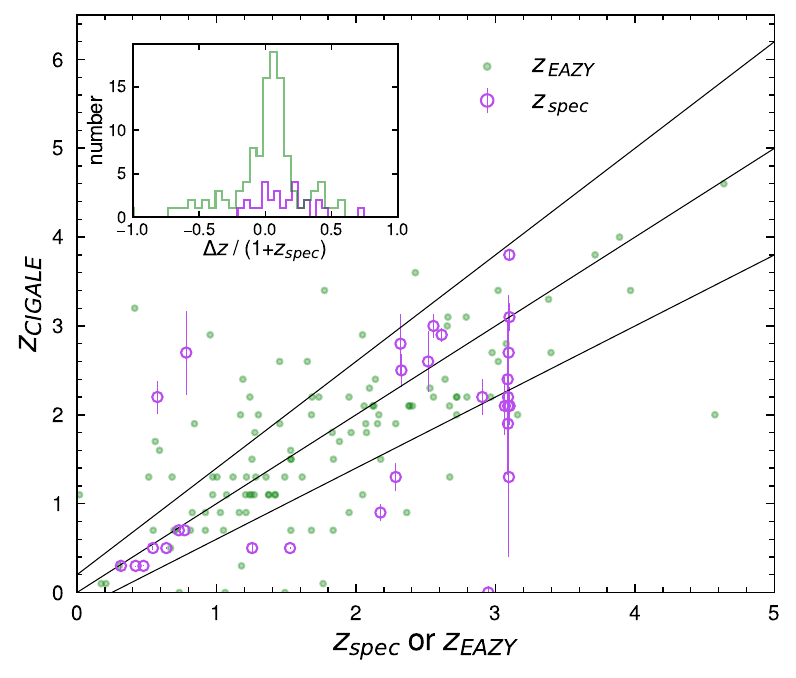}
    \caption{ The photometric redshifts derived from \texttt{CIGALE} SED fitting for SSA22 SMGs are compared against spectroscopic redshifts (purple dots) and \texttt{EAZY} photometric redshifts (green dots).
    The solid lines delineate the region where the normalized redshift residual $|\Delta z/(1+z_{spec})|$ $\leqslant$ 0.20.
    The inset panel presents histograms of the redshift residuals for both comparison methods, quantifying the agreement between the different redshift estimation techniques. }
    \label{fig:fig4-2_zcigale_vs_z}
    \end{minipage}
    
\end{figure}


\subsection{CIGALE SED Fitting}
\label{sec:sedfitting}

%

We perform SED fitting using Investigating GALaxy Emission \citep[CIGALE version 2022.1;][]{noll2009, boquien2019}.
By leveraging multi-wavelength photometry—from the optical/NIR through to radio—we derive key physical properties of our SMG sample. 
For galaxies lacking reliable $z_{phot}$ or $z_{spec}$ (Section~\ref{sec:eazy}), we additionally obtain their far-infrared (FIR) photometric redshifts.
This subsection describes the fitting configuration, and the resulting physical parameters.

\texttt{CIGALE} adopts a modular design, facilitating the invocation and combination of different physical component models within galaxies, including stellar populations, star formation histories, dust emission and attenuation, radio emission, AGNs, nebular emission, and more. 
The SED fitting process employs an energy balance approach, where the stellar and AGN emission absorbed by dust in the UV/optical regime is re-radiated in the infrared through self-consistent dust modeling \citep{casey2014}.
\texttt{CIGALE} generates comprehensive SEDs for galaxies by exploring all possible combinations of model parameters through an extensive grid (typically containing millions to billions of parameter sets). The optimal SED fit is determined by calculating and minimizing the reduced $\chi^2$ statistic for each template.

The multiple galaxy factors, such as stellar age, metallicity, and attenuation often exhibit degeneracies \citep{boquien2019}, which can lead to galaxies with diverse properties potentially exhibiting similar SEDs. Therefore, \texttt{CIGALE} employs a Bayesian-like approach that weights all templates based on the goodness of fit to estimate physical properties and uncertainties, as well as partially mitigates the degeneracies effects between physical parameters \citep{boquien2019}.

\begin{deluxetable*}{llll}[!h]
\tablecaption{CIGALE configuration}
\label{tab:tab4_cigale_set}

\addtocounter{table}{0}    
\tablewidth{0pt}
    \tablehead{ \multicolumn{1}{l}{Modules} & 
                \multicolumn{1}{l}{Description} & 
                \multicolumn{1}{l}{Parameters} & 
                \multicolumn{1}{l}{Values} }
    \startdata
    SFH: sfh2exp & age of the main stellar population    & age       & 500, 1000, 3000, 5000, 10000 \\
                 & e-folding time of the main stellar population   & tau\_main  & 1000, 3000, 5000, 10000 \\
                 & age of the burst population & burst\_age & 10, 50, 100, 300    \\
                 & e-folding time of the burst population    & tau\_burst & 30, 100, 300, 9000    \\
                 & mass fraction of the burst population    & f\_burst   & 0, 0.01, 0.1, 0.3, 0.5    \\
    \hline
    SSP: bc03    &  initial mass function   & imf            & 1 (\citet{chabrier2003})    \\
                 &                          & metallicity    & 0.02    \\
    \hline
    Nebular emission: nebular &    & $\textit{default}$ &  $\textit{default}$ \\                      
    \hline
    Dust attenuation:        & V-band attenuation in ISM   & Av\_ISM   & 0.3, 0.6, 1.0, 1.6, 2.3, 3.0, 3.8, 5.0, 8    \\
    dustatt\_modified\_CF00  & Av\_ISM / (Av\_BC+Av\_ISM)     & mu        & 0.44    \\
                             & attenuation power-law slope in the ISM & slope\_ISM & -0.7    \\
                             & attenuation power-law slope in the BC & slope\_BC  & -1.3        \\
    \hline
    Dust emission: dl2014  & mass fraction of PAH    & qpah  & 0.47, 1.12, 2.50, 3.90, 6.63    \\
                           & minimum radiation field   & umin  & 0.1, 0.5, 1.0, 5.0, 10.0, 25.0, 50.0    \\
                           & radiation field power-law slope  & alpha & 2.0, 2.5, 3.0        \\
    \hline
    AGN: fritz2006  & angle between equator and line of sight  & psy            & 0.001, 50.1, 89.990    \\
                    & AGN fraction                                     & fracAGN        & 0.0, 0.001, 0.01, 0.1, 0.3, 0.5, 0.7, 0.9    \\
                    & optical depth at 9.7 $\micron$                    & tau            & 1.0, 6.0    \\
                    & opening angle of the dust torus             & opening\_angle & 100.0    \\
                    & polar dust temperature                    & temperature    & 100.0    \\
                    & emissivity index of the polar dust               & emissivity    & 1.6    \\
    \hline
    Radio: radio  & FIR/radio correlation coefficient                                   & qir\_sf    & 2.15, 2.45, 2.78    \\
                  & power-law slope of emission related to stellar           & alpha\_sf  & 0.8    \\
                  & power-law slope of AGN radio emission                                & alpha\_agn & 0.7    \\
    \hline
    redshifting  & redshift grid   & redshift & 0.0001, \\
                 &                &          & 0.1 to 1.4 with a step of 0.2,  \\
                 &                &          &           1.5 to 3.9 with a step of 0.1, \\
                 &                &          &           4 to   6 with a step of 0.2  \\
    \enddata
\end{deluxetable*}

\subsubsection{Configuration}
\label{sec:cigale_config}

The CIGALE configuration models and parameter settings adopted in our work are summarized in Table~\ref{tab:tab4_cigale_set}.

We adopt the \textit{sfh2exp} model with a double exponential star formation history (SFH). 
By setting the mass fraction of the second burst, we determine whether the model experiences one or two star formation events, with the latter including one long-term star formation and one recent burst. %
Previous studies suggest that SMGs may undergo multiple intense star formation episodes, and an exponentially declining SFH also consistent with true values \citep{michalowski2014}. 
Both simulations and main-sequence models indicate that double SFHs provide a more likely description, regardless of whether galaxies evolve through mergers or in isolation \citep{speagle2014}.

We employ the initial mass function (IMF) model from \citet{chabrier2003} and adopt the simple stellar population (SSP) templates from \citet{bc03}.
The SSP models incorporate all phases of stellar evolution from zero-age main sequence to thermally pulsating asymptotic giant branch (TP-AGB) stages for low- and intermediate-mass stars, and core-carbon ignition for massive stars. 
The stellar models effectively constrain galaxy properties including star formation history, metallicity, and dust mass.

The continuum emission from thermally ionized gas in star-forming regions (e.g., HII regions) spans optical to radio wavelengths and may represent a non-negligible component for high-redshift galaxies \citep{boquien2019}. We implement the nebular emission template from \citep{inoue2011}.

Dust plays a fundamental role in the energy balance approach. 
For dust attenuation, we employ the physically motivated \textit{dusttt\_modified\_CF00} model, which accounts for dust grain mixing and geometric effects while distinguishing between different stellar environments (birth clouds and interstellar medium (ISM)). 
Tests on infrared-selected galaxy samples demonstrate that this model more accurately reproduces dust attenuation properties \citep{lofaro2017, barrufet2020}.

The \textit{dl2014} templates, an updated version of \citet{dl07}, are adopted to model dust emission, which incorporates a mixture of amorphous silicate and graphite grains with variable PAH abundances. 
The dust emission consists of two components: diffuse dust heated by the general interstellar radiation field and localized emission from star-forming regions.
Analysis of 850 $\micron$ sources by \citet{seo2018} indicates that \textit{dl2014}, with its more comprehensive physical parameters, provides superior fitting results. 
The 850 $\micron$ selection preferentially picks out colder SMGs, whose ISM emission peaks at longer wavelengths. Accordingly, we include lower radiation field intensities ($U_{min}$) in our parameter space.

\begin{figure*}[!h]
    \centering
    \includegraphics[width=0.98\textwidth]{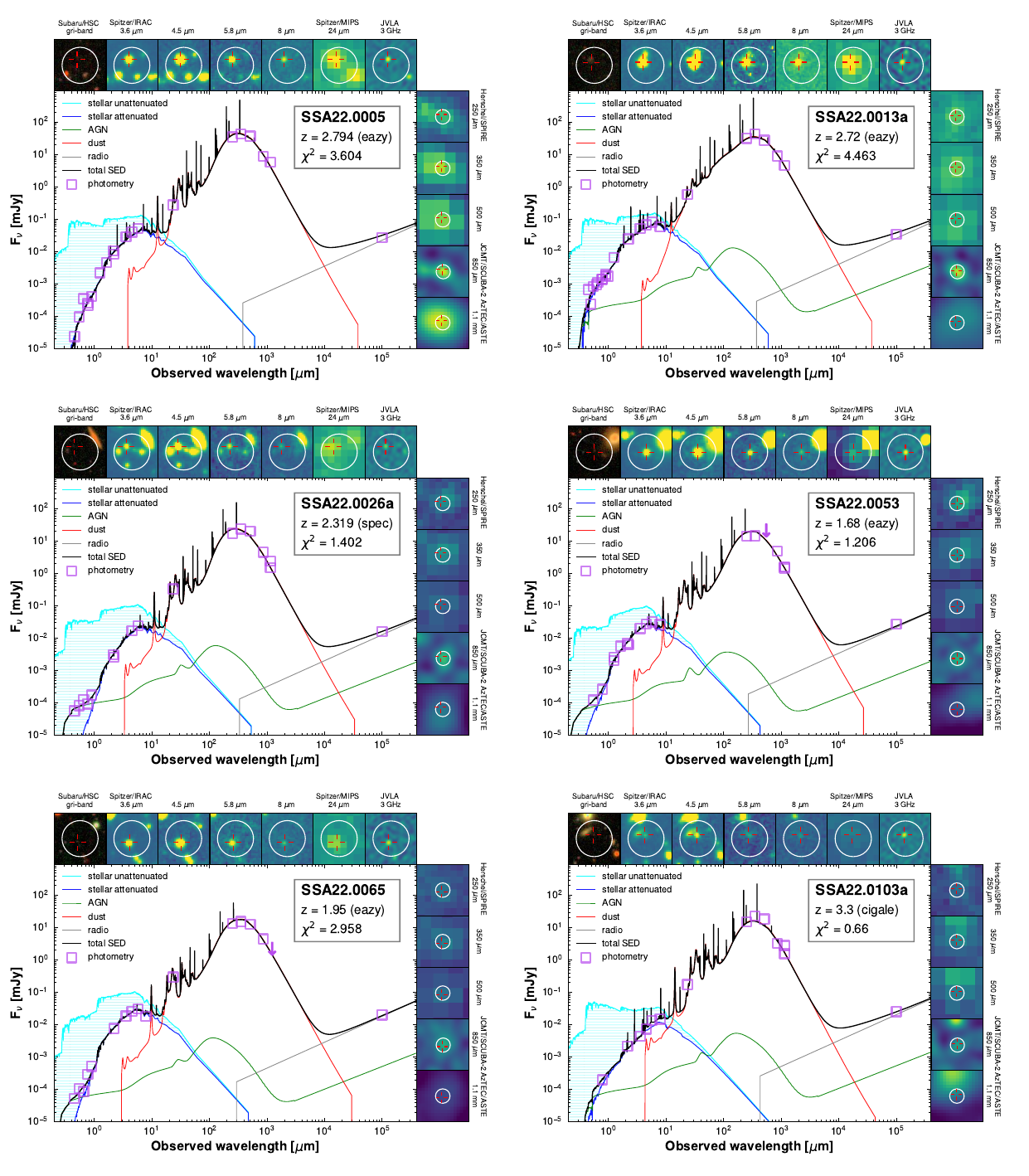}
    \caption{The best-fit SEDs and multiwavelength thumbnails for the SMGs.
    Various components are represented by colored lines, with cyan hatched regions indicating stellar light attenuated by dust.
    Based on our analysis, many SMGs exhibit a distinctive ``plateau" feature at bluer wavelengths. The fitting quality for these sources—such as SSA22.0026a, SSA22.0053, and SSA22.0065—is significantly improved by incorporating a minor AGN contribution. This phenomenon appears to be present in other surveys as well.
    Text boxes accompanying each SED provide the source name, redshift (including estimation method), and the reduced $\chi^2$ of the fit. Thumbnail images in the margins display observations from optical to radio wavelengths. White circles indicate the SCUBA-2 beam area, while red crosses mark the positions of identified counterparts.  }
    \label{fig:fig4-3_cigale_example}
\end{figure*}

AGNs may play a crucial role in driving galaxy evolution \citep{toft2014}, although not all SMGs necessarily undergo an AGN phase \citep{johnson2013}. Observations indicate that $\sim$ 20-40\% of SMGs show signs of AGN activity \citep{alexander2005, laird2010, an2019, stach2019}. In systems with strong AGN contributions, the radio emission from the AGN can compete with or even dominate over that from star formation, necessitating the inclusion of an AGN component in our analysis.
We adopt the AGN model from \citep{fritz2006}, which self-consistently treats the central active nucleus, dust scattering in the torus, and thermal re-emission while accounting for viewing angle effects. 
Our fitting experience reveals that the optical emission in many sources cannot be fully explained by dust-attenuated stellar radiation alone, often showing a smooth transition to a ``plateau" at bluer wavelengths (e.g., Figure~\ref{fig:fig4-3_cigale_example}, sources SSA22.0026a, SSA22.0053, and SSA22.0065). 
The inclusion of even minimal AGN contributions (fractions of 0.001-0.01) significantly improves the SED fits by better reproducing this blue excess emission.

The radio emission in most galaxies is predominantly powered by star formation activity, consisting of the non-thermal synchrotron radiation from relativistic electrons accelerated in supernova remnants, and  thermal bremsstrahlung emission (free-free emission) from HII regions surrounding massive stars. While bremsstrahlung exhibits a flat spectrum, its contribution becomes negligible compared to synchrotron emission at frequencies below 30 GHz \citep{helou1985, condon1992}.
The well-established radio-FIR correlation arises because both wavebands trace star formation activity (Section~\ref{sec:radio+mips24}). However, we allow for variations in the correlation coefficient to account for potential redshift evolution of the radio-FIR relation \citep{speagle2014}, and possible radio excess caused by AGN contamination. 

For sources lacking spectroscopic redshifts or those too optically faint for reliable \texttt{EAZY} photometric redshift estimates, we simultaneously derive their redshifts during the \texttt{CIGALE} SED fitting process. 
To maintain computational efficiency and optimize parameter space exploration, we adopt redshift bins with varying resolution: coarse sampling ($\Delta z$ = 0.2) for low/high-redshift range ($z =$ 0-1.5 and $z =$ 4-6), fine sampling ($\Delta z$ = 0.1) for the critical redshift range ($z =$ 1.5-4), where the majority of SMGs are expected to reside \citep{dud2020}.

\subsubsection{Fitting Result}
\label{sec:cigale_fitting}

For sources lacking `$z_{spec}$' and `$z_{EAZY}$' estimates, we derive their redshifts ($z_{CIGALE}$) simultaneously during the \texttt{CIGALE} SED fitting process using the photometric redshift mode.

Comparison with $z_{spec}$ and $z_{EAZY}$ reveals residuals of 0.091 ± 0.053 and 0.052 ± 0.022, respectively (Figure~\ref{fig:fig4-2_zcigale_vs_z}).
Given the typically higher accuracy of photometric redshifts derived from optical/NIR bands compared to longer wavelengths, we adopt the $z_{spec}$, if available, followed by the $z_{EAZY}$, and then the $z_{CIGALE}$ in analysis.


We conducted visual inspections of all SED fitting results, excluding sources with poor fits. Specifically, we removed sources with weak or no constraints in the FIR/submillimeter bands, as well as those lacking constraints in other critical wavelength ranges. This quality control process eliminated 27 sources ($\sim$ 10\% of the sample) from our final analysis.

Additionally, we identified six sources (SSA22.0113, SSA22.0193, SSA22.0196a, SSA22.0209, SSA22.0329, and SSA22.0374) that consistently produced poor fits when using their fixed $z_{EAZY}$ photometric redshifts. Our analysis suggests potential inaccuracies in the \texttt{EAZY} photometric redshifts for these sources. Allowing their redshifts to vary as free parameters yielded significantly improved fits, and we consequently adopted these revised solutions (marked in blue in the relevant figures).

Our final catalog contains 221 SMGs (corresponding to 186 SCUBA-2 sources) after excluding 27 unreliable fits and incorporating the 6 revised solutions. Notably, this includes 115 newly detected and analyzed SMGs associated with 95 SCUBA-2 sources, representing important additions to the known population of submillimeter galaxies.

\section{RESULTS}
\label{sec:results}

In this section, we present and discuss the physical properties of our identified sample, including redshift (Section~\ref{sec:redshift}), SFR/stellar mass (Section~\ref{sec:sfr_mstar}), infrared luminosity (Section~\ref{sec:lir}), dust mass (Section~\ref{sec:dust}), and dust attenuation (Section~\ref{sec:av}).

\citet{liao2024, uematsu2024} compared the physical properties derived from the \texttt{CIGALE} and \texttt{MAGPHYS}. They found that differences between dust modeling assumptions in these codes lead to modest systematic offsets in estimated dust masses and temperatures, as well as in stellar masses.
Our \texttt{CIGALE} configuration closely follows that of \citet{liao2024}; therefore, we expect good overall agreement between our derived physical parameters and those from \texttt{MAGPHYS}—except for dust mass, which we explicitly discuss in Section~\ref{sec:dust}. 
In the following sections, we will directly compare our estimates with published values.


\begin{figure*}[!ht]
    \centering
    \includegraphics[width=0.96\textwidth]{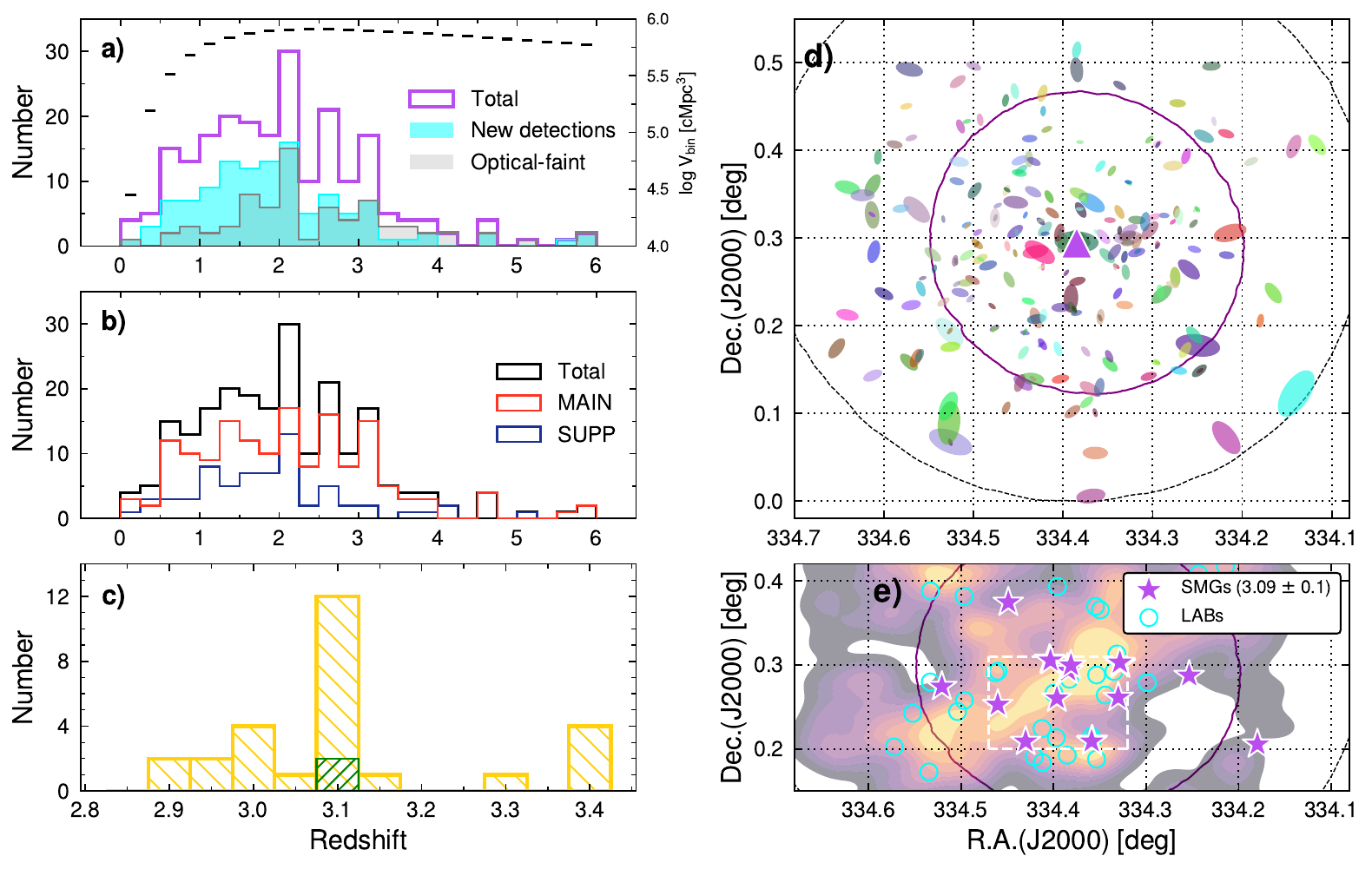}
    \caption{The redshift distribution of SSA22 SMGs.
    Panel (a): The purple line shows the redshift distribution of the full sample (median $z=2.00\pm0.08$). Cyan and grey shadings denote newly detected and optically faint SMGs, respectively. Optically faint SMGs—-two-thirds of the sample-—have a median redshift of $\sim$ 2.2, rising to $\sim$ 3.0 for those undetected in optical/NIR. The 115 new detections peak slightly lower at $\sim$ 1.9. Short bars on the right y-axis indicate comoving volume per redshift bin.
    Panel (b): The redshift distributions of sources identified via different methods. The ``IRAC8" and ``iK-color" selections yield distributions consistent with those derived from ``Radio" and ``MIPS24" counterparts, reaffirming the validity of the former methods for identifying high-redshift SMGs.
    Panel (c): Zoomed view of the redshift distribution for the total sample in the range $z =$ 2.8-3.4. The green shaded regions indicate two SMGs detected by the ADF22 survey that were not recovered by our identification method. 
    Nevertheless, the SMGs identified through our approach still provide a robust tracer of overdense regions, underscoring the reliability of SMGs as probes of cosmic structure formation and their utility in mapping the mass distribution within high-redshift overdensities.
    Panel (d): Projection distribution of the SMGs sample. Symbol sizes scale with 850 $\micron$ flux density, while colors and rotation angles are randomly assigned. The black dashed line outlines the SCUBA-2 850 $\micron$ footprint, the purple solid line marks the deep survey region of SSA22, and the purple triangle indicates the location of source SSA22.0000. 
    Panel (e): The projected distribution of SMGs within the redshift range of $3.09 \pm 0.1$.
    The background contours represent the surface density of LAEs, which trace the large-scale structure at $z = 3.09$; brighter regions correspond to higher overdensities. 
    The white dashed rectangle approximately corresponds to the “SSA22a” field \citep{steidel1998}.
    The majority of massive SMGs are spatially coincident with the cores of large-scale structures, closely associated with overdense regions and other galaxy populations.}
    \label{fig:fig5-1-1_redshift_distribution}
\end{figure*}


\subsection{Redshift Distribution}
\label{sec:redshift}

%


Our final sample comprises 221 SMGs (counterparts to 186 SCUBA-2 sources), with a median redshift of 2.00 $\pm$ 0.08 (bootstrap error) and 16th-84th percentiles of 0.97-3.08. 
Our results are consistent with previous multiwavelength identifications of SMGs, which typically report median redshifts of $\sim$ 2-2.3 \citep{zavala2018, an2019, barrufet2020, hyun2023}. However, larger samples of ALMA 870 $\micron$-selected SMGs show systematically higher median redshifts of $\sim$ 2.5-2.7 \citep{cunha2015, cowie2017, dud2020}. 
As noted by \citet{hyun2023}, the redshift distribution of SMGs identified through radio and near-infrared detection appears truncated at $z \sim$ 4, resulting in a lower median redshift of $\sim$ 2, a finding that aligns well with our results. This discrepancy highlights the limitations of multiwavelength identification methods, where finite survey depths and uneven wavelength coverage can lead to incomplete identification of high-redshift SMGs.

Figure~\ref{fig:fig5-1-1_redshift_distribution} panels (a) and (d) show the redshift histogram and spatial distribution of the full SMG sample, respectively.
The horizontal bars in panel (a) represent the comoving cosmological volume probed within each redshift bin. The data reveal a pronounced evolution in SMG number density: at z $\lesssim$ 4, the density increases sharply by approximately a factor of 6, while at $z \sim$ 1-3 it rises more gradually, maintaining values between $\sim$ 1.78-3.16 $\times$ 10$^{-5}$ cMpc$^{-3}$. This two-phase evolutionary pattern suggests distinct regimes of SMG formation and growth across cosmic time (see \citetalias{paperiv} of this series).
%
Notably, this includes 115 newly identified SMGs (corresponding to 95 SCUBA-2 sources with fluxes $\gtrsim$ 2.8-3.5 mJy) that we report for the first time, which exhibit a slightly lower median redshift of $\sim$ 1.9.

Our analysis reveals a clear dichotomy in the redshift distribution based on optical/NIR detectability: approximately two-thirds of the SMGs in our catalog qualify as optically bright sources \citep[with $\geqslant$ 4 available optical/NIR photometric bands;][]{cunha2015}, exhibiting a median redshift of $z \sim$1.75, significantly lower than the typical SMG redshift distribution. The remaining one-third, classified as optically faint sources ($\textless$ 4 available bands), show a higher median redshift of $z \sim$ 2.2.
Moreover, the 27 sources completely undetected in optical/NIR bands (excluding SSA22.0347 which lies outside the optical/NIR images coverage) demonstrate an even more extreme redshift distribution, with a median of $z \sim$ 3.0. This value not only exceeds that of the optically faint population but also lies above the characteristic redshift peak for SMGs. These results confirm that SMGs represent a population of heavily dust-obscured systems preferentially located in the high-redshift universe, with the most severely obscured members tending to reside at even greater cosmological distances.

When multiple redshift estimates are available for a source, we adopt them in order of reliability: $z_{spec} > z_{EAZY} > z_{CIGALE}$ (Section~\ref{sec:cigale_fitting}).
Our sample contains 30 sources with spectroscopic redshifts ($z_{spec}$), 115 with EAZY photometric redshifts ($z_{EAZY}$), and 76 relying solely on CIGALE long-wavelength photometric redshifts ($z_{CIGALE}$), exhibiting median redshifts of $\sim$ 2.42, 1.65, and 2.15, respectively.

The ``MAIN" sample of 155 sources identified through radio and ``$MIPS24$" methods shows a median redshift of 2.07, with those having SCUBA-2 SNR $\textgreater$ 4 displaying a slightly higher median of 2.2. These results are consistent with the S2TDF survey under comparable SCUBA-2 depths and identification methods \citep{hyun2023}. 
The supplementary (``SUPP") sample of 66 sources identified solely via ``$IRAC8$" and ``$iK\text{-}color$" methods exhibits a lower median redshift of 1.83 (Figure~\ref{fig:fig5-1-1_redshift_distribution} image (b)).
However, when considering all ``$IRAC8$/$iK\text{-}color$" selected galaxies (including those also detected in ``radio/$MIPS24$"), the median redshift rises to 2.00, matching the overall sample distribution.

When examining the results from each method individually (Figure~\ref{fig:fig3-2-3_misid_hist}): ``$VLA$" detection identifies 124 SMGs with median $z =$ 2.10; ``$MIPS24$" selects 62 SMGs at median $z =$ 1.92; ``$IRAC8$" recovers 94 SMGs with median $z =$ 2.12; and the ``$iK\text{-}color$" method identifies 91 SMGs with median $z =$ 1.80. 

The lack of significant differences in properties between the `MAIN’ and `SUPP’ samples (Table~\ref{tab:tab5_average_properties}), as well as across identification methods, validating the ``$IRAC8$" and ``$iK\text{-}color$" methods as effective tools for identifying high-redshift SMGs.
The higher median redshift of $IRAC8$-selected sources particularly highlighting the importance of observational depth for robust identification.

\begin{figure}
    \centering
    \includegraphics[width=0.45\textwidth]{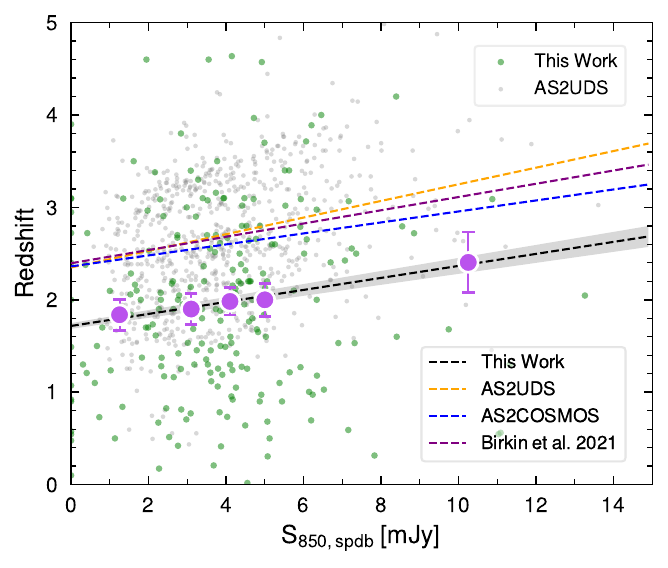}
    \caption{Redshift versus 850 $\micron$ flux density.
    The binned median redshifts represented by purple circles, with the linear fit is shown as a black solid line, accompanied by a shaded region indicating the 16th–84th percentile range.
    We compare our results with those from AS2UDS \citep{stach2019,dud2020}, AS2COSMOS \citep{simpson2020}, and \citet{birkin2021}. 
    Our analysis yields a slope consistent with these previous surveys, revealing a statistically significant positive correlation: brighter submillimeter sources tend to reside at higher redshifts.
    The lower intercept of our fit likely reflects incompleteness in high-redshift identifications.}
    \label{fig:fig5-1-3_red_f}
\end{figure}


\citet{cowie2017,cowie2018} reported that SMGs with brighter submillimeter fluxes tend to lie at higher redshifts.
Analyses of the ALESS and AS2UDS samples by \citet{simpson2014,simpson2017} found a similar trend but argued that it is not statistically significant and may be an artifact caused by incomplete redshift information at fainter flux levels.
\citet{stach2019}  demonstrated this phenomenon using a sample of 708 AS2UDS SMGs, attributing it to the ``downsizing" effect \citep{cowie1996}. Additionally, due to the negative $k$-correction, higher flux densities tend to be found at higher redshifts, indicating significant infrared luminosity evolution.
This trend is driven by the presence of higher gas fractions and stronger SFR in distant galaxies, resulting in greater luminosities \citep{simpson2020, birkin2021}.
Dividing our SMGs into five flux bins, we find:
\begin{equation}
    z_{\text{median}} = (1.717 \pm 0.031) + (0.065 \pm 0.006) \times S_{850} \ [\text{mJy}],
    \label{eq:z_vs_f}
\end{equation}
a slope consistent with AS2COSMOS, AS2UDS, and CO surveys (0.06–0.09 mJy$^{-1}$; \citealp{stach2019, simpson2020, birkin2021}) (Figure~\ref{fig:fig5-1-3_red_f}). 
In contrast, the median 850 $\micron$ flux shows no clear evolution with redshift, remaining near $\sim$ 4 mJy. This indicates that while the brightest SMGs favor higher redshift, the general SMG population shows no strong link between submillimeter flux and redshift.

\subsubsection{Overdensity}
\label{sec:over}
\citet{steidel1998, steidel2000} identified a large-scale structure at $z = 3.09$ in the central region of the SSA22 deep field using low-mass LAEs.
Our identified SMGs also exhibit significant overdensity within this redshift slice (Figure~\ref{fig:fig5-1-1_redshift_distribution} (c)).

When considering the broader redshift interval $z = 2.8$–$3.4$ the overdense region contains 17 SMGs, corresponding to a $\sim$8.4$\sigma$ (7.8$\sigma$) overdensity for bin widths of 0.05 (0.1).
The ADF22 survey \citep{umehata2017a, umehata2018} contributes five additional counterparts, two of which lie within the overdense region (green shaded area), confirming that our identification methods already robustly trace this large-scale structure at $\sim$7$\sigma$ significance.

Figure~\ref{fig:fig5-1-1_redshift_distribution} plot (e) shows the spatial distribution of SMGs within the structure; nearly all reside in the dense core. Within the central $3.6 \times 2.7$ pMpc$^2$ region at $z = 3.09 \pm 0.1$—approximately corresponding to the “SSA22a” region (white dashed rectangle)—we find 13 SMGs. 
This yields a surface density of $\sim$1.3 pMpc$^{-2}$, consistent with the AzTEC SMG density reported by \citet{umehata2014}. Assuming these galaxies reside within the cosmic web node traced by LAEs (with a characteristic depth of $\sim$4.9 pMpc; \citealt{matsuda2005}), their comoving volume density is $\sim$0.27 pMpc$^{-3}$ (13/47.5).

The 850 $\micron$ fluxes of these SMGs are fairly uniformly distributed between 1 and 10 mJy, peaking at 4–6 mJy. Their total stellar mass sums to $\sim 3.1 \times 10^{12}$ M$_\odot$, and their combined SFR reaches $\sim 9100$ M$_\odot \, \text{yr}^{-1}$. This corresponds to volume-averaged contributions of $6.6 \times 10^{10}$ M$_\odot \, \text{pMpc}^{-3}$ in stellar mass density and 193 M$_\odot \, \text{yr}^{-1} \, \text{pMpc}^{-3}$ in SFR density.
The ultra-deep survey of the core region \citep[ADF22+, 5$\sigma$;][]{huang2025} yields an even higher stellar mass and SFR volume density, reaching 1.2 $\times 10^{11}$ M$_\odot \, \text{pMpc}^{-3}$ and $\sim$ 296 M$_{\odot}$ yr$^{-1}$ pMpc$^{-3}$, respectively.

We further estimate the overdensity using angular surface densities. From the SCUBA-2 source perspective, these 13 SMGs originate from 9 distinct SCUBA-2 detections, yielding an angular density of $\sim$700 deg$^{-2}$—comparable to the S2CLS field average of $\sim$570 deg$^{-2}$ \citep{geach2017}. However, as noted by \citet{wangg2025}, our 9 sources are confined to an extremely narrow redshift slice ($z = 3.09 \pm 0.1$, i.e., $\Delta r \approx 186$ pMpc), whereas S2CLS is sensitive to a much broader redshift range (e.g., $z = 1.8$–$3.4$, or $\sim$1837 pMpc; \citealt{dud2020}). Accounting for this, the number density in the overdense region is enhanced by a factor of $\sim$12 relative to the blank field.

Using deblended SMG catalogs for a fairer comparison, we compute the angular surface densities of SMGs in the same redshift window ($z = 3.09 \pm 0.1$) from AS2UDS \citep{dud2020} and JWST-TDF \citep{hyun2023}, finding $\sim$66 deg$^{-2}$ and 34 deg$^{-2}$, respectively. The overdensity represents a factor of $\sim$15–29 enhancement over the field.
Thus, overall, the SMG overdensity exceeds blank field average by more than an order of magnitude.

As shown in the figure~\ref{fig:fig5-1-1_redshift_distribution} (e), Lyman-alpha Blobs (LABs)—objects exhibiting extreme galactic activity \citep{geach2005, geach2014}—are co-spatial with our SMG sample within the large-scale structure of SSA22.
Moreover, studies have shown that SMGs are frequently associated with other populations of galaxies within large-scale structures \citep{tamura2009, umehata2014, wangg2025} and co-evolve with these cosmic environments \citep{umehata2019}.
These results highlight the role of large-scale structures in driving galaxy evolution, while conversely underscoring the potential of massive dusty galaxies as tracers of cosmic structure formation—particularly for mapping mass distributions in high-redshift overdense regions.

\movetabledown=60mm
\begin{rotatetable*}

\begin{deluxetable*}{lccccccccc}
\tablecaption{Average properties of the SSA22 SMGs}
\label{tab:tab5_average_properties}
\tablewidth{0pt}

\tablehead{
    \colhead{Porperties} & \colhead{Total sample} &\colhead{New detections} \tablenotemark{a} & \colhead{`MAIN'} \tablenotemark{b} & \colhead{`SUPP'} & \colhead{Conservative Sample} \tablenotemark{c} & \colhead{$z <$ 2.0} &\colhead{$z \geqslant$ 2.0} & \colhead{Opt/NIR faint} \tablenotemark{d}  &\colhead{Opt/NIR bright} \\
    \colhead{} & \colhead{221 ids}     &\colhead{115 ids}       & \colhead{155 ids}      & \colhead{66 ids}           & \colhead{133 ids}             & \colhead{110 ids}        &\colhead{111 ids}                  &\colhead{73 ids}                  &\colhead{148 ids} }

\startdata
Redshift                                      & 2.00$^{+1.08}_{-1.03}$  & 1.90$^{+0.98}_{-0.89}$  & 2.07$^{+1.02}_{-1.12}$  & 1.83$^{+0.77}_{-0.81}$  & 2.24$^{+0.86}_{-0.82}$  & 1.25$^{+0.47}_{-0.60}$  & 2.70$^{+0.74}_{-0.57}$  & 2.20$^{+1.25}_{-0.70}$  & 1.75$^{+1.01}_{-0.93}$  \\ 
log M$_{star}$ [$M_{\odot}$]                  & 11.19$^{+0.49}_{-0.82}$  & 11.20$^{+0.42}_{-0.76}$  & 11.23$^{+0.45}_{-0.68}$  & 11.03$^{+0.65}_{-0.75}$  & 11.28$^{+0.43}_{-0.60}$  & 10.93$^{+0.58}_{-0.93}$  & 11.32$^{+0.48}_{-0.56}$  & 11.29$^{+0.54}_{-0.65}$  & 11.14$^{+0.48}_{-0.84}$  \\ 
log SFR [$M_{\odot}$ yr$^{-1}$]               & 2.22$^{+0.63}_{-0.97}$  & 2.07$^{+0.64}_{-0.84}$  & 2.25$^{+0.62}_{-1.05}$  & 2.06$^{+0.69}_{-0.71}$  & 2.38$^{+0.52}_{-0.71}$  & 1.58$^{+0.76}_{-0.70}$  & 2.55$^{+0.43}_{-0.50}$  & 2.25$^{+0.65}_{-0.80}$  & 2.12$^{+0.60}_{-0.94}$  \\ 
log L$_{IR}$ [$L_{\odot}$])                   & 12.35$^{+0.49}_{-0.69}$  & 12.25$^{+0.50}_{-0.63}$  & 12.40$^{+0.43}_{-0.73}$  & 12.32$^{+0.59}_{-0.68}$  & 12.51$^{+0.36}_{-0.49}$  & 11.91$^{+0.44}_{-0.59}$  & 12.69$^{+0.28}_{-0.35}$  & 12.42$^{+0.44}_{-0.48}$  & 12.31$^{+0.52}_{-0.76}$  \\ 
log M$_{dust}$ [$M_{\odot}$]                  & 9.29$^{+0.43}_{-0.55}$  & 9.22$^{+0.46}_{-0.58}$  & 9.22$^{+0.44}_{-0.55}$  & 9.48$^{+0.41}_{-0.45}$  & 9.22$^{+0.39}_{-0.46}$  & 9.35$^{+0.44}_{-0.81}$  & 9.19$^{+0.38}_{-0.39}$  & 9.35$^{+0.37}_{-0.42}$  & 9.27$^{+0.44}_{-0.61}$  \\ 
log M$_{star}$/$L_H$ [$M_{\odot}/L_{\odot}$]  & 0.21$^{+0.52}_{-0.32}$  & 0.20$^{+0.40}_{-0.30}$  & 0.21$^{+0.48}_{-0.32}$  & 0.21$^{+0.53}_{-0.33}$  & 0.20$^{+0.41}_{-0.41}$  & 0.29$^{+0.49}_{-0.27}$  & 0.12$^{+0.48}_{-0.41}$  & 0.34$^{+0.64}_{-0.54}$  & 0.17$^{+0.37}_{-0.28}$  \\
$A_V$ [mag]                                     & 3.09$^{+1.40}_{-1.42}$  & 2.82$^{+1.18}_{-1.28}$  & 3.08$^{+1.14}_{-1.41}$  & 3.16$^{+1.93}_{-1.51}$  & 3.09$^{+0.89}_{-1.36}$  & 3.10$^{+1.97}_{-1.53}$  & 3.06$^{+0.92}_{-1.30}$  & 3.87$^{+1.33}_{-1.62}$  & 2.82$^{+1.14}_{-1.20}$  \\ 
log Age$_m$ [yr] \tablenotemark{e}            & 8.75$^{+0.57}_{-0.33}$  & 9.04$^{+0.28}_{-0.49}$  & 8.75$^{+0.58}_{-0.27}$  & 9.05$^{+0.27}_{-0.64}$  & 8.70$^{+0.57}_{-0.28}$  & 9.17$^{+0.15}_{-0.51}$  & 8.62$^{+0.58}_{-0.25}$  & 8.91$^{+0.41}_{-0.53}$  & 8.75$^{+0.57}_{-0.31}$  \\ 
log sSFR [yr$^{-1}$]                          & -9.05$^{+1.19}_{-0.86}$  & -9.18$^{+1.26}_{-0.71}$  & -9.08$^{+1.21}_{-0.82}$  & -9.03$^{+1.18}_{-1.00}$  & -9.00$^{+1.13}_{-0.72}$  & -9.37$^{+1.50}_{-0.84}$  & -8.83$^{+0.97}_{-0.52}$  & -9.15$^{+1.30}_{-0.86}$  & -9.00$^{+1.13}_{-0.85}$  \\ 
$\Delta$ MS \tablenotemark{f}                   & -0.16$^{+0.98}_{-0.61}$  & -0.21$^{+0.91}_{-0.51}$  & -0.19$^{+0.98}_{-0.52}$  & -0.03$^{+0.97}_{-0.84}$  & -0.20$^{+1.00}_{-0.49}$  & -0.20$^{+1.16}_{-0.81}$  & -0.15$^{+0.93}_{-0.47}$  & -0.27$^{+1.03}_{-0.65}$  & -0.08$^{+0.92}_{-0.60}$  \\
\enddata

\tablecomments{Sample grouping is based on various classification schemes. Average properties represent median values of individual parameters, with uncertainties representing the 16th-84th percentile range.}
\tablenotetext{a}{Counterparts of submillimeter sources detected in our deeper observations.}
\tablenotetext{b}{“MAIN” subsample refers to counterparts identified via $VLA$, $MIPS24$, or direct ALMA detections. “SUPP” refers to counterparts identified using $IRAC8$ or the $iK\text{-}color$ method.}
\tablenotetext{c}{To minimize contamination, we adopt a conservative sample selection, retaining only sources with redshift $z >$ 1 that have $VLA$ or $MIPS24$ counterparts.}
\tablenotetext{d}{The optically faint subsample consists of identifications with detections in fewer than four opt/NIR filters, whereas the optically bright subsample includes sources detected in four or more opt/NIR filters.}

\tablenotetext{e}{The mass-weighted ages of the galaxies, computed as $\frac{\int^t_0 t' \,\text{SFR}(t') \, dt'}{\int^t_0 \text{SFR}(t') \, dt'}$\citep{noll2009}, are derived from \texttt{CIGALE}.}
\tablenotetext{f}{The offset of SMGs from the main sequence, defined as $\Delta$MS = $\log_{10}$(sSFR/sSFR$_\text{MS}$), where the main sequence relation is adopted from \citet{speagle2014}. See also \citetalias{paperiv}.}

\end{deluxetable*}
\end{rotatetable*}

\subsection{SFRs and Stellar Mass}
\label{sec:sfr_mstar}

\begin{figure*}
    \centering
    \includegraphics[width=\textwidth]{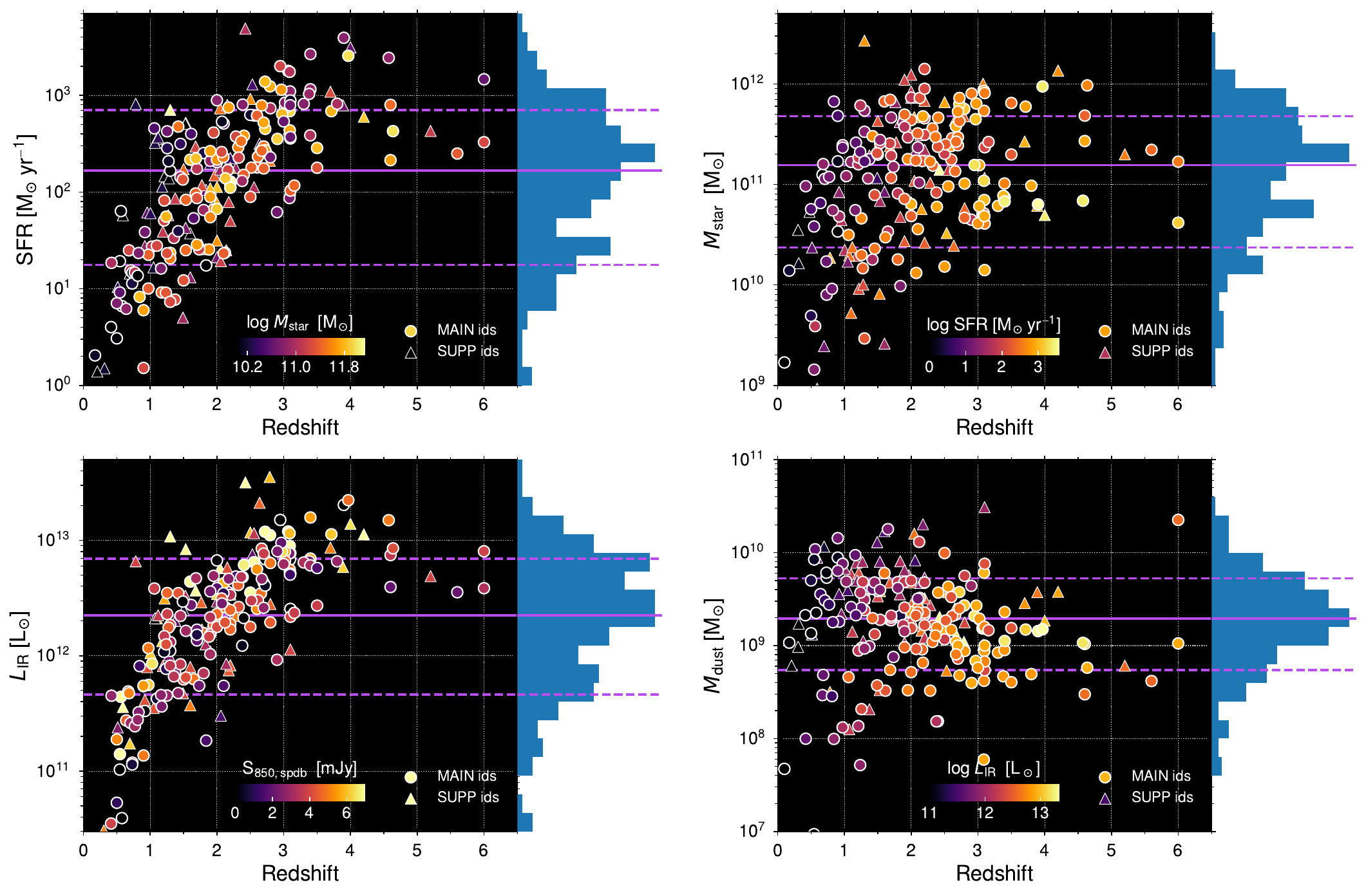}
    \caption{The evolution of fundamental properties for the 221 SMGs is examined as a function of redshift.
    In each panel, circles denote sources identified via the ``Radio/MIPS24" method, while triangles represent those selected through the ``IRAC8/iK-color" approach. Both symbol types are color-coded according to additional characteristics.
    The median values are represented as solid lines, bounded by two dashed lines indicating the 16th–84th percentile range.
    A horizontal histogram inset on the right side of each panel illustrates the distribution of the respective property.  }
    \label{fig:fig5-2-1_props}
\end{figure*}

The distributions of SFRs and stellar masses ($M_\text{star}$) for the 221 SMGs identified in the SSA22 deep field are presented in Figure~\ref{fig:fig5-2-1_props}, with the histogram shown to the right. 

The overall sample exhibits a median SFR of $166 \pm 25$ M$_\odot$ yr$^{-1}$, with the 16th–84th percentile range spanning 18–703 M$_\odot$ yr$^{-1}$. Over 90\% of the sources have SFRs exceeding 10 M$_\odot$ yr$^{-1}$, and the highest SFR reaches $\gtrsim 4000$ M$_\odot$ yr$^{-1}$. The median stellar mass is $(1.55 \pm 0.22) \times 10^{11}$ M$_\odot$, with 68\% of the sample falling within the range $0.24–4.77$ $\times$ 10$^{11}$ M$_\odot$. Approximately 92\% of the sources have stellar masses $\gtrsim 10^{10}$ M$_\odot$, and the most massive galaxy reaches $2.7 \times 10^{12}$ M$_\odot$.

The stellar mass distribution of SMGs in the SSA22 field is consistent with previous studies, where median masses typically range from $0.8$–$1.7 \times 10^{11}$ M$_\odot$ \citep{cunha2015, zavala2018, dud2020, shim2022, hyun2023}. Although the values are similar in magnitude, the median SFR in our SSA22 sample is $\sim$ 0.1–0.2 dex lower than in the literature. 

From the AS2UDS sample \citep{dud2020}, we selected sources at $z > 1$ with radio or 24 $\micron$ counterparts. Their median stellar mass and star formation rate (SFR) are $1.45 \pm 0.07 \times 10^{11}$ M$_{\odot}$ and $258 \pm 19$ M$_{\odot}$ yr$^{-1}$, respectively. Applying the same redshift cut to the multi-band identified S2TDF sample \citep{hyun2023} yields median values of $1.53 \pm 0.22 \times 10^{11}$ M$_{\odot}$ for stellar mass and $224 \pm 36$ M$_{\odot}$ yr$^{-1}$ for SFR.
Under identical criteria, we selected a conservative sample at $z > 1$ with identifications from VLA or MIPS 24 $\micron$ (Table~\ref{tab:tab5_average_properties}). This sample has a median stellar mass of $1.90 \pm 0.21 \times 10^{11}$ M$_{\odot}$ and a median SFR of $241 \pm 29$ M$_{\odot}$ yr$^{-1}$. While the SFR is consistent with the literature values cited above, the stellar mass is somewhat larger, a discrepancy that can likely be attributed to differences in SED fitting methodologies and results (see Section~\ref{sec:results}).
Our radio or 24 $\micron$ identified sample also includes a number of “quenched” systems, yet their SFRs are consistent with those reported in literature catalogs. 
If we further restrict the selection to sources at $z > 1$ with radio or 24 $\micron$ counterparts that lie on or above the star-forming main sequence, the median stellar mass and SFR become $1.55 \times 10^{11}$ M$_{\odot}$ and $321$ M$_{\odot}$ yr$^{-1}$, respectively. Under the same conditions, the corresponding median values for the AS2UDS subsample are $1.33 \times 10^{11}$ M$_{\odot}$ and $300$ M$_{\odot}$ yr$^{-1}$, and for the S2TDF subsample are $1.48 \times 10^{11}$ M$_{\odot}$ and $228$ M$_{\odot}$ yr$^{-1}$.
Therefore, we suggest that this may arise from relatively weak constraints in the far-infrared during SED fitting, which could lead to greater uncertainty in derived properties and consequently their misclassification.

The ADF22+ survey \citep{huang2025} has identified a population of relatively faint dusty star-forming galaxies (DSFGs) in the core of the protocluster. 
Within the redshift range $z = 3.09 \pm 0.1$, these galaxies have median stellar masses and SFRs of $4.6 \times 10^{10} \, M_\odot$ and $86 \, M_\odot \, \text{yr}^{-1}$, respectively—significantly lower than the corresponding medians for our SMGs in the same redshift interval, which are $\sim 1.1 \times 10^{11} \, M_\odot$ and $538 \, M_\odot \, \text{yr}^{-1}$.
Nevertheless, their stellar mass and SFR volume density are nearly twice ours (Section~\ref{sec:over}). This implies that conventional, shallower submillimeter/millimeter surveys miss a substantial population of faint dusty galaxies, yet these galaxies collectively contribute more to the star formation activity than the bright SMGs alone.
Moreover, these DSFGs all lie on the star-forming main sequence, indicating that the majority of protocluster members are undergoing steady, gas-fueled star formation sustained by inflows from the cosmic web—rather than being driven by major mergers \citep{huang2025}.

The median SFR and $M_\text{star}$ for the `MAIN' and `SUPP' samples are 178 M$_\odot$ yr$^{-1}$, $1.68 \times 10^{11}$ M$_\odot$ and 116 M$_\odot$ yr$^{-1}$,  $1.06 \times 10^{11}$ M$_\odot$, respectively. However, we emphasize that the physical properties of sources identified through different methods show no significant differences. The median values of SFR, $M_\text{star}$, $L_\text{IR}$, and $M_\text{dust}$ are in close agreement, spanning ranges of 140-190 M$_{\odot}$ yr$^{-1}$, $1.7-1.8 \times 10^{11}$ M$_{\odot}$, $2.2-2.7 \times 10^{12}$ L$_{\odot}$, $1-2 \times 10^9$  M$_{\odot}$, respectively.


The median deblended 850  fluxes for counterparts of newly detected sources ($\sim$ 2.8-3.5 mJy) and S2CLS-detected sources  ($\gtrsim$ 3.5–4.2 mJy) are 3.2 mJy and 4.9 mJy, respectively. These two SMG groups show median SFR and $M_{\rm star}$ values of 117 M$_{\odot}$ yr$^{-1}$, 1.59 $\times$ 10$^{11}$ M$_{\odot}$ and 186 M$_{\odot}$ yr$^{-1}$, 1.55 $\times$ 10$^{11}$ M$_{\odot}$, correspondingly.
These results reveal systematic differences in SFR distributions across flux bins, the relation was derived by least-squares fitting: log$_{10}$(SFR) = (0.49 $\pm$ 0.21) $\times$ log$_{10}$(S$_\text{850}$) + (1.78 $\pm$ 0.13).
When restricting the sample to sources brighter than 4 (or 5) mJy, the median SFR increases to 192 (296) M$_{\odot}$ yr$^{-1}$. 
However, lower SNR sources do not significantly impact the estimated physical properties. For instance, sources with SNR $> 4\sigma$ exhibit median SFR and $M_\text{star}$ of 175 M$_\odot$ yr$^{-1}$ and $1.57 \times 10^{11}$ M$_\odot$, while lower-SNR sources (3.5-4$\sigma$) show values of 130 M$_{\odot}$ yr$^{-1}$, 1.39 $\times$ 10$^{11}$ M$_{\odot}$.


We adopt the main sequence relation from \citet{speagle2014}. We classify the SMGs exhibiting star formation rates three times above this relation as starburst systems, while those with values three times below are categorized as quenched systems (see \citetalias{paperiv}  of this series).
SMGs located within the main sequence have median SFR and $M_\text{star}$ values of 176 M$_\odot$ yr$^{-1}$ and $1.77 \times 10^{11}$ M$_\odot$, respectively. 
The `starburst system' exhibit higher SFR but lower galaxy masses, with median values of SFR $\sim$ 510 M$_\odot$ yr$^{-1}$ and $M_\text{star}$ $\sim$  $2.85 \times 10^{10}$ M$_\odot$.
In contrast, the `quench system' are on average more massive, with a median $M_\text{star}$ of $3.12 \times 10^{11}$ M$_\odot$ — approximately twice the average of full sample — but a median SFR of only 28 M$_\odot$ yr$^{-1}$.

Overall, the SFR of SMGs evolves strongly with redshift, decreasing by nearly an order of magnitude over a $\sim$ 1 Gyr period when the universe was 2–3 Gyr old. Though this behaviour of increasing SFR ($L_\text{IR}$) with redshift is very likely influenced by a well-know selection effect \citep{miettinen2017, dud2020}. Fitting sources within $z < 4$, we obtain $d(\log \text{SFR})/dz = 0.70 \pm 0.04$. 
In contrast, the stellar mass $M_\text{star}$ distribution remains relatively stable over a broad redshift range ($z > 1.5$), with a median mass of $\sim 10^{11}$ M$_\odot$. 
At redshift above $z \sim 3$, SMGs exhibit median SFR $\gtrsim 700$ M$_\odot$ yr$^{-1}$ and stellar mass $\gtrsim 10^{11}$ M$_\odot$. In the redshift range $z =$ 1.5-3, median SFR decrease to approximately $200$ M$_\odot$ yr$^{-1}$ and stellar mass remains $2 \times 10^{11}$ M$_\odot$, respectively.

Following cosmic noon, massive galaxies gradually exhaust their gas reservoirs through duty cycles of intense star formation, transitioning out of the `SMG phase' toward quiescent evolution. 
As a result, the number density of massive SMGs declines sharply at $z \lesssim$ 2, rendering them increasingly difficult to detect. 
Concurrently, a notable reversal emerges at $z =$ 1–2 (see Figure~\ref{fig:fig5-2-1_props}): lower-mass SMGs exhibit elevated SFRs and begin to dominate cosmic star formation activity. 
This intriguing phenomenon, known as the ``downsizing" effect \citep{cowie1996}, where star formation shifts from massive galaxies at earlier epochs to lower-mass systems at later times.
Further discussion of these trends will be presented in the subsequent paper of this series (see \citetalias{paperiv}).

Below $z \sim 1$, 850 $\micron$ SMGs become increasingly challenging to detect,  with the observed number density decline rapidly.
Their median SFR and $M_\text{star}$ decline to $\sim$ 10 M$_{\odot}$ yr$^{-1}$ and $3 \times 10^{10}$ M$_{\odot}$, respectively. Cosmic star formation activity progressively shifts to other galaxy populations.

%
%

\subsubsection{Mass-to-light Ratio}

\begin{figure*}
    \centering
    \includegraphics[width=1\textwidth]{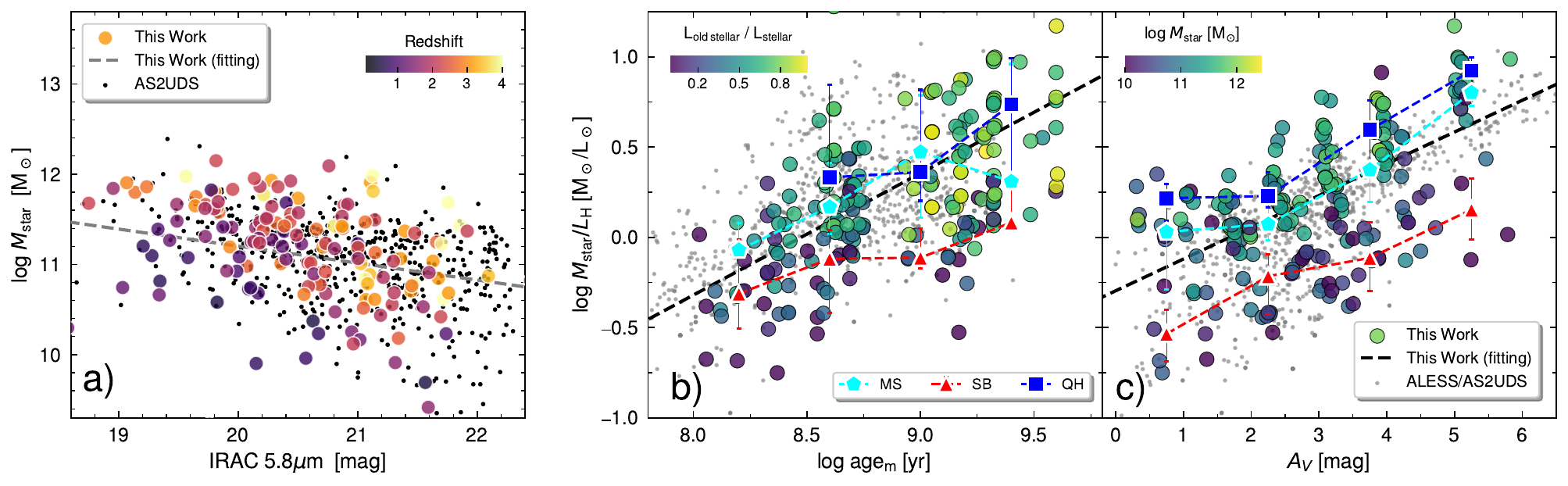}
    \caption{Panel (a) displays the relation between stellar mass and observed 5.8 $\micron$ magnitude, which corresponds approximately to the rest-frame H-band.
    Panel (b) and (c) illustrate how the stellar mass-to-H-band luminosity ratio ($M_{\text{star}}/L_\text{H}$) varies with mass-weighted age and dust attenuation, respectively.
    As galaxies age, the proportion of optical emission from older stellar population increases, leading to higher $M_{\text{star}}/L_\text{H}$ values. Dust attenuation also significantly influences this ratio, with higher extinction reducing observed luminosity and elevating age inferred mass-to-light ratio. More massive and evolutionarily mature galaxies systematically exhibit larger $M_{\text{star}}/L_\text{H}$.  }
    
    \label{fig:fig5-2-2_mlh}
\end{figure*}

At typical redshifts for SMGs, the IRAC 5.8 $\micron$ band corresponds approximately to the rest-frame H-band, which serves as a reliable tracer of the dominant stellar population—low-mass stars \citep{dud2020}. As shown in Figure~\ref{fig:fig5-2-2_mlh}, the 151 SMGs with 5.8 $\micron$ detections follow a distribution and trend consistent with the AS2UDS sample: brighter 5.8 $\micron$ emission corresponds to higher stellar mass. A linear fit yields: 
$\text{log} \, M_{\text{star}}$ [M$_{\odot}] = (-0.19 \pm 0.06) \times m_{5.8}$ [mag] + (14.97 $\pm$ 1.16).

The mass-to-light ratio characterizes the relationship between stellar mass and luminosity. 
We derive the rest-frame 1.6 $\micron$ luminosity from the best-fit SED for each source and estimate the H-band mass-to-light ratio, $M_{\text{star}}/L_\text{H}$. 
Our sample exhibits a median $M_{\text{star}}/L_\text{H}$ of $1.61 \pm 0.13$ M$_\odot$/L$_\odot$, consistent with values from ALESS and AS2UDS.

\citet{cunha2015} suggest that mass-to-light ratio correlates with stellar mass, as the most massive galaxies tend to host older stellar populations and/or experience higher dust attenuation, resulting in larger mass-to-light ratio values. 
This implies a connection between galaxy mass and evolutionary state—more massive galaxies generally exhibit larger mass-to-light ratios.
For instance, SMGs on the star-forming main sequence and those in quiescent phases show similar distributions, with relatively high stellar masses and elevated mass-to-light ratios, having median values of $\sim$ 1.61 and 3.83 M$_\odot$/L$_\odot$, respectively.
In contrast, SMGs undergoing intense starburst activity—dominated by recently formed, young stars—exhibit significantly lower mass-to-light ratios, with a median $M_{\text{star}}/L_\text{H}$ $\sim$ 0.75 M$_\odot$/L$_\odot$,

Furthermore, $M_{\text{star}}/L_\text{H}$ shows strong correlations with both mass-weighted age (age$_\text{m}$) and dust attenuation ($A_V$) (Figure~\ref{fig:fig5-2-2_mlh}). The linear relations are:
log $M_{\text{star}}/L_{\text{H}} = (0.67 \pm 0.06) \times \text{log age}_{\text{m}} + (-5.72 \pm 0.56)$, 
log $M_{\text{star}}/L_{\text{H}} = (0.18 \pm 0.01) \times A_{V} + (-0.30 \pm 0.04)$.

Our median mass-weighted age, $566 \pm 211$ Myr, is consistent with that reported in AS2UDS. Older galaxies have a greater proportion of their light emitted by older stars, leading to a higher $M_{\text{star}}/L_\text{H}$ due to the lower luminosity efficiency of evolved stars. This trend underscores that not only the presence of low-mass stars, but also increased stellar evolutionary age, reduces luminous efficiency and raises the mass-to-light ratio.

Although the H-band is less affected by dust attenuation, dust obscuration still significantly influences the apparent $M_{\text{star}}/L_\text{H}$. Among SMGs with $A_V > 3.09$ mag (the sample average), the median $M_{\text{star}}/L_\text{H}$ is $\sim 2.87$ M$_\odot$/L$_\odot$, compared to $\sim 1.17$ M$_\odot$/L$_\odot$ for sources with lower extinction.


\subsection{Infrared Luminosity}
\label{sec:lir}

\begin{figure*}
    \centering
    \includegraphics[width=0.9\textwidth]{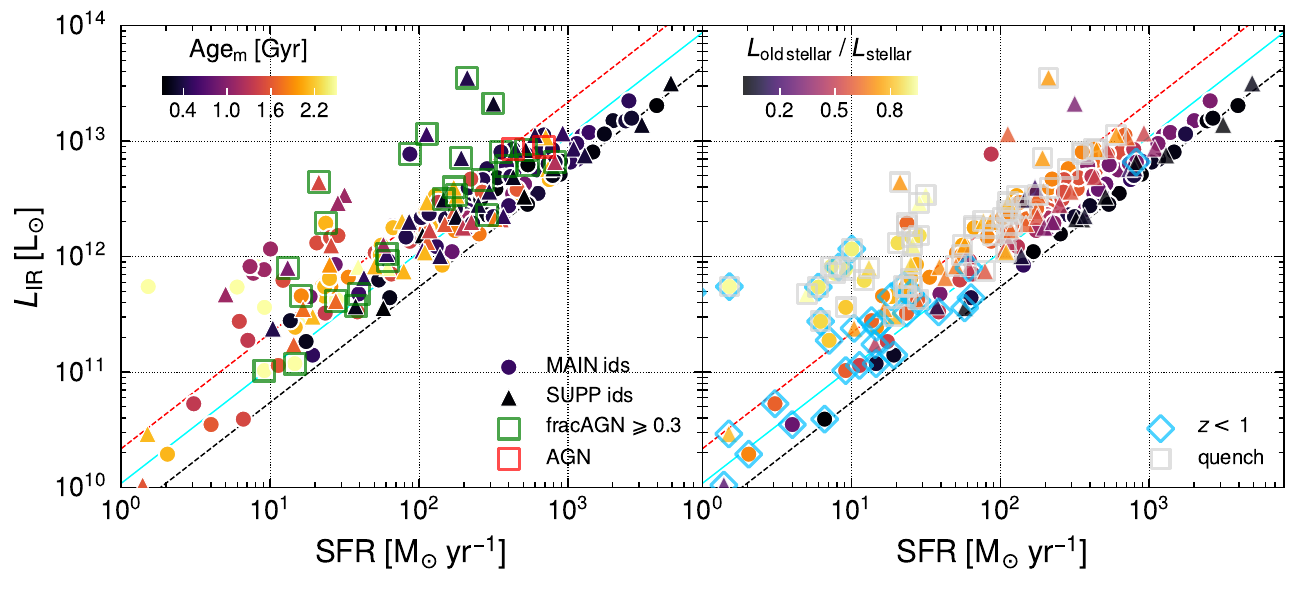}
    \caption{The infrared luminosity versus SFR relation for SMGs. 
    Both quantities trace star formation activity, exhibiting a tight linear correlation: SFR [M$_{\odot}$ yr$^{-1}$] $\simeq$ 1.09 $\times$ 10$^{-9}$ $L_\text{IR}$ [L$_{\odot}$] under an assumption of Chabrier IMF. The cyan solid line delineates the relation, with dashed lines enclosing the 0.3 dex scatter.
    We note that certain SMGs exhibit infrared luminosities significantly elevated above this relation.
    A portion of these outliers can be attributed to AGN contribution, while others correspond to older galaxies where the observed ``shift" in the $L_\text{IR}$-SFR relation is explained by excess infrared emission from evolved stellar populations.
    The green rectangles denote sources with an AGN contribution fraction $\geqslant$ 0.3 as derived from SED fitting. The red rectangles highlight two confirmed X-ray AGNs: AGN7 and ADF22.A1 (numbered following \citealp{alexander2016, monson2023}).  
    In the right panel, blue and gray symbols mark sources at $z <$ 1 and quenching systems, respectively.    }
    
    \label{fig:fig5-3-1_lir}
\end{figure*}

The median infrared luminosity ($L_\text{IR}$, integrated over 8–1000 $\micron$ in the rest frame) of the sample is $(2.25 \pm 0.25) \times 10^{12}$ L$_\odot$, with a 1$\sigma$ range of $0.46$–$6.93 \times 10^{12}$ L$_\odot$, and the most luminous source reaches $\gtrsim 3 \times 10^{13}$ L$_\odot$ (lower-left panel of Figure~\ref{fig:fig5-2-1_props}). %
While slightly lower than literature values, the infrared luminosities of SSA22 SMGs remain comparable in order of magnitude to previous studies.

Similar to the SFR results discussed earlier, when we conservatively restrict our sample to sources with $z >$ 1 that have radio or 24 $\micron$ counterparts, the median $L_\text{IR}$ is 3.24 $\times 10^{12}$ L$_\odot$.
For comparison, selecting AS2UDS sources that have multiwavelength counterparts and satisfy $z >$ 1 yields a median infrared luminosity of 2.88 $\times 10^{12}$ L$_\odot$. Applying the same redshift cut to the S2TDF sample gives a median value of 3.63 $\times 10^{12}$ L$_\odot$, in excellent agreement with our result.

The majority of the SSA22 sample consists of ultra-luminous infrared galaxies (ULIRGs, $L_{\text{IR}} \sim 10^{12-13}$ L$_\odot$) or hyper-luminous infrared galaxies (HLIRGs, $L_{\text{IR}} \sim 10^{13-14}$ L$_\odot$), with $\sim$ 63\% ($\sim$ 8\%) classified as ULIRGs (HLIRGs). At $z > 2$, SMGs predominantly fall into  ULIRG category, while the rare, exceptionally bright HLIRGs primarily appear before $z > 2.5$, hosting some of the most intense star formation episodes in the universe.
The remaining quarter of sources are luminous infrared galaxies (LIRGs, $L_{\text{IR}} \sim 10^{11-12}$ L$_\odot$). These become increasingly prevalent at $z \lesssim 2$ and represent nearly all SMGs below $z \sim 1$.

Due to the tight correlation between infrared luminosity and star formation activity, $L_\text{IR}$ exhibits a significant evolution with the cosmic star formation history in Figure~\ref{fig:fig5-2-1_props}. 
\citet{dud2020} found that over the redshift range $z = 0.5$–4, $L_\text{IR}$ scales approximately as $(1+z)^4$, but they argue this trend may be biased by the dependence of SED fitting on FIR data. 
Nevertheless, it is evident that the most luminous SMGs show a clear decline in brightness: HLIRGs are predominantly found at $z \gtrsim 2.5$, and at $z < 1$, SMG infrared luminosities typically fall below $10^{12}$ L$_{\odot}$.

$L_\text{IR}$ is excellent indicator of star formation activity, and always shows a tight correlation with SFR, follows the relation (scaled to \citet{chabrier2003} IMF): SFR [M$_{\odot}$ yr$^{-1}$] $\simeq$ 1.09 $\times$ 10$^{-10}$ $L_\text{IR}$ [L$_{\odot}$] \citep{kennicutt1998} , as shown by the cyan solid line in Figure~\ref{fig:fig5-3-1_lir}. Our sample exhibits a scatter of approximately 0.3 dex around this relation.

We note several outliers with higher luminosities than predicted. 
At SFR $\gtrsim 100$ M$_\odot$ yr$^{-1}$, deviations from this relation occur in systems with elevated AGN fractions ($\geqslant 0.3$, derived from \texttt{CIGALE} fitting), with AGN luminosities typically 0.5–3 times, and up to 9 times, the dust luminosity. 
Coloring the data by mass-weighted age (age$_\text{m}$), we find that these AGN-dominated systems tend to be younger systems.

However, we note that sources residing in the large-scale structure, and those with significant AGN contributions do not systematically deviate from the infrared–SFR relation. 
We mark two confirmed X-ray AGNs (AGN7 and the quasar ADF22.A1; \citealp{lehmer2009a, alexander2016, monson2023, umehata2025b}), yet both lie well within the linear range of the relation between infrared luminosity and SFR. This is consistent with the findings of \citet{monson2023}, who concluded that the majority of AGN-hosting SMGs are indistinguishable from normal star-forming galaxies.

We also identify a group of counterparts with SFRs around 10 M$_\odot$ yr$^{-1}$ that lie significantly below the relation. Most sources at $z <$ 1 do not show strong deviations; instead, the outliers are predominantly quiescent identifications. These typically exhibit mass-weighted stellar ages of $\gtrsim$ 1–1.5 Gyr and a high fractional contribution from old stellar populations to the infrared luminosity.
As noted by \citet{cunha2015, dud2020}, MAGPHYS allows heating of dust by older stellar populations, which can produce excess infrared emission—particularly in galaxies with high mass-weighted ages. This effect may contribute to the apparent offset of these quiescent systems from the canonical IR–SFR relation.

Finally, we find that sources with multiple counterparts do not systematically lie outside the expected relation. Given the low multiplicity fraction in our sample, we conclude that flux misallocation due to source blending does not significantly bias our results.

Therefore, we believe that both AGN activity and older stellar populations can cause galaxies to deviate from the canonical $L_\text{IR}$–SFR relation. AGN-dominated SMGs tend to be younger and more luminous, whereas systems with significant old stellar populations are generally older and located around $z \sim 1$.
%
If these outliers are excluded, the median $L_\text{IR}$ and SFR of the remaining sample are $2.80 \times 10^{12}$ L$_\odot$ and 253 M$_\odot$ yr$^{-1}$, respectively.


\subsection{Dust and Dust fraction}
\label{sec:dust}

Dust in galaxies originates from various processes including supernova ejection of massive star, stellar wind from evolved star, and infalling of intergalactic medium (IGM). While the dust is destructed through supernova shocks, feedback from stellar and AGN \citep{liao2024}. 
Dust grains efficiently absorb ultraviolet to near-infrared radiation and re-emit this energy in the infrared, it serves as a crucial tracer of star formation activity in DSFGs and a key constraint for galaxy evolution models. The dust-to-stellar mass fraction, $\mu_\text{dust}$, reflects the balance between dust production and destruction, and thus provides important insights into feedback processes in galaxies \citep{dud2021, liao2024}.

The median dust mass ($M_\text{dust}$) for SSA22 SMGs is $(1.95 \pm 0.14) \times 10^9$ M$_\odot$, with the 16th–84th percentile range of $0.54$–$5.31 \times 10^9$ M$_\odot$. Over 73\% of the sources contain at least $10^9$ M$_\odot$, with the most extreme reaching $\gtrsim 3 \times 10^{10}$ M$_\odot$. 

The dust mass estimates for our sample appear systematically higher than those reported in previous studies \citep{cunha2015, dud2020, hyun2023}.
We attribute these differences to the distinct dust models employed: \texttt{MAGPHYS} utilizes modified blackbody spectra to model warm and cold dust components, whereas our \texttt{CIGALE} implementation adopts the \textit{dl2014} model, which accounts for dust heating by both star-forming regions and the diffuse interstellar radiation field.
This interpretation is supported by \citet{liao2024}, who report that \texttt{CIGALE} yields dust masses $\sim$ 1.3–1.5 times greater than \texttt{MAGPHYS} estimates.
Furthermore, the inclusion of shorter FIR bands in SED fitting introduces systematic biases. Wavelength coverage near the far-infrared emission peak preferentially samples warmer dust components, thereby reducing sensitivity to colder dust and potentially underestimating total dust mass \citep{tacconi2018}. Although this effect will be less at high redshifts \citep{tacconi2018}, studies suggest a potential underestimate by a factor of 1.2–2 \citep{birkin2021}.
Additionally, contamination during the identification process and potential misallocation of flux may also introduce uncertainties (Section~\ref{sec:caveat}).
While methodological variations exist in both approach and results, they do not materially impact subsequent analyses.

In general, at fixed infrared luminosity, SMGs with higher 850 $\micron$ flux densities tend to exhibit greater dust masses and lower temperatures \citep{swinbank2014, cunha2015}. 
As noted by \citet{dud2020, hyun2023}, the 850 $\micron$ flux is tightly correlated with cold dust mass. 
We derive the following empirical relation for our sample:  
log$_{10}$($M_\text{dust}$ [$M_\odot$]) = $(0.76 \pm 0.12) \times$ log$_{10}$(S$_{850}$ [mJy]) + $(8.85 \pm 0.07)$.  
However, despite the correlation between $S_{850}$ and $M_\text{dust}$, and between $M_\text{dust}$ and $\mu_\text{dust}$, no significant correlation is found between $\mu_\text{dust}$ and $S_{850}$.

Sources with log$_{10}$($M_\text{dust}$/M$_\odot$) $< 9$ predominantly have log$_{10}$(SFR/M$_\odot$ yr$^{-1}$) $\gtrsim 2.2$, while those with log$_{10}$(SFR/M$_\odot$ yr$^{-1}$) $\lesssim 2.0$ generally exhibit log$_{10}$($M_\text{dust}$/M$_\odot$) $\gtrsim$ 9.2, suggesting an apparent anti-correlation between $M_\text{dust}$ and SFR. However, given the strong evolution of SFR with redshift, this trend likely reflects the cumulative buildup of dust mass over time rather than a direct physical dependence.
Overall, $M_\text{dust}$ (and $\mu_\text{dust}$) shows only a weak correlation with SFR (thus with $L_\text{IR}$), and no clear relation with V-band attenuation ($A_V$). 
This suggests that dust geometry and heating efficiency—rather than total dust mass—are more critical in determining the observed infrared luminosity of SMGs, as compact dust configurations yield higher temperatures and luminosities \citep{cunha2015}.
Additionally, the diversity of SED and associated uncertainties in dust mass estimates further weaken any significant correlation between $M_\text{dust}$ (or $\mu_\text{dust}$) and $L_\text{IR}$/SFR.

\subsubsection{Dust fraction}

\begin{figure*}
    \centering
    \includegraphics[width=0.9\textwidth]{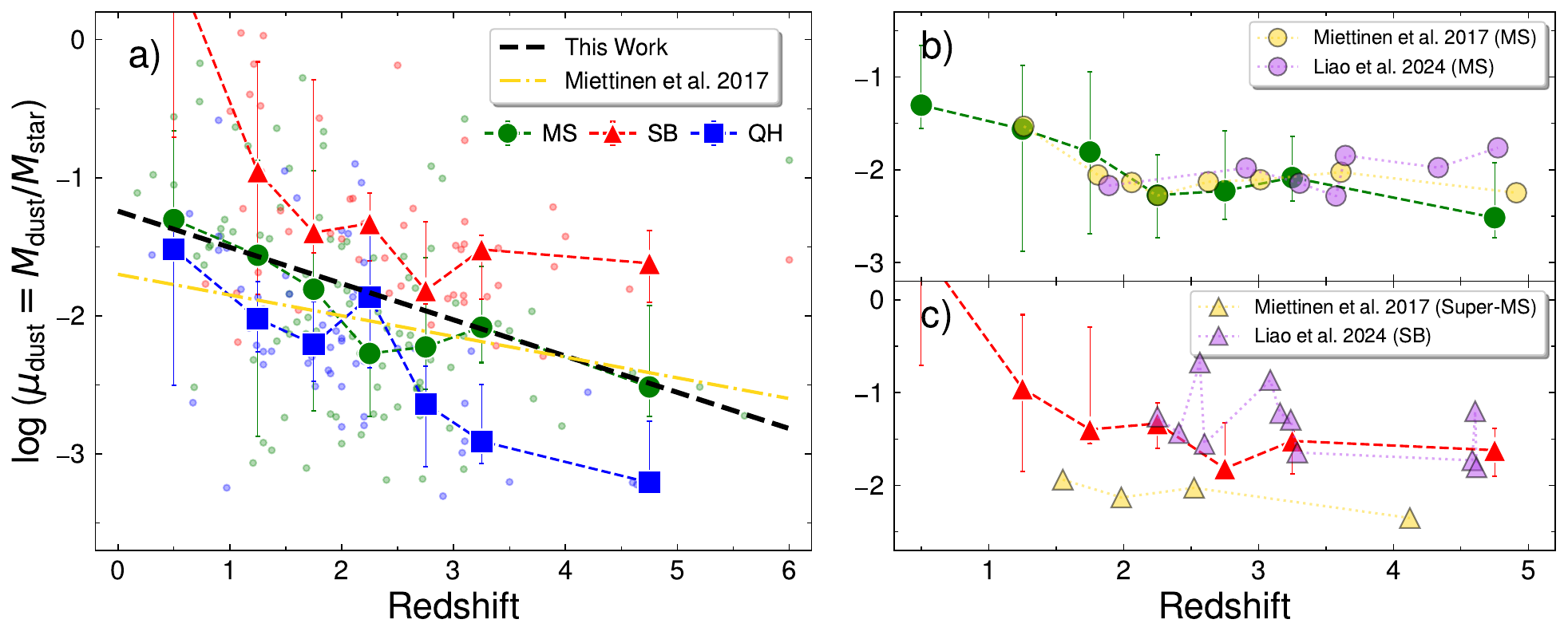}
    \caption{Panel (a) illustrate the evolution of the dust mass fraction $\mu_\text{dust}$ with redshift, showing a general decreasing trend.
    Significant disparities are observed among galaxies in different evolutionary phases, with systems undergoing more intense star formation exhibiting higher dust fraction. We suggest that this pattern is partially driven by the distinct stellar mass distributions characterizing SMGs at different evolutionary stages.
    A noticeable turnover in the dust fraction of main-sequence and quenched SMGs occurs at $z \sim 2-2.5$, which we attribute to the ``downsizing" effect.
    Panel (b) and (c) compare dust fractions from different SMG surveys.
    When accounting for incompleteness in high-redshift samples, the dust fraction of SSA22 SMGs at $z \gtrsim 4$ appears to remain flat, consistent with values reported in the literature \citep{miettinen2017, liao2024}.}
    
    \label{fig:fig5-4-1}
\end{figure*}

We define the dust mass fraction as $\mu_{\text{dust}} = M_{\text{dust}}/M_{\text{star}}$. The median $\mu_\text{dust}$ for the sample is $(1.4 \pm 0.2) \times 10^{-2}$, with a 16th–84th percentile range of $2.8 \times 10^{-3}$–$8.6 \times 10^{-2}$.

Figure~\ref{fig:fig5-4-1} left panel  illustrates the redshift evolution in $\mu_\text{dust}$: median values are $7.6 \times 10^{-3}$ at $z = 2.5–4$, $1.1 \times 10^{-2}$ at $z = 1.5–2.5$, and $3.7 \times 10^{-2}$ at $z = 0.5–1.5$. A linear fit to the full sample yields log $\mu_{\text{dust}} = (-0.26 \pm 0.05) \times z - (1.24 \pm 0.12)$.
The slope is steeper than that reported by \citet{miettinen2017} for ALMA-detected SMGs in the COSMOS field. We believe this discrepancy arise from the incompleteness of high-redshift sources in our sample. When restricting our analysis to $z < 4$, the trend becomes consistent with previous results: $\mu_\text{dust}$ decreases at $z < 2$ and remains approximately flat at $z \gtrsim 2$ \citep{miettinen2017, liao2024}.

The dust mass fraction $\mu_\text{dust}$ correlates with the evolutionary state of galaxies. While main-sequence SMGs follow a trend consistent with the overall sample, starburst and quenched systems deviate by approximately 0.5 dex.
We note that SMGs in different evolutionary states possess distinct stellar mass distributions, indicating that the apparent correlation between $\mu_\text{dust}$ and evolutionary phase is partially driven by its underlying relationship with stellar mass.

A notable transition in $\mu_\text{dust}$ occurs at $z \sim 2$–2.5 for main-sequence SMGs (Figure~\ref{fig:fig5-4-1} (b)), in good agreement with previous studies. Concurrently, the average $\mu_\text{dust}$ of quenched systems shows an opposite trend, increasing across the same redshift range.
We interpret this behavior as a signature of ``downsizing'': a population of intermediate-mass, dust-rich main-sequence galaxies gradually transitions to quiescence at $z \sim 2-2.5$, causing a decline in the dust content of the main-sequence sample while increasing the average $\mu_\text{dust}$ of the quiescent population. Over time, lower-mass starburst galaxies transition onto the main sequence, leading to a recovery in the dust fraction of the main-sequence population. This evolution proceeds dynamically, by $z \sim 1$, most intermediate-mass galaxies in the 850 $\micron$ field have become quiescent, and their dust content rises to levels comparable to those of main-sequence galaxies.
Figure~\ref{fig:fig5-4-1} panel (c)) shows that the $\mu_\text{dust}$ values for starburst SMGs in SSA22 are higher than those reported by \citet{miettinen2017}, but consistent with \citet{liao2024}, particularly when accounting for differences in the definition of starburst galaxies.

\begin{figure}
    \centering
    \begin{minipage}{\columnwidth}
    \includegraphics[width=0.9\textwidth]{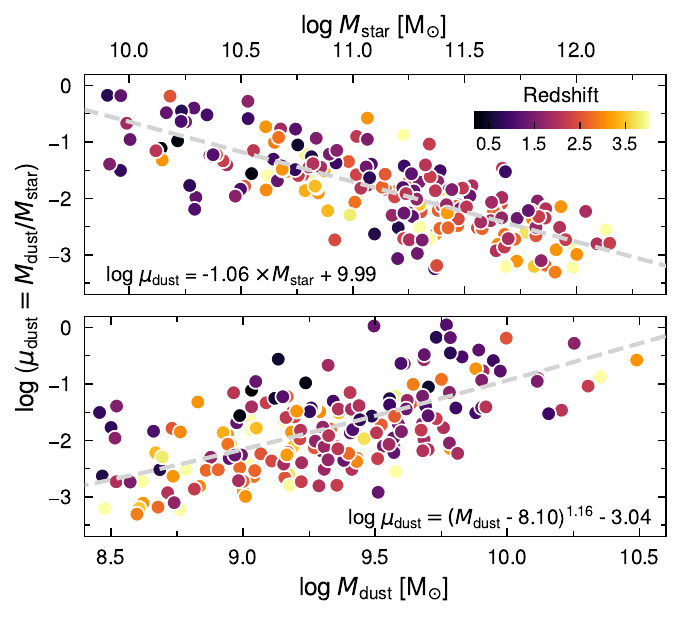}
    \caption{The dust fraction of SMGs exhibits a strong correlation with both stellar mass and dust mass.
    The observed redshift-dependent evolution in dust fraction is primarily driven by variations in $M_\text{star}$ and $M_\text{dust}$ across cosmic time.
    The evolution of stellar mass is governed by the ``downsizing" scenario, wherein more massive galaxies complete their star formation earlier.
    Concurrently, the evolution of dust mass is modulated by the increase in gas-phase metallicity \citep{bethermin2015, liao2024} and shifts in the balance metal enrichment, grain growth, and dust destruction processes\citep{miettinen2017, dud2021}.}
    \label{fig:fig5-4-2}
    
    \end{minipage}
\end{figure}

We find strong correlations between $\mu_\text{dust}$ and both $M_\text{star}$ and $M_\text{dust}$ (Figure~\ref{fig:fig5-4-2}):  
log$_{10}$$\mu_{\text{dust}} = (-1.06 \pm 0.05) \times \log_{10}M_{\text{star}} + (9.99 \pm 0.58)$,  
log$_{10}$$\mu_{\text{dust}} = (\log_{10}M_{\text{dust}} - 8.10 \pm 0.43)^{1.16 \pm 0.10} - (3.04 \pm 0.48)$.  
Dust fractions increase with decreasing stellar mass, with median $\mu_\text{dust}$ of 0.005, 0.014, and 0.047 for $M_{\text{star}} = 10^{11.4-12}$, $10^{10.9-11.4}$, and $10^{10-10.9}$ M$_\odot$, respectively.
Moreover, at fixed stellar mass, galaxies at lower redshifts consistently exhibit higher dust fractions, further supporting the picture of dust content evolving with cosmic time.

We suggest that the observed trend of decreasing $\mu_\text{dust}$ with redshift is jointly driven by the increase in dust mass and the decrease in stellar mass at lower redshifts.
The decreasing evolve of stellar mass arise from the ``downsizing'' scenario.
The increasing dust is deeply influenced by rising gas-phase metallicity \citep{bethermin2015, liao2024}, and shifts in the balance between metal enrichment, grain growth, and dust destruction \citep{miettinen2017, dud2021, liao2024}. 
The gas-phase metallicity increases toward lower redshift \citep{bethermin2015}, implying that $M_{\text{dust}}$ ($\propto Z_{\text{gas}} \times M_{\text{gas}}$) and dust formation efficiency also rise in the same direction \citep{miettinen2017, liao2024}. 
Additionally, the rate of dust accumulation regulated by cold gas surface density, while AGN feedback potentially suppressing dust growth by reducing gas density \citep{liao2024}.

\subsection{Dust attenuation}
\label{sec:av}

\begin{figure*}[!ht]
    \centering
    \includegraphics[width=0.9\textwidth]{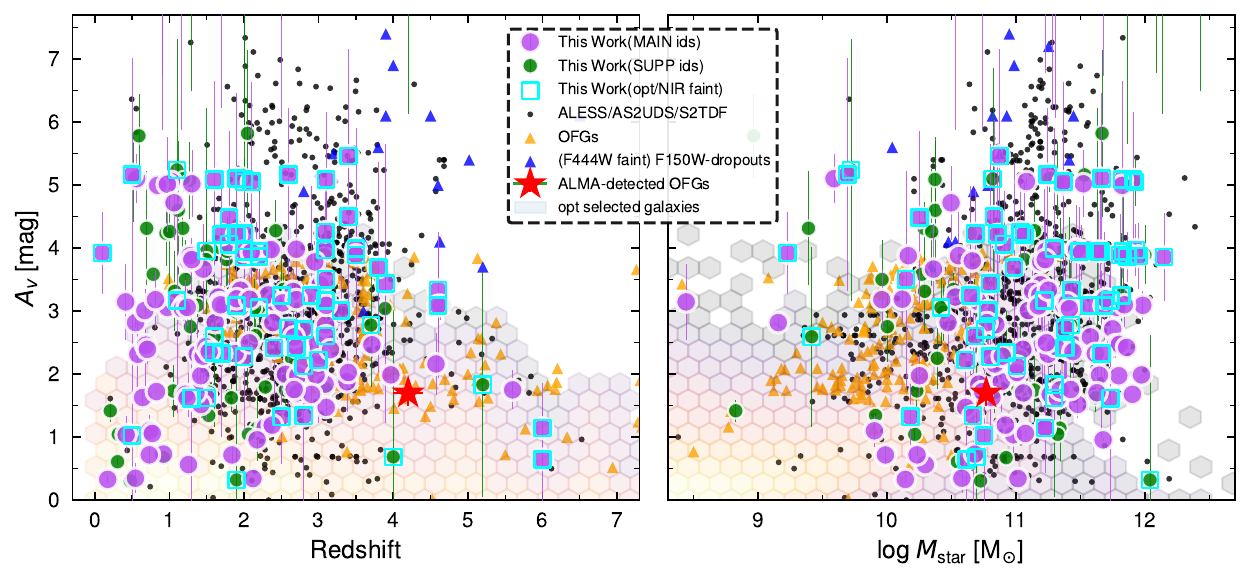}
    \caption{Dust attenuation as a function of redshift and stellar mass.
    The attenuation of our sample align with previous studies and show no significant variation with redshift or stellar mass.
    We also compare our results with the ALMA survey \citep[black dots;][]{cunha2015, dud2020, hyun2023}, OFG samples \citep[red triangles;][]{got2024, williams2024}, F150W-dropouts \citep[blue triangles;][]{man2025}, ALMA-detected OFGs \citep[red star;][]{xiao2023} and UV-selected galaxies (transparent hexagons).
    OFGs display a distribution similar to SMGs, representing a population of massive, dusty systems with moderate attenuation levels. 
    A notable break in attenuation properties occurs at $z \sim 4$ across galaxy populations–including SMGs, OFGs, and UV-bright systems–suggesting a transition in dominant star-formation modes from unobscured to dust-obscured activity.
    This shift may reflect the large-scale accumulation of metallicity in the interstellar medium during this epoch.}
    
    \label{fig:fig5-5-1}
\end{figure*}

In the rest frame, ultraviolet to near-infrared photons from galaxies are attenuated by dust in both birth clouds and the diffuse interstellar medium. 
However, the estimated $A_V$ represents an average attenuation across the entire galaxy and does not account for the spatial distribution of stars and dust, and the actual attenuation become increasing severe at shorter wavelengths. Thus, $A_V$ should be regarded as a lower limit to the true attenuation \citep{dud2020}. Nevertheless, $A_V$ remains a useful indicator for assessing the level of dust obscuration in these highly active DSFGs.

We measure the median $A_V$ for the sample to be $3.09 \pm 0.07$ mag, with 68\% of sources lying within the range 1.69–4.44 mag. On average, this corresponds to a factor of $\sim$20 attenuation in the optical and near-infrared bands. Combined with the $k$-correction at high redshift, this implies that constructing a mass-complete sample based solely on UV/optical observations is highly challenging, as heavily obscured galaxies are likely to be missed \citep{wang2019, dud2020}.

Our median $A_V$ and its distribution are consistent with those reported for AS2UDS \citep[$\sim 2.9$ mag, ][]{dud2020} and S2TDF \citep[$\sim 3.3$ mag, ][]{hyun2023}, and slightly higher than SCUBADive \citep[$\sim 2.6$ mag, ][]{mck2025}, although the overall distributions remain comparable. 
Our value is also higher than that of ALESS\citep[$\sim$ 1.9 mag, ][]{cunha2015}, which may be due, as suggested by \citet{dud2020}, to cosmic variance leading to an underrepresentation of highly obscured SMGs at $z < 3$ in their sample.

The median $A_V$ values for all VLA counterparts, 24 $\micron$ counterparts, 8 $\micron$ counterparts, and color-selected counterparts are 3.10, 2.65, 3.06, and 3.13 mag, respectively. Since most 8 $\micron$ and color-selected counterparts are also identified in the VLA or 24 $\micron$ samples—and radio-detected sources dominate the overall counterpart population—their dust attenuation remains consistent with that of the radio-identified sample.
Color-selected galaxies do indeed exhibit redder colors and higher dust attenuation. Considering only those color-selected galaxies without counterparts from other identification methods, the median $A_V$ rises to 3.82 mag. 
However, this result could be influenced by contamination from lower-redshift sources. When restricted to $z > 1$, the median $A_V$ values for the respective counterpart samples become 3.10, 3.04, 3.08, and 3.09 mag.

Dividing the sample into optically bright and faint sources (based on whether the number of detected optical/NIR bands exceeds four), we find the median $A_V$ values are $2.82 \pm 0.20$ mag (bright) and $3.87 \pm 0.24$ mag (faint), respectively. As expected, optically faint sources experience significantly stronger dust obscuration, attenuating V-band light by a factor of $\sim 35$ on average—about three times more than bright sources.

The distribution of $A_V$ with redshift (Figure~\ref{fig:fig5-5-1}) shows no significant evolutionary trend. 
We confirm that the `MAIN' and `SUPP' subsamples exhibit statistically indistinguishable $A_V$ distributions, with no significant differences between them (Table~\ref{tab:tab5_average_properties}).
We note that SMGs' attenuation decrease significantly at $z \gtrsim 4$, the median $A_V$ is $\sim 2.00$, consistent with other SMG surveys (e.g., AS2UDS, S2TDF), which also show a decline in $A_V$ with increasing redshift—reaching $A_V \sim 1$ at $z \sim 6$.

Recent studies have analyzed high-redshift optically faint/dark galaxies (OFGs), which are defined as H-band dropouts and comparable optically faint populations \citep{williams2024}. 
These investigations reveal that the majority of OFGs are extended dusty star-forming systems, with only a minor fraction being quiescent galaxies \citep{barrufet2025}. Their SFRs typically range from 10 to 100 M$_\odot$ yr$^{-1}$,  while stellar masses are approximately 1-2 dex lower than those of SMGs \citep{got2024, williams2024}. 
These findings highlight OFGs as a critical population bridging extreme DSFGs and the broader population of massive galaxies, warranting further investigation.

We compare the sample with those OFGs. Although OFGs generally have lower SFRs, stellar masses, and $A_V$ compared to SMGs, they are still more dust-obscured, more actively star-forming, and more massive than typical optically selected galaxies.
The OFG sample of CEERS and GOODS-S field (orange triangles in Figure~\ref{fig:fig5-5-1}; \citealp{got2024,williams2024}) has an average $A_V$ of 2.7 mag, rising to 3.1 mag at $z < 4$, but decreasing significantly at $z > 4$ to a median of 1.9 mag. This behavior closely resembles the trend seen in SMGs. 
Interestingly, JWST-selected optical galaxies\footnote{https://s3.amazonaws.com/aurelien-sepp/primer-cosmos-east-grizli-v7.0/catalog/primer-cosmos-east-grizli-v7.0\_morpho-phot.fits.gz} (transparent hexagons in Figure~\ref{fig:fig5-5-1}) at $z = 4$–5 exhibit a similar trend. 


Not only SMGs, but also OFGs and UV-selected galaxies exhibit a similar trend. If this observed trend is not an artifact of sample selection or small-number statistics, we interpret it as evidence for a fundamental shift in the dominant mode of cosmic star formation—from an unobscured to a dust-obscured regime \citep{bourne2017, dunlop2017, bouwens2020, zavala2021}. The underlying physical driver may be the gradual buildup of metallicity in galaxies over cosmic time, which eventually reaches a threshold enabling efficient dust production and obscuration (see Section~\ref{sec:dust}).

\citet{simpson2014, dud2020} found that OFGs in the ALESS and AS2UDS samples tend to lie at higher redshifts.  
\citet{smail2021} noted that, despite having dust masses and stellar masses comparable to those of other SMGs, these near-infrared–faint SMGs exhibit more compact dust emission and higher star formation rate surface densities. This suggests that such systems may be linked to the formation of compact, spheroidal structures observed in high-redshift galaxies.

The GOODS-S field OFGs sample \citep{xiao2023} has a median redshift of $\sim 4.3$ and relatively low average $A_V$ ($\sim 0.7–1.7$ mag). Among these, sources detected by ALMA (red star Pattern in Figure~\ref{fig:fig5-5-1}) exhibit higher $A_V$, dust mass, and SFR, along with more abundant and colder dust, suggesting that more heavily obscured systems tend to have colder dust temperatures.
In contrast, \citet{man2025} identify a population of MIR-faint 1.5 $\micron$-dropouts (i.e., F444W faint F150W-dropouts) in the COSMOS field (blue triangles in Figure~\ref{fig:fig5-5-1}),
characterized by high redshifts and substantial dust attenuation, with median values of $z \approx 3.95$ and $A_V \approx 5.45$ mag, respectively. Most of these sources lie on the star-forming main sequence.
However, due to the lack of FIR and (sub)millimeter constraints, estimated physical properties for such sources  are subject to non-negligible uncertainties \citep{williams2024, mck2025}.

The brightest 850 $\micron$ SMGs ($S_{850} \gtrsim 8$ mJy) have an average $A_V \sim 4.55 \pm 0.87$ mag, which seem suggesting a link between high submillimeter flux and extreme dust obscuration. However, this may reflect small-number bias, as no significant correlation is found between $A_V$ and $S_{850}$ across the full sample. In the large AS2UDS sample, the brightest SMGs show $A_V \gtrsim 2$ mag but remain consistent with the median $A_V$.
Galaxies in starburst or quiescent phases exhibit $A_V$ values close to the sample average, while main-sequence SMGs show slightly lower attenuation ($A_V \sim 2.72 \pm 0.25$ mag). 
Although more massive SMGs (log$_{10}$($M_\text{star}$/M$_{\odot}$) $\gtrsim 11$) tend to have higher $A_V$ (typically $\gtrsim 1$ mag), and those with $-1 < \log_{10}\mu_{\text{dust}} < 0$ have a slightly lower median $A_V$ ($\sim 1.97 \pm 0.72$ mag), we find no significant correlation between $A_V$ and stellar mass, SFR, dust mass, or dust fraction.

The lack of correlation with key physical properties or star formation state suggests that dust geometry—rather than intrinsic galaxy properties—dominates the observed attenuation.

\subsubsection{Attenuation and Optical Color}

\begin{figure}
    \centering
    \begin{minipage}{\columnwidth}
    \includegraphics[width=0.9\textwidth]{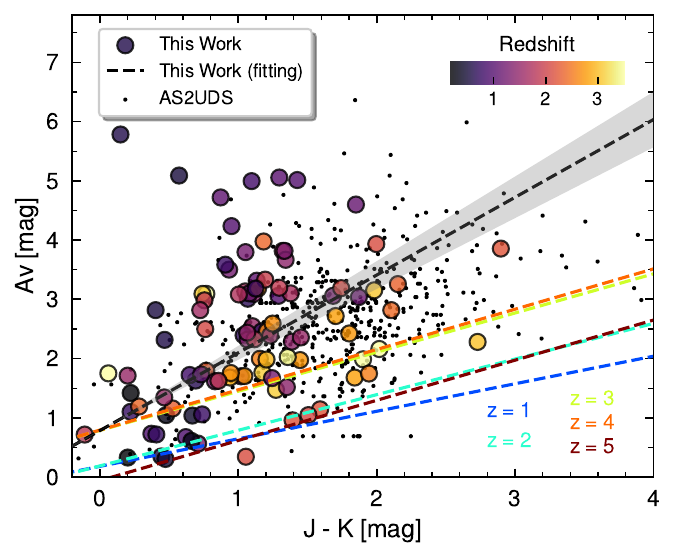}
    \caption{The relationship between dust attenuation and ($J-K$) color is examined, where this observed color corresponds approximately to the rest-frame ($U-R$) band.
    Our sample is color-coded according to redshift, and the black dashed line represents the best-fit relation when fix the intercept, with the grey shaded area indicating the 1$\sigma$ uncertainty range.
    The colored dashed lines depict predictions from the Calzetti reddening law at various redshifts, which systematically underestimate the observed attenuation.
    These results indicate tat a single reddening law derived from local UV-bright star-forming galaxies is insufficient to describe the dust attenuation properties of high-redshift SMGs, suggesting the need for steeper attenuation curves or alternative dust grain properties in these systems.  }
    \label{fig:fig5-5-2}
    
    \end{minipage}
\end{figure}

In this section, we briefly examine the relationship between $A_V$ and optical colors in SMGs. 
Given their substantial dust attenuation, we focus on the correlation between ($J–K$) color and $A_V$. 

Among the 104 sources in our sample with detections in both $J$ and $K$ bands, the ($J–K$) color ranges from 0 to 2 mag. Although the UDS deep field enables AS2UDS to probe significantly redder sources—up to ($J-K$) $\sim$ 3–4 mag—our ($J-K$) distribution is in good agreement with AS2UDS.
The median redshift and $A_V$ for the $J/K$-detected SMGs in SSA22 are 1.53 and 2.37 mag, respectively, while the remaining sources, as expected, exhibit higher median redshift and $A_V$ (2.16 and 3.86 mag). 
These sources undetected in the $J$ or $K$ bands likely occupy the regime of higher $A_V$ and redder ($J-K$) colors in the color–magnitude space.

At the median redshift of our sample ($z \sim 2$), the $J$- and $K$-bands correspond approximately to the rest-frame $U$- and $R$-bands, respectively.
The attenuation in the rest-frame $U$- and $R$-bands, $A_U$ and $A_R$ can be expressed as:
\[
    A_U = E(B-V) \times k(U) = \dfrac{A_V}{R_V} \times k(U),  
\]
\[
    A_R = E(B-V) \times k(R) = \dfrac{A_V}{R_V} \times k(R).
\]
From the color-excess relation:
\[
    E(U-R)=(U-R)-(U-R)_0 = A_U - A_R,
\]
we derive the linear relation between observed color and A$_V$:
\begin{equation}
    (U-R)=(U-R)_0 + (\dfrac{k(U)-k(R)}{R_V}) \times A_V.
\end{equation}
where (U-R) is the observed color, (U-R)$_0$ is the intrinsic color, $k(\lambda) = \dfrac{A_{\lambda}}{E(B-V)} = R_V \dfrac{A_{\lambda}}{A_V}$, describes the reddening law.

For redshifts $z =$ 1–4, we compute the rest-frame wavelengths corresponding to $J$ and $K$ bands and estimate the intrinsic $(\lambda_1-\lambda_2)_0$ and $(k(\lambda_1)-k(\lambda_2))$ using an average stellar spectrum.
As highly star-forming galaxies, SMGs have strong intrinsic optical emission, resulting in modest intrinsic colors: $(\lambda_1-\lambda_2)_0 \sim 0.2–0.8$, with a median of 0.78. Assuming the Calzetti extinction law \citep{calzetti2000}, $(k(\lambda_1)-k(\lambda_2))$ ranges from 0.5 to 0.7, with a value of 0.60 at $z =$ 2.

However, as shown in Figure~\ref{fig:fig5-5-2}, the observed $A_V$ values for SMGs are systematically higher than those predicted by the Calzetti law, indicating that this low-redshift prescription does not accurately describe the dust attenuation in high-redshift SMGs.
The Calzetti reddening law is derived from low-redshift UV-bright star-forming galaxies, which differ fundamentally from the dust-rich SMGs. 
Furthermore, \citet{menendez2009} note that SMGs exhibit more extended star formation and dust distributions compared to local starbursts and ULIRGs, leading to different extinction properties.

Additionally, SMGs are typically selected based on flux in specific bands, resulting in a heterogeneous population \citep{hayward2011, johnson2013}. 
For instance, galaxies selected in this way span over two orders of magnitude in both stellar mass and SFR, exhibit a wide range of dust masses and spatial distributions, and encompass diverse evolutionary states.
The sample commonly includes both merging or interacting systems and isolated galaxies, with some hosting AGN while others do not.
This diversity is evident in their SEDs and radio properties (see \citetalias{paperiv} for the discussions of SED and IR-radio correlation). Attenuation curves can vary significantly between different galaxy types, and even within a single galaxy, they may differ substantially along different lines of sight \citep{uematsu2024}. This intrinsic diversity further limits the applicability of a single, universal extinction law to the SMG population.

As noted earlier, the intrinsic rest-frame colors $(\lambda_1-\lambda_2)_0$ of SMGs are relatively modest; fixing this value at the median of 0.78 and fitting the data yields a slope of $(k(\lambda_1)-k(\lambda_2))/R_V \sim 1.31$, corresponding to $(k(\lambda_1)-k(\lambda_2)) \sim 5.3$ \citep[for $R_V$ = 4.05;][]{calzetti2000}.
This implies that highly obscured high-redshift star-forming galaxies require significantly steeper attenuation in the blue optical regime than predicted by the Calzetti law. A steeper reddening curve—relative to the canonical Calzetti form—is necessary to reproduce the observed dust attenuation in SMGs.
This enhanced attenuation may be attributed to the fact that newly formed stars in SMGs are still embedded within dense molecular clouds, leading to more severe extinction of their strong UV emission, consistent with the two-component extinction model employed in our SED fitting (Section~\ref{sec:sedfitting}). 
However, the substantial scatter in the data also highlights the diverse SED properties of SMGs and/or variations in dust geometry, which may collectively produce a wide range of dust attenuation patterns.


\section{Summary}
In this work, we present a multiwavelength analysis of 850 $\micron$ SMGs in the SSA22 deep field.

Using four identification method (Figure~\ref{fig:fig3-0_ids}), we associate 248 SMGs ($\gtrsim 84\%$ with a deblended 850 $\micron$ flux density of $\geqslant 1$ mJy) with 192 SCUBA-2 sources, of which 136 counterparts to 101 SCUBA-2 sources are reported here for the first time (Section~\ref{sec:ids}). We confirm the effectiveness of identification techniques based on IRAC 8 $\micron$-band and optical/NIR color selection (Figure~\ref{fig:fig3-2-4_reliablity}). The average multiplicity of SCUBA-2 sources is $\sim$ 26\%, with brighter sources exhibiting higher multiplicity (Figure~\ref{fig:fig3-3_multiplicity}).

We apply the ``super-deblending" technique (Section~\ref{sec:super_deblend}, Appendix~\ref{sec:app_spdb}) to obtain deblended FIR and submillimeter fluxes, and combine these with panchromatic photometry from optical to radio wavelengths to perform SED fitting using \texttt{CIGALE} (Section~\ref{sec:sedfitting}). After applying quality cuts based on SED fitting reliability, our final sample comprises 221 SMGs ($\gtrsim 89\%$ with a deblended S$_{\rm 850}$ of $\geqslant 1$ mJy) associated with 186 SCUBA-2 sources. Our main results are summarized as follows:

\begin{itemize}
\item[1.] 
The median redshift of SSA22 SMGs is $z = 2.0 \pm 0.08$, with consistent distributions across identification methods (Figure~\ref{fig:fig5-1-1_redshift_distribution}). 
This is slightly lower than literature values, likely due to incompleteness in our multiwavelength data at the highest redshifts.
The comoving volume density of SMGs increases by a factor of $\sim 6$ at $z \lesssim 4$, plateauing at $\sim$ 1.78-3.16 $\times$ 10$^{-5}$ cMpc$^{-3}$ over $z \sim 1-3$.
Newly identified SMGs show a similar redshift distribution (median 1.9) to the overall sample. 
Approximately two-thirds of sources are classified as optically bright (median redshift $\sim 1.75$), while optically faint sources exhibit higher median redshift ($\sim 2.2$), with completely undetected optical/NIR sources reaching $\sim 3.0$.
Brighter 850 $\micron$ SMGs are preferentially found at higher redshifts.

\item[2.]
SMGs in large-scale structures show significant overdensity $\sim 8\sigma$ (Figure~\ref{fig:fig5-1-1_redshift_distribution} image (c)), demonstrating their reliability as traces of cosmic structure formation at high redshift.
Additionally, the SMG overdensity exceeds blank field average by more than an order of magnitude (Section~\ref{sec:over}; Figure~\ref{fig:fig5-1-1_redshift_distribution} image (e)).

\item[3.]
The median stellar mass and SFR of our SMG sample are $(1.55 \pm 0.22) \times 10^{11}$ M$_\odot$ and $166 \pm 25$ M$_\odot$ yr$^{-1}$, with 16-84\% ranges of $0.24-4.77 \times 10^{11}$ M$_\odot$ and $18-703$ M$_\odot$ yr$^{-1}$, respectively (Figure~\ref{fig:fig5-2-1_props}).
Newly detected SMGs have comparable stellar masses but lower SFRs, consistent with their fainter submillimeter fluxes.
Starburst-phase SMGs exhibit high SFRs and relatively lower stellar masses, whereas quenched systems are more massive but have suppressed star formation.

\item[4.]
The stellar mass distribution of SMGs remains remarkably stable across cosmic time, while SFR evolution remains debated due to selection effects \citep{miettinen2017, dud2020}. 
We observe a clear ``downsizing" signature: at early epochs ($z \gtrsim$ 2), both high- and low-mass SMGs undergo intense star formation. After cosmic noon ($z \lesssim 2$), massive SMGs exhaust their gas reservoirs and transition to quiescence, while lower-mass SMGs continue forming stars and dominate the cosmic SFR density.

\item[5.]
Older galaxies exhibit higher mass-to-light ratios due to increased contributions from evolved stellar populations and reduced luminosity efficiency (Figure~\ref{fig:fig5-2-2_mlh}).
Dust attenuation and the presence of low-mass stars also significantly affect $M_{\text{star}}/L_\text{H}$. 
Quiescent SMGs show systematically higher ratios than star-forming ones.

\item[6.]
The median infrared luminosity is (2.25 $\pm$ 0.25) $\times$ 10$^{12}$ L$_{\odot}$, with 68\% range of 0.46-6.93 $\times$ 10$^{12}$ L$_{\odot}$ (Figure~\ref{fig:fig5-2-1_props}).
Approximately $\sim$ 63\% ($\sim$ 8\%)  of the sample are classified as ULIRGs (HLIRGs). 
A clear decline in brightness is observed among the most luminous SMGs, indicating significant evolution of infrared luminosity with cosmic time.
At $z > 2$, nearly all SMGs are ULIRGs, while HLIRGs primarily appear before $z \sim 2.5$.
Although $L_\text{IR}$ is tightly correlated with SFR, deviations from this relation can occur: AGN activity in young SMGs and excess emission from evolved stellar populations in older systems can drive $L_\text{IR}$ significantly above the canonical relation (Figure~\ref{fig:fig5-3-1_lir}).

\item[7.]
The median dust mass is (1.95 $\pm$ 0.14) $\times$ 10$^{9}$ M$_{\odot}$ (16-84\% range of 0.54-5.31 $\times$ 10$^{9}$ M$_{\odot}$), slightly higher than literature values–possibly due to differences in dust modeling assumptions.
The 850 $\micron$ flux strongly traces cold dust mass, whereas $L_\text{IR}$ is governed more by dust geometry, compactness, and heating efficiency than by total dust mass or dust fraction.

\item[8.]
The median dust fraction is (1.4 $\pm$ 0.2) $\times$ 10$^{-2}$, with 68\% value range of $2.8 \times 10^{-3} - 8.6 \times 10^{-2}$.
Dust fraction decrease at $z < 2$ but remains roughly flat at $z \gtrsim 2$ (Figure~\ref{fig:fig5-4-1}). This evolution is shaped by the combined redshift trends of dust and stellar mass (Figure~\ref{fig:fig5-4-2}).
Dust mass growth is regulated by rising gas-phase metallicity, cold gas surface density, and the balance between grain growth and destruction. Stellar mass evolution follows the ``downsizing" scenario, which also drives a turnover in $\mu_\text{dust}$ at $z \sim 2-2.5$, with decreases for main-sequence SMGs and increases for quiescent systems.

\item[9.] The median $A_V$ is 3.09 $\pm$ 0.07 mag (68\% range: 1.69-4.44 mag). Optically bright and faint SMGs have $A_V = 2.82 \pm 0.20$ mag and $3.87 \pm 0.24$ mag, respectively.
While $A_V$ shows no strong evolution over $z \sim 1-4$, we detect a significant drop in attenuation at $z > 4$–not only for SMGs, but also for OFGs and UV-selected galaxies (Figure~\ref{fig:fig5-5-1}). 
This phenomenon possibly reflecting a transition from unobscured to dust-obscured star formation modes \citep{bourne2017, dunlop2017, bouwens2020, zavala2021}, potentially driven by metallicity accumulation thresholds.

\item[10.] $A_V$ shows no clear correlation with stellar mass or evolutionary state, suggesting that dust geometry–not just total dust content–dominates attenuation.
Dust attenuation with observed ($J-K$) color, but the scatter and offset from predictions based on the Calzetti law indicate that a single, local reddening law is insufficient for high-redshift SMGs (Figure~\ref{fig:fig5-5-2}). Their diverse SEDs and composite nature further complicate attenuation modeling.

\end{itemize}


\begin{acknowledgements}
\section*{Acknowledgments}
We sincerely thank Ian Smail for his constructive comments and insightful suggestions, which significantly improved the rigor and clarity of this manuscript.

Y.A. acknowledges the support from the National Key R\&D Program of China (2023YFA1608204), the National Natural Science Foundation of China (NSFC grant 12173089), the China Manned Space Program with grant no. CMS-CSST-2025-A10, and the Strategic Priority Research Program of the Chinese Academy of Sciences (grant No.XDB0800301).

K.M. acknowledges the Waseda University Grant for Special Research Projects (Project number: 2025C-484).

The James Clerk Maxwell Telescope is operated by the East Asian Observatory on behalf of The National Astronomical Observatory of Japan; Academia Sinica Institute of Astronomy and Astrophysics; the Korea Astronomy and Space Science Institute; the National Astronomical Research Institute of Thailand; Center for Astronomical Mega-Science (as well as the National Key R\&D Program of China with No. 2017YFA0402700). Additional funding support is provided by the Science and Technology Facilities Council of the United Kingdom and participating universities and organizations in the United Kingdom and Canada. Additional funds for the construction of SCUBA-2 were provided by the Canada Foundation for Innovation.

This work is based on observations taken with the JCMT under project codes M15AI91 and MJLSC02. The MJLSC02 data are part of the SCUBA-2 Cosmology Legacy Survey.
This research has made use of the NASA/IPAC Infrared Science Archive, which is funded by the National Aeronautics and Space Administration and operated by the California Institute of Technology.
This research has made use of the SVO Filter Profile Service “Carlos Rodrigo”, funded by MCIN/AEI/10.13039/501100011033/ through grant PID2023-146210NB-I00.

\end{acknowledgements}

%
%

\vspace{5mm}
\facilities {JCMT(SCUBA-2), JVLA(S-band), ASTE(AzTEC), ALMA(band 6),
CFHT(Megaprime), Subaru(Suprime, MOIRCS and HSC), UKIRT(WFCAM), 
IRSA, $Spitzer$(IRAC and MIPS), $Herschel$(SPIRE) }


\vspace{5mm}
\software {SExtractor \citep{bertin1996}, PyBDSF \citep{mohan2015}, EAZY \citep{brammer2008}, CIGALE v2022.1 \citep{boquien2019}, dustmap \citep{green2018}, galfit \citep{peng2010}, astropy \citep{astropy2013, astropy2018, astropy2022}, pandas \citep{reback2020pandas}, numpy \citep{harris2020array}, matplotlib \citep{Hunter2007}, scipy \citep{2020SciPy-NMeth,mckinney-proc-scipy-2010}, }




\appendix
\clearpage


\section{Optical/Near-infrared Data Reduction}
\label{sec:app_opt_reduction}

We use \texttt{SExtractor} \citep{bertin1996} to detect sources and perform photometry on the optical/near-infrared images. The full set of configuration parameters is listed in Table~\ref{tab:tab6_se_conf}.
This section primarily references \citet{bertin1996, sedummy} and the \texttt{SExtractor} official documentation \footnote{https://sextractor.readthedocs.io/en/latest/}.

Source detection and photometry are carried out on sky-background-subtracted images, following the background estimation procedure described in \citet{bertin1996, sedummy}. \texttt{SExtractor} divides the image into a grid of cells with size defined by \texttt{BACK\_SIZE}. Within each cell, it iteratively computes the median and standard deviation of pixel values, rejecting outliers beyond $\pm 3\sigma$ from the median until convergence. The final background value for each cell is determined based on local source density: in sparse regions—identified when the change in $\sigma$ during iteration is less than 20\%—the mean is adopted; otherwise, the mode is estimated as $2.5 \times median - 1.5 \times mean$. A median filter of size \texttt{BACK\_FILTERSIZE} is then applied to suppress residuals from bright or extended objects. Finally, a full-resolution background map is generated via spline interpolation and subtracted from the original image.

To improve photometric accuracy, we set \texttt{BACKPHOTO\_TYPE = LOCAL} and specify the annulus thickness with \texttt{BACKPHOTO\_THICK}. This allows \texttt{SExtractor} to estimate the local background residual from a rectangular ring surrounding each source and apply a correction to aperture photometry.

The RMS (background noise) map is constructed simultaneously from the per-cell standard deviations. We identify candidate sources as connected groups of at least 5 pixels exceeding $2\sigma$ above the local background \citep{hayashino2004, umehata2014}. The final source catalog is further refined by requiring that the peak pixel of each source exceeds $3\sigma$. During detection, we employ the default conical convolution kernel to enhance the detectability of faint sources, although \citet{sedummy} suggests that a top-hat kernel may be more effective for very low-surface-brightness objects.

Each detected source is represented as a hierarchical tree of nested isophotal levels (typically $\sim$30, set by \texttt{DEBLEND\_NTHRESH}). Deblending proceeds top-down: at each branching node with threshold $i$, the object is split if (1) the flux in one branch exceeds a fraction (\texttt{DEBLEND\_MINCONT}) of the total flux, and (2) at least one other branch also satisfies condition (1) \citep{bertin1996}. Additionally, \texttt{SExtractor} performs source “cleaning” by evaluating whether neighboring objects would still be detected if the current source were removed—a critical step for recovering faint sources near bright stars or galaxies.

For photometry, we adopt \texttt{MAG\_AUTO} as our fiducial flux estimator \citep{uchimoto2012}. This method implements an adaptive Kron-like elliptical aperture that robustly captures the majority of a source’s flux across a wide dynamic range, even in crowded fields or in the presence of nearby contaminants. It effectively balances systematic and random photometric errors and provides a reliable estimate of the total magnitude \citep{bertin1996, sedummy}, making it the standard choice in extragalactic surveys.

The \texttt{MAG\_AUTO} algorithm is based on the ``first-moment" technique introduced by \citet{kron1980}. An initial ellipse is defined from the second moments of the light distribution and scaled by a factor of 6 along both axes. Within this enlarged region, the Kron radius is computed as
\[
r_{\text{Kron}} = \frac{\sum r I(r)}{\sum I(r)}.
\]
A Kron factor $k$ is then applied to define the photometric aperture semi-axes as $k r_{\text{Kron}} / \epsilon$ and $\epsilon k r_{\text{Kron}}$, where $\epsilon$ is the ellipticity derived from the initial moment analysis. For sources with Gaussian-like profiles (e.g., stars and typical galaxies), \citet{kron1980} and \citet{infante1987} showed that $k = 2$ encloses $\sim$90\% of the total flux, nearly independent of brightness. Increasing $k$ to 2.5 raises this fraction to $\sim$94\%, albeit at the cost of reduced signal-to-noise. We adopt $k = 2.5$ to maximize flux recovery while accepting a modest decrease in photometric precision.

To prevent underestimation of flux for faint sources in high-noise regions, we enforce a minimum Kron radius. Furthermore, to mitigate contamination from overlapping or nearby bright/extended sources, we set \texttt{MASK\_TYPE = CORRECT}, which replaces contaminated pixels within the aperture with symmetric, uncontaminated counterparts from the opposite side of the source center.

\begin{deluxetable}{lll}
\tablecaption{SExtractor Configuration}
\label{tab:tab6_se_conf}

\tablewidth{0pt}

\addtocounter{table}{0}    
\tablehead{
    \colhead{Procedure} & \colhead{Parameter} & \colhead{Value} }

\startdata
Background & BACK\_TYPE       & AUTO          \\
           & BACK\_SIZE       & 64            \\
           & BACK\_FILTERSIZE & 3             \\
           & BACKPHOTO\_TYPE  & LOCAL         \\
           & BACKPHOTO\_THICK & 24            \\
\hline
Detection  & THRESH\_TYPE     & RELATIVE      \\
           & DETECT\_TYPE     & CCD           \\
           & DETECT\_MINAREA  & 5             \\
           & DETECT\_THRESH   & 2             \\
           & FILTER          & Y               \\
           & FILTER\_NAME     & default.conv  \\
           & DEBLEND\_NTHRESH & 32            \\
           & DEBLEND\_MINCONT & 0.005         \\
           & CLEAN           & N               \\
           & CLEAN\_PARAM     & 1.0           \\
           & MASK\_TYPE       & CORRECT       \\
\hline
Photometry & PIXEL\_SCALE     & 0.2           \\
           & PHOT\_AUTOPARAMS & $2.5$, $3.5$  \\
           & PHOT\_AUTOAPERS  & $0.0$, $0.0$  \\
           & SATUR\_LEVEL     & 50000.0       \\
           & SATUR\_KEY       & SATURATE      \\
\enddata

\end{deluxetable}

\section{Super-deblending}
\label{sec:app_spdb}
\begin{figure}
  \centering
  \begin{minipage}{\columnwidth}
    \includegraphics[width=\textwidth]{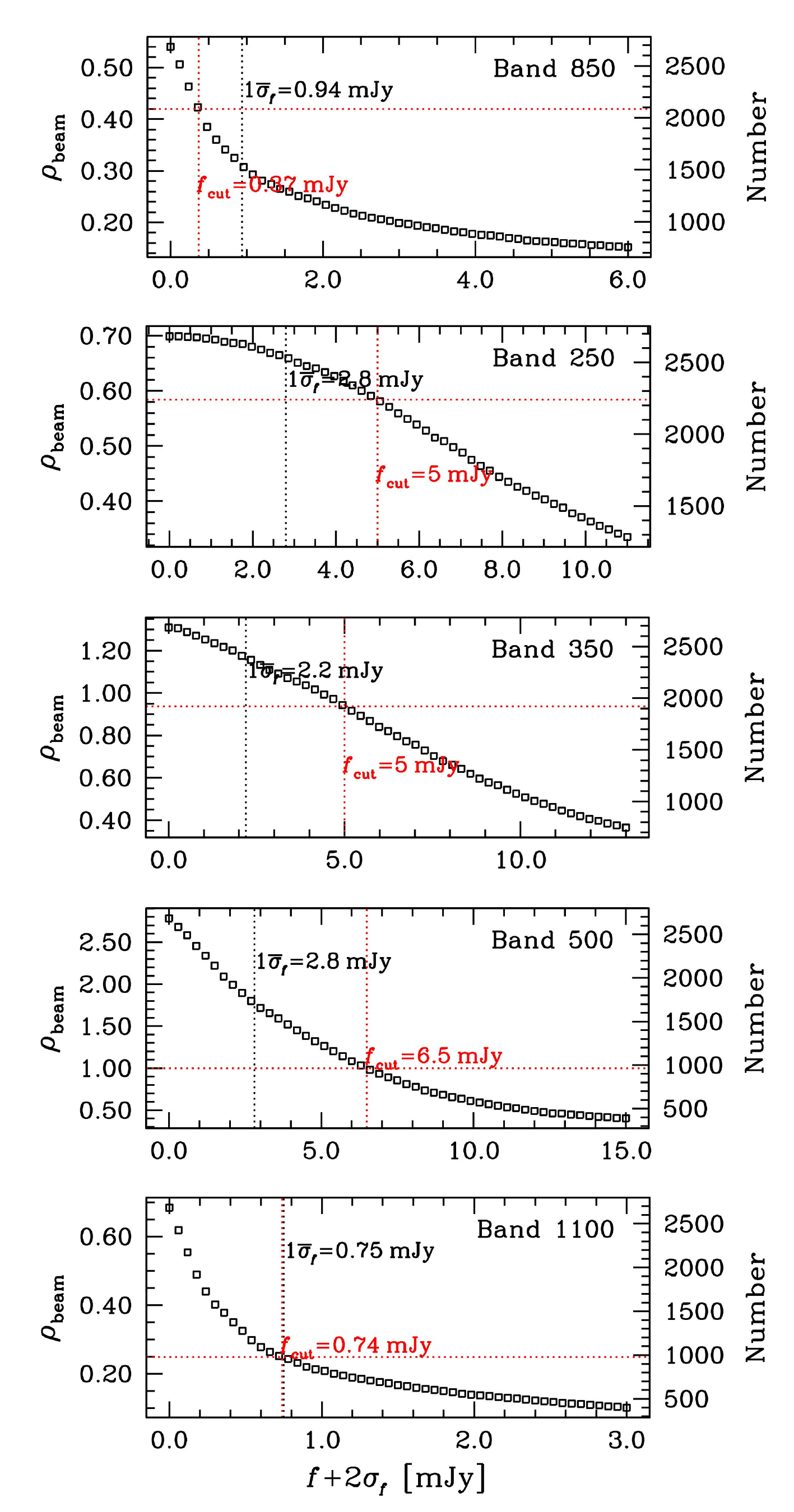}
    \caption{The cumulative number density of prior sources as a function of flux threshold for each band.
    The black vertical line marks the empirical median noise per beam at this band, and the red line indicates the final flux cut $f_{\text{cut}}$.
    To efficiently deblend FIR/(sub)mm sources, we require a moderate prior source density within the beam area ($\rho_{\text{beam}} \lesssim 1$), while ensuring the flux cutting threshold satisfies $1\sigma \lesssim f_{\text{cut}} \lesssim 3\sigma$.
    Priors are selected for PSF fitting and source deblending when their flux density exceeds the threshold, i.e., $f_{\text{SED}}+2\sigma_{\text{SED}} \geqslant f_{\text{cut}}$.
    }
  \end{minipage}
  \label{fig:fig2-5-1_fluxcut}
\end{figure}


\begin{figure*}
    \centering
    \includegraphics[width=0.9\textwidth]{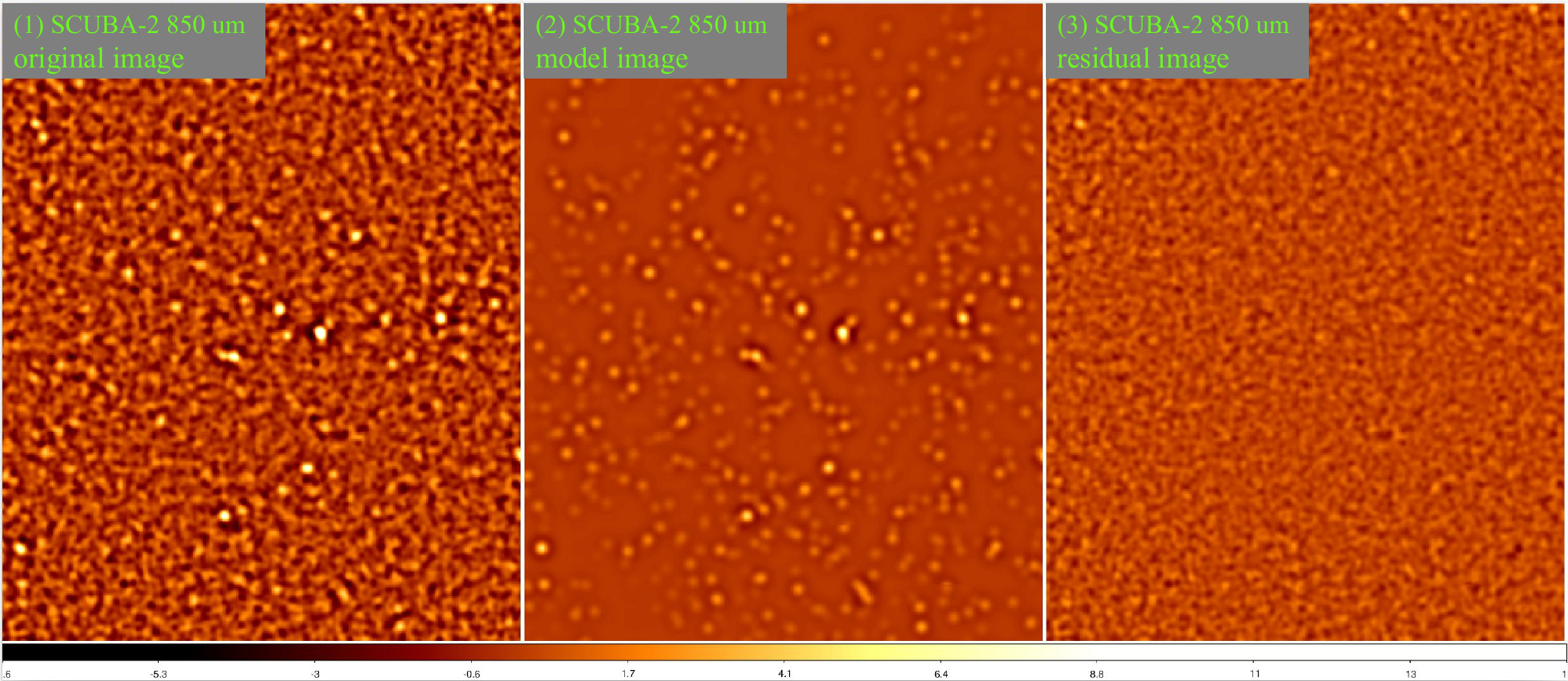}
    \caption{The super-deblending processes is demonstrated using the SCUBA-2  850 $\micron$ images.
    Panel (1) shows the original  850 $\micron$ images. Panel (2) displays the best-fitting PSF models constructed using \texttt{galfit} at the positions of prior sources. Panel (3) presents the residual image obtained after subtracting the model sources.    }
    \label{fig:fig2-5-2_spdb_850}
\end{figure*}

The FIR information of galaxies is crucial for measuring the star formation rates and dust properties of galaxies \citep{liu2018, zavala2018}. However, FIR/mm images have coarse spatial resolution ranging from $\sim$ 14$\arcsec$ - 35$\arcsec$, leading to significant confusion, which not only blurs our observations of the deep universe but also hinders our understanding of the characteristics of high-redshift star-forming galaxies.
To obtain deblended source fluxes from 250 $\micron$ to 1.1 mm, we adopted an innovative technique proposed by \citet{liu2018, jin2018}, ``super-deblended", to deblend these images and obtain accurate FIR fluxes for SMGs. We will provide a brief introduction to the super-deblended process, while detailed principles and steps can be found in \citet{liu2018, jin2018}.

The galaxy main sequence ensures that FIR/mm detected objects always have counterparts in IRAC images \citep{liu2018}. Due to the high sensitivity of \textit{Spitzer}/IRAC imaging and the availability of deeper IRAC advanced productions, we used MIPS 24$\micron$ and radio sources with IRAC-detected counterparts to create a prior source catalog.
We combined the SEIP and SpIES catalogs to obtain the IRAC source catalog and matched it with the VLA 3 GHz catalog. We selected 24 $\micron$ or radio sources with SNR greater than 3 as prior sources. 
Additionally, we utilized ``the representative redshift catalog'' in the SSA22 field compiled by \citet{mawatari2023} and available photometric redshift information (see Section~\ref{sec:eazy}) to identify sources with low SNR ($\geqslant$ 2.5$\sigma$) but located at higher redshifts ($z \textgreater$ 2) as part of the prior sources. The final source catalog includes 2683 ``24+radio" prior sources.

The SED templates employed a combination of four component models, including stellar, mid-infrared AGN, dust continuum, and radio components. Additionally, the SED fitting  also considering starburst populations. We introduce three flags to differentiate the fitting process for different types of galaxies: the MS/SB (main sequence or starburst galaxies) flag, the radio-excess flag, and the FIR photometry high-quality flag. The latter is used to prevent fitting failures when dealing with radio-loud or AGN sources. For detailed template information and the SED template fitting process, please refer to \citet{liu2018}.

The beam area of each FIR/mm image contains a significant number of star-forming galaxies, making it difficult to obtain an effective deblended flux when considering the contribution of all sources within the beam. Therefore, we set a threshold flux S$_{\text{cut}}$, by considering the average density of prior sources, i.e. the number of prior sources within each beam. The threshold flux S$_{\text{cut}}$ is set within the range of 1$\sigma$ to 3$\sigma$, ensuring that the number of fitted sources in each band remains at a reasonable value of about or less than one per PSF beam area, see also Figure~\ref{fig:fig2-5-1_fluxcut}. If the sum of the prior source SED fitting flux and twice the fitting uncertainty, S$_{\text{SED}}$ + 2S$\sigma_{\text{SED}}$, is less than S$_{\text{cut}}$, we exclude the prior source in this band due to the faintness of the sources. We subtract these faint sources from the observed images and then use \texttt{galfit} \citep{peng2010} for photometry measurement of the prior sources in the subtracted images (Figure~\ref{fig:fig2-5-2_spdb_850}).

Before obtaining photometry of the prior sources in the current band using \texttt{galfit}, we need to consider that although the 24+radio catalog provides good FIR deblending priors, there are still sources that contribute to the FIR flux but are not included in the prior catalog due to different galaxy properties. Therefore, for the residual images after prior source extraction, we perform ``blind-extraction'' using \texttt{SExtractor} with a detection threshold set to 2.5-3.0. After visual inspection, these sources are added to the prior source catalog, and photometry measurement is redone using \texttt{galfit} for the FIR observed images.
The deblending process proceeds from the shortest to the longest wavelength band, one band at a time. During the super-deblending process for each band, we flag the prior sources and use the deblended photometry measured in the previous band to optimize the prior sources in the current band, and re-perform FIR/mm SED fitting for each source. This is a crucial step in super-deblended photometry, helping us obtain robust deboosted fluxes for the prior sources.

Finally, we use Monte Carlo simulation to verify whether super-deblended photometry has significant bias and correct the flux based on the simulation results. Furthermore, we provide uncertainty estimates for deblended photometry using a highly optimized and nearly ideal uncertainty estimation method, which is a key advantage of super-deblending approach. In fact, in the presence of highly correlated noise and confusion noise in FIR/mm images, \texttt{galfit} cannot provide accurate photometry uncertainty.
In the faint source-excluded image mentioned above, we inject simulated sources at random positions, simulating one source each time. Then, sources are extracted for the simulated image, and this simulation process is repeated approximately 3000 times. Thus, we obtain input and output fluxes of the simulated sources, and \texttt{galfit} photometric errors. 
Additionally, we measure the local rms noise value of the simulated sources, the local absolute flux density and local scatter in the residual images, and the crowdedness parameter. These simulated parameters are used to correct the bias in super-deblended photometry and the final flux uncertainty.


\bibliography{allbibtext}

\bibliographystyle{aasjournal}



\end{document}